\documentclass[twocolumn]{aastex61}
\pdfoutput=1 
\usepackage{amsmath,amstext}
\usepackage[T1]{fontenc}
\usepackage{apjfonts} 
\usepackage[figure,figure*]{hypcap}

\shorttitle{APOKASC 2}
\shortauthors{Pinsonneault et al.}

\begin{document}

\title{The Second APOKASC Catalog: The Empirical Approach}

\author{Marc H. Pinsonneault}
\affiliation{Department of Astronomy, The Ohio State University, Columbus, OH 43210, USA}
\author{Yvonne P. Elsworth}
\affiliation{University of Birmingham, School of Physics and Astronomy, Edgbaston, Birmingham B15 2TT, UK}
\affiliation{Stellar Astrophysics Centre, Department of Physics and Astronomy, Aarhus University, Ny Munkegade 120, DK-8000 Aarhus C, Denmark}
\author{Jamie Tayar}
\affiliation{Department of Astronomy, The Ohio State University, Columbus, OH 43210, USA}
\author{Aldo Serenelli}
\affiliation{Institute of Space Sciences (IEEC-CSIC), Campus UAB, E-08193 Bellaterra, Spain}
\author{Dennis Stello}
\affiliation{School of Physics, University of New South Wales, NSW 2052, Australia}
\author{Joel Zinn}
\affiliation{Department of Astronomy, The Ohio State University, Columbus, OH 43210, USA}
\author{Savita Mathur}
\affiliation{Instituto de Astrofsica de Canarias (IAC), C/Va Lactea, s/n, E-38200 La Laguna, Tenerife, Spain}
\affiliation{Universidad de La Laguna (ULL), Departamento de Astrof\'isica, E-38206 La Laguna, Tenerife, Spain}
\affiliation{Space Science Institute, 4750 Walnut street, Suite 205, Boulder, CO 80301, USA}
\author{Rafael A. Garc\'ia}
\affiliation{Universit\'e Paris Diderot, AIM, Sorbonne Paris Cit\'e, CEA, CNRS, F-91191 Gif-sur-Yvette, France}
\affiliation{IRFU, CEA, Universit\'e Paris-Saclay, F-91191 Gif-sur-Yvette, France}
\author{Jennifer A. Johnson}
\affiliation{Department of Astronomy, The Ohio State University, Columbus, OH 43210, USA}
\affiliation{Center for Cosmology and Astroparticle Physics, The Ohio State University, Columbus, OH 43210, USA}
\author{Saskia Hekker}
\affiliation{Max-Planck-Institut f\"ur Sonnensystemforschung, Justus-von-Liebig-Weg 3, D-37077 Gottingen, Germany}
\affiliation{Stellar Astrophysics Centre, Department of Physics and Astronomy, Aarhus University, Ny Munkegade 120, DK-8000 Aarhus C, Denmark}
\author{Daniel Huber}
\affiliation{Institute for Astronomy, University of Hawaii, 2680 Woodlawn Drive, Honolulu, HI 96822, USA}
\author{Thomas Kallinger}
\affiliation{Institute for Astronomy, University of Vienna, T\"urkenschanzstrasse 17, A-1180 Vienna, Austria}
\author{Szabolcs M{\'e}sz{\'a}ros}
\affiliation{ELTE E\"otv\"os Lor\'and University, Gothard Astrophysical Observatory, Szombathely, Hungary}
\affiliation{Premium Postdoctoral Fellow of the Hungarian Academy of Sciences}
\author{Benoit Mosser}
\affiliation{LESIA, Observatoire de Paris, PSL Research University, CNRS, Universit\'e Pierre et Marie Curie, Universit\'e Paris Diderot,  92195 Meudon, France}
\author{Keivan Stassun}
\affiliation{Department of Physics and Astronomy, Vanderbilt University, Nashville, TN 37235, USA}
\author{L\'eo Girardi}
\affiliation{Osservatorio Astronomico di Padova, INAF, Vicolo dell Osservatorio 5, I35122 Padova, Italy}
\author{Tha\'ise S. Rodrigues}
\affiliation{Osservatorio Astronomico di Padova, INAF, Vicolo dell Osservatorio 5, I35122 Padova, Italy}
\author{Victor Silva Aguirre}
\affiliation{Stellar Astrophysics Centre, Department of Physics and Astronomy, Aarhus University, Ny Munkegade 120, DK-8000 Aarhus C, Denmark}
\author{Deokkeun An}
\affiliation{Department of Science Education, Ewha Womans University, 52 Ewhayeodae-gil, Seodaemun-gu, Seoul 03760, Korea}
\author{Sarbani Basu}
\affiliation{Department of Astronomy, Yale University, PO Box 208101, New Haven CT, 06520-8101}
\author{William J. Chaplin}
\affiliation{University of Birmingham, School of Physics and Astronomy, Edgbaston, Birmingham B15 2TT, UK}
\affiliation{Stellar Astrophysics Centre, Department of Physics and Astronomy, Aarhus University, Ny Munkegade 120, DK-8000 Aarhus C, Denmark}
\author{Enrico Corsaro}
\affiliation{INAF - Osservatorio Astrofisico di Catania, via S. Sofia 78, 95123 Catania, Italy}
\author{Katia Cunha}
\affiliation{University of Arizona, Tucson, AZ 85719, USA}
\affiliation{Observat\'orio Nacional, S\~ao Crist\'ov\~ao, Rio de Janeiro, Brazil}
\author{D. A. Garc\'ia-Hern\'andez}
\affiliation{Instituto de Astrofsica de Canarias (IAC), C/Va Lactea, s/n, E-38200 La Laguna, Tenerife, Spain}
\affiliation{Universidad de La Laguna (ULL), Departamento de Astrof\'isica, E-38206 La Laguna, Tenerife, Spain}
\author{Jon Holtzman}
\affiliation{Department of Astronomy, MSC 4500, New Mexico State University, P.O. Box 30001, Las Cruces, NM 88003, USA}
\author{Henrik J\"onsson}
\affiliation{Lund Observatory, Department of Astronomy and Theoretical Physics, Lund University, Box 43, SE-22100 Lund, Sweden}
\author{Matthew Shetrone}
\affiliation{University of Texas at Austin, McDonald Observatory, 32 Fowlkes Road, TX 79734-3005, USA}
\author{Verne V. Smith}
\affiliation{National Optical Astronomy Observatories, Tucson, AZ 85719 USA}
\author{Jennifer S. Sobeck}
\affiliation{Department of Astronomy Box 351580, U.W. University of Washington
Seattle, WA 98195-1580}
\author{Guy S. Stringfellow}
\affiliation{Center for Astrophysics and Space Astronomy, Department of Astrophysical and Planetary Sciences, University of Colorado, 389 UCB, Boulder Colorado 80309-0389}
\author{Olga Zamora}
\affiliation{Instituto de Astrofsica de Canarias (IAC), C/Va Lactea, s/n, E-38200 La Laguna, Tenerife, Spain}
\affiliation{Universidad de La Laguna (ULL), Departamento de Astrof\'isica, E-38206 La Laguna, Tenerife, Spain}
\author{Timothy C. Beers}
\affiliation{Dept. of Physics and JINA Center for the Evolution of the Elements
University of Notre Dame, Notre Dame, IN 46556  USA}
\author{J. G. Fern\'andez-Trincado}
\affiliation{Departamento de Astronom\'i a, Casilla 160-C, Universidad de Concepci\'on, Concepci\'on, Chile}
\affiliation{Institut Utinam, CNRS UMR6213, Univ. Bourgogne Franche-Comt\'e, OSU THETA , Observatoire de Besan\c{c}on, BP 1615, 25010 Besan\c{c}on Cedex, France}
\author{Peter M. Frinchaboy}
\affiliation{Department of Physics \& Astronomy, Texas Christian University, Fort Worth, TX 76129, USA}
\author{Fred R. Hearty}
\affiliation{Department of Astronomy and Astrophysics Pennsylvania State University
525 Davey Laboratory University Park PA 16802 USA}
\author{Christian Nitschelm}
\affiliation{Unidad de Astronom\'ia, Facultad de Ciencias B\'asicas, Universidad de Antofagasta, Antofagasta, Chile}

\begin{abstract}
We present a catalog of stellar properties for a large sample of 6676 evolved stars with APOGEE spectroscopic parameters and \textit{Kepler} asteroseismic data analyzed using five independent techniques.  Our data includes evolutionary state, surface gravity, mean density, mass, radius, age, and the spectroscopic and asteroseismic measurements used to derive them. We employ a new empirical approach for combining asteroseismic measurements from different methods, calibrating the inferred stellar parameters, and estimating uncertainties. With high statistical significance, we find that asteroseismic parameters inferred from the different pipelines have systematic offsets that are not removed by accounting for differences in their solar reference values. We include theoretically motivated corrections to the large frequency spacing ($\Delta \nu$) scaling relation, and we calibrate the zero point of the frequency of maximum power ($\nu_{\rm max}$) relation to be consistent with masses and radii for members of star clusters. For most targets, the parameters returned by different pipelines are in much better agreement than would be expected from the pipeline-predicted random errors, but 22\% of them had at least one method not return a result and a much larger measurement dispersion. This supports the usage of multiple analysis techniques for asteroseismic stellar population studies. The measured dispersion in mass estimates for fundamental calibrators is consistent with our error model, which yields median random and systematic mass uncertainties for RGB stars of order 4\%. Median random and systematic mass uncertainties are at the 9\% and 8\% level respectively for RC stars.

\end{abstract}


\section{Introduction}

Stellar astrophysics is in the midst of a dramatic transformation.  We are moving from a domain defined by small, local and disjoint data sets into an era where we have rich time domain information, complemented by  spectroscopic, photometric, and astrometric surveys for large populations of stars across the Milky Way galaxy.
In this paper we present the second release of the joint APOKASC asteroseismic and spectroscopic survey. Our targets have high-resolution H-band spectra from the Apache Point Observatory Galactic Evolution Experiment (APOGEE) project \citep{maj17} which  were obtained during the third Sloan Digital Sky Survey, hereafter SDSS-III \citep{eis11} and analyzed during the fourth Sloan Digital Sky Survey, hereafter SDSS-IV \citep{bla17}. Our asteroseismic data was obtained by the \textit{Kepler} mission \citep{bor10}, analyzed by members of the Kepler Asteroseismology Science Consortium (KASC), and interpreted by the team using both asteroseismic and spectroscopic data.

The primary scientific goal of the APOGEE project is reconstructing the formation history of the Milky Way galaxy through detailed studies of its stellar populations. This is frequently referred to as Galactic archeology. The relatively high resolution (R$\sim$22,000) of the spectra permits detailed stellar characterization. The infrared spectra from APOGEE can reach targets that would be heavily obscured in the optical, and the combination of a relatively large field of view (6 square degrees) and multi-plexing (300 fibers per plate) can yield large samples of representative Galactic stellar populations.  Evolved low-mass stars (both H-shell burning, or red giant stars, and He-core burning, or red clump stars) are the primary targets for APOGEE because they are intrinsically luminous, relatively common, and their H-band spectra are information-rich. 

Despite these attractive features, there are drawbacks associated with using red giant and clump stars for population studies. Using spectra alone, it is difficult to infer ages, crucial for tracing the evolution of populations, because stellar evolution transforms stars with a wide range of main sequence temperatures and luminosities into cool giants with a relatively narrow range of properties. As a consequence, indirect age proxies - for example, kinematics, or abundance mixtures associated with youth or age, have to be employed by spectroscopic surveys working alone.

The combination of spectroscopic and asteroseismic data is powerful, however, and both can now be measured for thousands of evolved cool stars. Large space-based planet transit surveys such as CoRoT and \textit{Kepler} naturally produce detailed information on stellar variability with a cadence ideally suited to detecting oscillations in giants \citep{der09, bed10}. These oscillation patterns encode detailed information about their structure and global properties. A major application for stellar population studies is the discovery that the frequency pattern can be used to distinguish between shell H-burning (or red giant) stars, with degenerate cores, and core He-burning (or red clump) stars, whose cores are larger and much less dense \citep{bed11}. For some targets, detailed studies of the measured frequencies can also be used to study features such as internal stellar rotation \citep{bec12, deh12}. However, for  bulk stellar populations, there is still powerful information in two key measures of the oscillation pattern which can be measured for large samples: the frequency of maximum power, $\nu_{\rm max}$, and the large frequency spacing $\Delta \nu$.  

The well-studied solar oscillation frequency pattern serves as a benchmark, with a $\nu_{\rm max}$ of order 3100 $\mu$Hz (five minutes) and $\Delta \nu$ around 135 $\mu$Hz.  Because the acoustic cut-off frequency is related to the surface gravity \citep{kje95}, we can adopt a semi-empirical scaling relation of the form

\begin{equation}
f_{\nu_{\rm max}}\left( \frac{\nu_{\rm max}} {\nu_{\rm max,\sun}} \right) =  \left( \frac{M} {M_{\sun}} \right) \left( \frac{R} {R_{\sun}} \right)^{-2} \left( \frac{\rm T_{\rm eff}} {\rm T_{\rm eff,\sun}} \right) ^{-0.5}
\end{equation}

In this equation the factor $f_{\nu_{\rm max}}$ can be a scalar or a function that captures deviations from the scaling relation. It can be shown analytically that the square of the large frequency spacing $\Delta \nu$ is proportional to the mean density in the limiting case of homology and large radial order n \citep{ulr86}. We can therefore define an analogous scaling relation for $\Delta \nu$, 

\begin{equation}
f_{\Delta \nu}\left( \frac{\Delta \nu} {\Delta \nu_{\sun}} \right) =  \left( \frac{M} {M_{\sun}} \right)^{0.5} \left( \frac{R} {R_{\sun}} \right)^{-1.5}
\end{equation}

The term $f_{\Delta \nu}$ can be computed from a detailed stellar model, and is in general a function of both the initial conditions and the current evolutionary state. In simple scaling relations $f_{\Delta \nu}=f_{\nu_{\rm max}}=1 $, and the mass and radius ($M_{sc}$ and $R_{sc}$) are defined by 

\begin{equation}
\left( \frac{M_{sc}} {M_{\sun}} \right)= \left( \frac{\nu_{\rm max}} {\nu_{\rm max,\sun}} \right)^{3} \left( \frac{\rm T_{\rm eff}} {\rm T_{\rm eff,\sun}} \right)^{1.5} \left( \frac{\Delta \nu} {\Delta \nu_{\sun}} \right)^{-4}
\end{equation}

and

\begin{equation}
\left( \frac{R_{sc}} {R_{\sun}} \right) =  \left( \frac{\nu_{\rm max}} {\nu_{\rm max,\sun}} \right) \left( \frac{\rm T_{\rm eff}} {\rm T_{\rm eff,\sun}} \right) ^{0.5} \left( \frac{\Delta \nu} {\Delta \nu_{\sun}} \right) ^{-2}
\end{equation}

In \citet{p14}, which we will refer to as APOKASC-1, we presented the first major catalog using both asteroseismic and spectroscopic data for a large sample of red giants. There are two natural applications of this approach: detailed studies of stellar physics and studies of stellar populations. The availability of simultaneous mass and composition data can be used to search for correlations between mass, age and spectroscopic observables. This is an especially exciting prospect because the set of stars with spectroscopic data from large surveys greatly exceeds the sample with asteroseismic data, which can be used to calibrate such relationships. For example, the surface [C/N] abundance is a product of the first dredge-up on the red giant branch, which is both expected on theoretical grounds to be mass and composition dependent \citep{sal15} and observed to be so in open cluster data \citep{tau15}. Data sets prior to APOGEE, however, were sparse and the samples were small. APOKASC-1 data was used to calibrate mass using both [C/N] \citep{mar16} and the full APOGEE spectra \citep{nes16} using the CANNON methodology. This approach has also been used for stellar population studies \citep{sil17}.

Another early result from the APOKASC-1 data was the discovery of a significant population of high-mass stars with high [$\alpha$/Fe] by \citet{mar15}; this was discovered independently by \citet{chi15} using a combination of CoRoT and APOGEE data in the related CoRoGEE project. This is a striking result because high-[$\alpha$/Fe] stars are typically regarded as a purely old, and by extension low-mass, population.  Some of these objects are evolved blue stragglers, or merger products \citep{jof16}, but explaining all of them with this channel would require a very high merger rate \citep{izz18}.  The alternative is an unusual nucleosynthetic origin; see \citet{chi15} for a discussion.  The discovery and characterization of unusual chemical stellar populations is a major prospect for Galactic archeology in general, as is the understanding of the products of binary star interactions. The joint data set has also enabled detailed studies of stellar physics, including tests of models of extra-mixing on the red giant branch \citep{mas17} and of the structure of core-He burning stars \citep{con15,bos17}.

However, there are recognized drawbacks to the approach used in the initial paper.  Important populations, such as members of open clusters, very metal-poor stars, and luminous giants were relatively sparsely sampled. Of more import, the APOKASC-1 effort did not attempt to calibrate the masses, radii, and uncertainties against fundamental data.  This is not a priori unreasonable, as initial checks of asteroseismic radii against interferometric values \citep{hub12} and those inferred from Hipparcos parallaxes combined with $\rm T_{\rm eff}$ \citep{vsa12} found encouraging agreement at the 5\% level.  However, even early on there was a recognized tension between masses derived from simple scaling relations and those expected for red giants in the old open cluster NGC 6791 \citep{bro12}. With the advent of the APOKASC-1 catalog, larger field star samples could be obtained and additional tests were possible.  The masses for halo stars derived from scaling relations in APOKASC-1 were found to be well above astrophysically reasonable values for old stellar populations \citep{eps14}. Offsets between fundamental and asteroseismic mass and radius values were also found for eclipsing binary stars \citep{gau16}. These results highlighted the need for improvements in the overall approach, which we now describe.

\subsection{Differences with Prior Work and the Grid Modeling Effort.}

The APOKASC-1 catalog contained asteroseismic and spectroscopic data for 1916 stars.  Since that time there has been both a substantial increase in the sample size and a change in the data analysis techniques.  The APOKASC-1 approach used spectroscopic data from the tenth data release (hereafter DR10) of the Sloan Digital Sky Survey \citep{dr10}; two different temperature scales were considered to account for scale shifts in spectroscopic data. The asteroseismic analysis was based on standard scaling relations. Measurements and theoretically estimated random uncertainties were taken from a single analysis pipeline with average results close to the mean of the measurements from all methods. Differences between pipelines were then used to infer systematic uncertainties and added in quadrature to the random ones to derive a total error budget. Our final stellar properties were derived including constraints from both the asteroseismic parameters and stellar interior models (a procedure usually called grid-based modeling).  In our revised catalog we critically examine each of these assumptions.

The spectroscopic pipeline has been extensively tested and modified since DR10 (see Section 2.2 below); the key ingredient for our purposes is $\rm T_{\rm eff}$, which enters directly into the formulas for asteroseismic surface gravities, masses and radii.  If grid modeling is being performed, $\rm T_{\rm eff}$, [Fe/H] and [$\alpha$/Fe] are needed to predict stellar parameters from evolutionary tracks.  The effective temperature is a defined quantity that can be measured in stars with known radius and total luminosity; such stars define a true fundamental $\rm T_{\rm eff}$ reference system. Because the revised APOGEE effective temperatures are tied to the IRFM fundamental scale \citep{holtz01}, we do not explicitly consider different overall temperature scales in the current effort. We have, however, assessed the impact of systematic changes in the underlying methodology by comparing results from the same stars for different SDSS data releases; the differences in derived masses arising from adopting DR13 as opposed to DR14 parameters are less than 1\% in mass with small scatter, which is well below other identified error sources.

We employ multiple methods for measuring the asteroseismic parameters $\nu_{\rm max}$ and $\Delta \nu$. In APOKASC-1 we adopted what the solar-scaled hypothesis, assuming that the measurements themselves are all scaled relative to a method-specific solar reference value.  So, if a given analysis method returns a solar $\nu_{\rm max}$ 10\% lower than the norm, all of the $\nu_{\rm max}$ measurements would be expected to be systematically 10\% lower than other techniques.  In this paper we replace the solar-scaled hypothesis with a data-driven approach for comparing the measurements; we have also revised our techniques for estimating both random and systematic measurement uncertainties.

Once we have a set of asteroseismic and spectroscopic observables, we then convert them to inferred masses and radii via scaling relations. The $\Delta \nu$ scaling relation is theoretically well-motivated but not expected to be exact \citep{ste09, whi11}. In a detailed study, \citet{bel13} studied the physics of the asteroseismic scaling relation for $\Delta \nu$, emphasizing how departures from homology in the structures of evolved stars perturb the scaling relation. We therefore explore theoretically motivated corrections to the $\Delta \nu$ scaling relation, which are known to improve agreement between asteroseismic stellar parameters and fundamental ones \citep{sha16,rod17,han17}. These corrections are sensitive to the internal structure, so knowledge of the evolutionary state is essential; evolutionary state is also important for ages.  We therefore also include asteroseismic and spectroscopic evolutionary state measurements in this paper. This was not done in APOKASC-1, which did not report ages or use corrections.

The empirical $\nu_{\rm max}$ scaling relation has a weaker theoretical basis than the $\Delta \nu$ scaling relation, although there have been detailed physical studies of its basis \citep{bel11}.  Despite this concern, it performs well when compared with empirical data. However, adjustments in the zero point for evolved stars are certainly reasonable, and different methods also yield different values even for the Sun.  We therefore treat the absolute zero point for the $\nu_{\rm max}$ scaling relation as a free parameter which can be calibrated against fundamental mass data.

Finally, we consider the impact of adopting grid-based modeling for evolved red giant stars. Grid-based modeling is in principle powerful, because it includes all of the constraints from observables and theory on the derived properties of the star. For stars on or near the main sequence, precisely measured $\rm T_{\rm eff}$, log g and abundances can set stringent constraints on mass and radius that complement asteroseismic measurements; see \citet{ser17} for our discussion in the dwarf context. Unfortunately, one cannot test the validity of the underlying models if their accuracy is assumed in the solution, and \citet{tay17} found significant offsets between theoretical expectations from solar-calibrated isochrones and APOKASC data. The origin of these differences may be in the treatment of the mixing length, as noted in that paper and by \citet{li18}, or it may be tied to other choices of input physics as discussed \citet{sal18}. In either case, there is no guarantee that solar calibrated models agree in the mean with data for evolved stars. A direct consequence is that there will be systematic offsets between stellar properties inferred from the tracks alone and stellar properties inferred from asteroseismology alone, which can inject complex systematic differences in the derived stellar properties unless the models are explicitly calibrated to remove such differences. As a result, there is benefit in choosing to test the asteroseismic scale itself directly against fundamental data, rather than doing so with a hybrid grid-modeling value.  In this paper we therefore do not imposed grid-based modeling constraints on our observables.  A companion paper \citep{s18} investigates asteroseismic parameters from our data including grid-based modeling. Finally, for usage in stellar population studies, we have taken our data and used it to infer ages and extinctions. 


In summary, the improvements and changes in our APOKASC-2 analysis are:
\begin{enumerate}
\item Our spectroscopic parameters and uncertainties are taken from the fourteenth data release (hereafter DR14) of the Sloan Digital Sky Survey \citep{dr14} instead of DR10.
\item We have inferred evolutionary states for virtually all of the stars in our sample for APOKASC-2, either from asteroseismology or from spectroscopic diagnostics calibrated on asteroseismic observables.
\item The relative zero points for $\nu_{\rm max}$ and $\Delta \nu$ from different pipelines are inferred from the data, and not assumed to be strictly defined by their relative solar reference values.
\item With zero point differences accounted for, the scatter of the individual pipeline values about the ensemble mean is used to infer the random uncertainty for each star, rather than relying on formal theoretical error estimates.
\item $\Delta \nu$- and $\nu_{\rm max}$- dependent differences between individual pipeline values and the ensemble mean are treated as systematic error sources.
\item The $\Delta \nu$ scaling relation is corrected with the same theoretically motivated approach as that in \citet{s18}, rather than being treated as exact.
\item The absolute zero point of the $\nu_{\rm max}$ scaling relation is set by requiring agreement with fundamental radii and masses in star clusters with asteroseismic data, as opposed to adopting a solar reference value.
\item We do not use grid-based modeling in APOKASC-2.
\item We provide ages and extinction estimates.
\end{enumerate}

The outcome of this exercise is tabulated for the full joint sample, and the sample properties are then discussed.  The remainder of the paper is organized as follows.  We discuss the sample selection in Section 2 and present our basic data there.  The relative mean asteroseismic parameters and the absolute calibration from open cluster members are derived in Section 3.  The catalog itself is presented in Section 4 and the conclusions are given in Section 5.

\section{Sample Properties: Selection, Unusual Stars and Evolutionary State}

Our basic data is drawn from two sources: time domain data derived from the \textit{Kepler} satellite during the first four years of operation and spectroscopic data from the APOGEE survey of the Sloan Digital Sky Survey.  In addition, we employed additional photometric data for the calibrating star clusters NGC 6791 and NGC 6819  to test the absolute radius scale.  The photometric data and adopted cluster parameters are discussed in Section 3.4.

\subsection{\textit{Kepler} Data}

The details of the \textit{Kepler} data itself and the light curve reduction procedures used are described in \citet{e18}.  We employed five distinct pipelines for asteroseismic analysis of the reduced light curves known in the literature by three-letter acronyms: A2Z, CAN, COR, OCT, and SYD. We briefly reference each method below. For a more detailed discussion of the different approaches see \citet{s18}. The same data preparation method is not used in all cases. Two different methods were used with A2Z preparing their own datasets following \citet{gar11} and CAN, COR, OCT and SYD all using data prepared using the \citet{han14} method. A comparison and review of the methods is given in  \citet{hek11} and further discussed in \citet{hek12}, where they looked at the impact of data duration on the detectability of the oscillations and the precision of the parameters. For this paper, the precise method used to determine the average asteroseismic parameters is not of major importance because here we seek to show how the differences can be mitigated. Nevertheless, we give basic references to the method of operation of each pipeline. The A2Z pipeline was first described in \citet{mat10} and, together with their method of data preparation is updated in \citet{gar14}. The CAN pipeline is described in \citet{kal10}; the COR pipeline is described in \citet{mos09}; the OCT pipeline is described in \citet{hek10}; and the SYD pipeline is described in \citet{hub09}.







\subsection{Spectroscopic Data}

We collected the spectroscopic data	using the 2.5-meter Sloan Foundation
telescope \citep{gun06} and the APOGEE near-infrared spectrograph
at Apache Point Observatory. These spectra were	
obtained during SDSS-III, and the target selection criteria for stars in the
{\it Kepler} field are described in \citet{zas13}. All spectra are
re-reduced and re-analyzed for each data release. The procedures used
to flat-field, co-add, extract, and calibrate the spectra are described in \citet{nid15}. The spectra were then processed through the
APOGEE Stellar Parameters and Chemical Abundances Pipeline, or ASPCAP \citep{gar16}, which derives T$_{\rm eff}$, log g, metallicity and other properties through
a $\chi^2$ minimization of differences with a grid of theoretical spectra as described below.

The APOGEE survey has presented data in four SDSS data releases.  The first set of results, in Sloan DR10, was described in \citet{meszaros01}.  The subsequent DR12 data analysis technique was documented in \citet{holtz01}, while the data released in DR13 (as well as the subsequent DR14) is discussed in \citet{hol18} and \citet{jons01}. Each data release contained both 'raw' and 'calibrated' atmospheric parameters.  The 'raw' values reflect the output of the automated pipeline analysis, while the 'calibrated' values can include corrections to bring the results into agreement with external standards.  

As the survey has progressed, the corrections inferred from the calibration process have in general become smaller, because improvements implemented in ASPCAP allowed the APOGEE team to produce more accurate and precise atmospheric parameters. The first APOKASC catalog was compiled using DR10 parameters, while results presented in this paper use DR14 parameters, the latest SDSS-IV release. In this section, we detail the most important improvements to ASPCAP and changes in the calibration of effective temperature, [Fe/H] and [$\alpha$/Fe] between DR10 and DR14; these ingredients are the ones relevant to the data presented in this paper. There have been important changes made in the reduction techniques, the line list, model atmospheres and spectrum synthesis.

Data reduction in DR13 and DR14 included improved line spread function (LSF) characterization, telluric and persistence correction. ASPCAP pipeline results are benchmarked against the solar spectrum and that of Arcturus, with less secure line strengths empirically adjusted to match specified values, using the line list from \citet{she15}.  
A new set of Arcturus abundances has also been adopted for tuning the line strengths. The solar reference abundances table was changed from \citet{grevesse01} (DR10) to \citet{asplund01} in DR12 and onwards.
Abundances of nearly 23 elements are determined in DR13 and DR14, instead of 3 broad indices being reported, as was the case in DR10.

New ATLAS model atmospheres \citep{meszaros02} were computed for DR12 and are still in use in DR14 using the solar reference from \citet{asplund01}. A new set of synthetic spectra were included covering the range 2500 < $\rm T_{eff}$ < 4000 K, based on custom MARCS atmospheres. All synthetic spectra were calculated using Turbospectrum \citep{alv98}; previous syntheses were done using ASSeT in DR10. From DR12 onwards, a finer grid spacing was adopted in metallicity ([M/H]), with 0.25 dex steps instead of the 0.5 dex spacing used in DR10. The grid of model atmospheres was also extended to a higher metallicity of [M/H]=+0.75. A macroturbulent velocity relation was determined based on a fit with a subset of data and a macroturbulence dimension, rather than using a fixed value.

DR13 and DR14 use a multi-step analysis through multiple grids to determine the main atmospheric parameters.
Initial characterization was carried out using F, GK, and M coarse grids.  Once stars have passed quality control steps, the ASPCAP pipeline is then used to do a full solution in 6D or 7D space depending on the location of the star in the HR diagram.  This high dimensionality is required because the APOGEE spectral region is heavily influenced by CNO molecular features.  Therefore, in addition to the 4D ingredients typically considered in model atmosphere fits ($\rm T_{eff}$, surface gravity, overall metallicity, and microturbulence), ASPCAP also includes three additional dimensions: alpha-element enhancement (including O), C and N. The final step is the derivation of individual abundances, which were not included in DR10, and which use spectral windows rather than additional dimensions in the atmospheres grid.

For DR10, effective temperatures were calibrated to be in agreement with color temperatures for stars belonging to open and 
globular clusters. This comparison sample was improved in subsequent data releases by replacing the limited cluster calibration set with field stars that have low extinction, which have the advantage of providing many more calibrators in a larger metallicity and surface gravity phase space. In DR10, the effective temperature correction was fairly large (around 110-200 K, depending on $\rm T_{eff}$ and metallicity). As ASPCAP improved, spectroscopic temperatures showed a better agreement with photometric ones. This resulted in no correction applied in the DR13 data, as published.  However, a modest metallicity-dependent offset was discovered post-release; a similar metallicity-dependent temperature correction was therefore introduced again for DR14. The uncertainty was estimated from the scatter between spectroscopic and photometric temperatures for a subsample of targets.

Metallicities in DR13 and DR14 have been calibrated to remove $\rm T_{eff}$ trends using members of star clusters; the underlying assumption is that any systematic trends in inferred abundance within a cluster sample are analysis artifacts, as cluster stars share the same true metallicity. This is a significant departure from DR10, where [M/H] was calibrated to mean literature abundances for open and globular clusters as a whole, not star by star. This external calibration for [M/H] has been introduced again for DR14, but was not done in DR12 and DR13. It is important to point out that these calibrations induce changes generally smaller than 0.1 dex, and become larger than that only for the most metal poor stars below [M/H] < -1.0. The DR14 metallicity calibration effects are also smaller than those of DR10.

We illustrate the net impact of these changes in two figures.  Figure 1 compares the spectroscopic parameters for stars in APOKASC-1 between DR10 and Dr13. Systematic shifts are more important than random scatter, and the differences largely reflect changes in the choice of calibrators for the spectroscopic solution and improvements in the ASPCAP spectroscopic pipeline.  By comparison, the differences between DR13 and DR14, illustrated in Figure 2, are milder, although there are still clear zero-point offsets in the metallicity and scatter in the inferred carbon to nitrogen ratio, a diagnostic of the first dredge-up in evolved stars.


\begin{figure}


\plotone{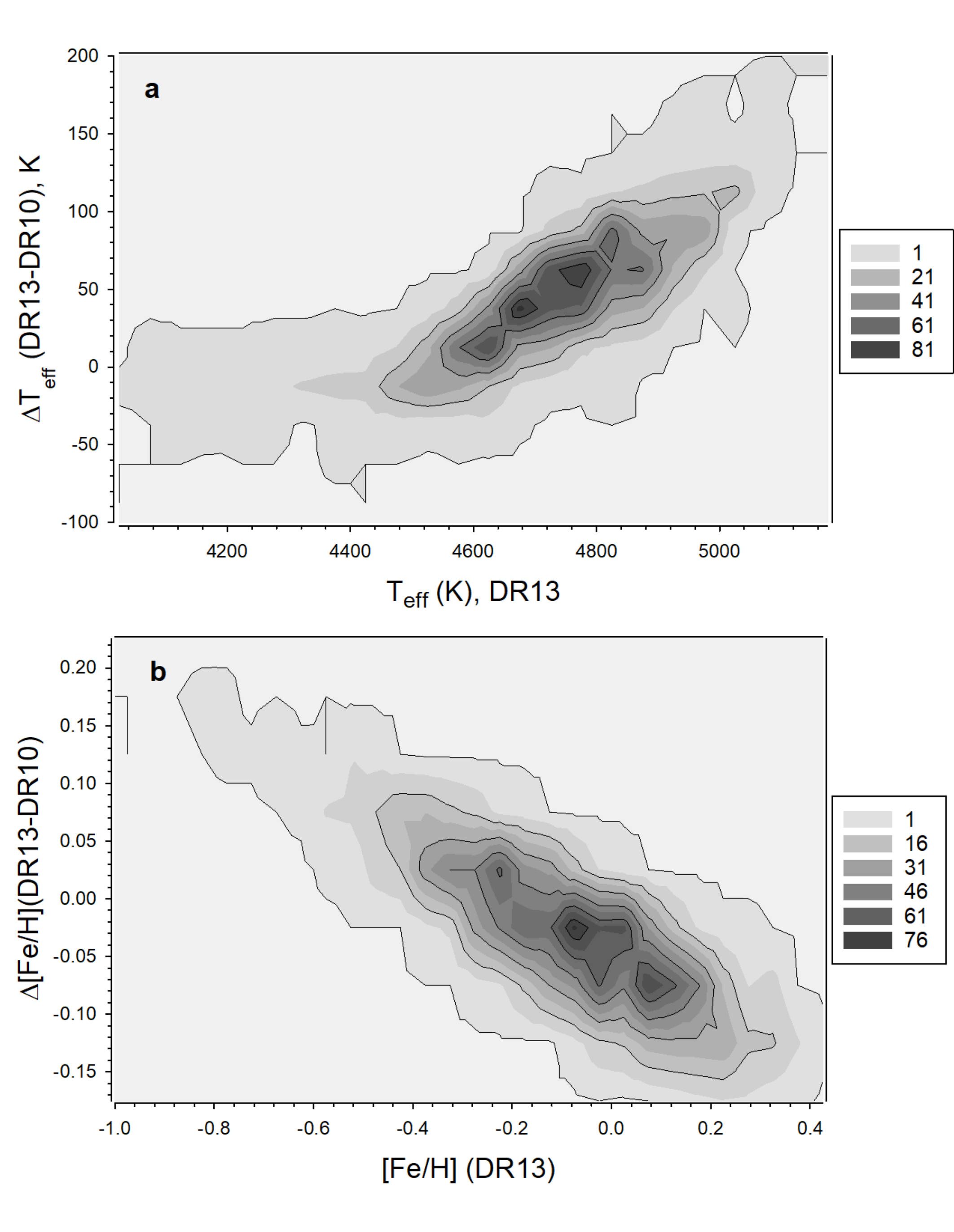}

\caption{Spectroscopic properties in our 2014 catalog compared with the current values for stars in common between the two data sets. Differences are in the sense DR13-DR10 and the color reflects the density of points. We compare $\rm T_{eff}$ in panel a, and [Fe/H] in panel b.}

\end{figure}

\begin{figure}


\plotone{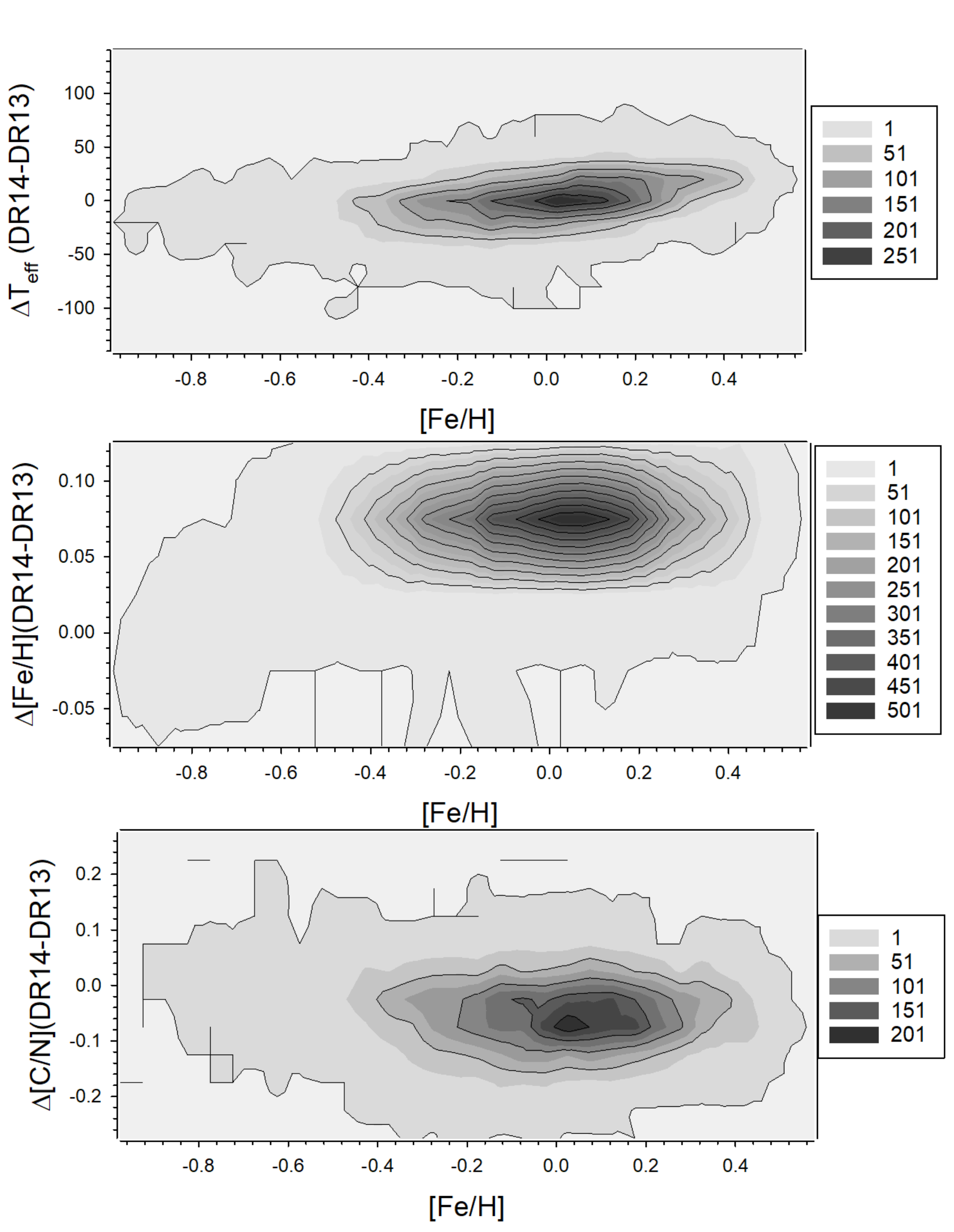}

\caption{Differences in DR13 and DR14 spectroscopic properties in APOKASC-2 are illustrated as a function of [Fe/H]. Differences are in the sense DR14 minus DR13 and the color reflects the density of points. We compare $\rm T_{eff}$ in panel a, [Fe/H] in panel b, and [C/N] in panel c.}

\end{figure}

\subsection{The SDSS-IV and APOKASC-2 Samples} The full APOGEE data sample we use was observed in SDSS-III (but analyzed in SDSS-IV) and contains 11,877 stars.  Many of these targets were not explicitly observed for asteroseismology, however, and some of the remainder turned out to be subgiants.  A total of 8604 of these stars had calibrated spectroscopic log g < 3.5 and were therefore potential red giant asteroseismic targets. Target selection for this sample was discussed in APOKASC-1. However, not all light curves were sufficiently long to detect asteroseismic signals; some had data artifacts; and a substantial number in the high surface gravity domain (3.3 < log g < 3.5) are technically challenging to analyze because their oscillation frequencies are close to the \textit{Kepler} 30-minute sampling for long-cadence data.  

\subsubsection{The Asteroseismic Parameter Calibration Sample} As discussed above, we employed 5 independent pipelines to detect and characterize oscillations. A subset of 4706 stars had data from all 5 pipelines and asteroseismic evolutionary states reported by \citet{els17}, and we use this subset of the sample for our empirical calibration of the asteroseismic measurements.  As there are known differences between the asteroseismic properties of core He-burning and shell H-burning stars \citep{m12} we analyze them separately. In our sample of targets with results from all pipelines, 2833 objects were classified as first ascent red giants (RGB) or as possible asymptotic giant branch stars (RGB/AGB).  For the purposes of this paper we defined any star in one of these two asteroseismically similar \citep{ste13} shell-burning categories, as RGB stars, a notation that we will use for the remainder of the paper. A total of 1873 targets are identified as either red clump (RC) stars, higher mass secondary clump (2CL) ones, or as intermediate between the two (RC/2CL).  For the remainder of the paper we refer to objects in this class as RC stars.

\subsubsection{The Catalog Sample} The APOKASC-2 sample analyzed in this paper contains 6676 targets with reduced light curves which were selected for asteroseismic analysis. There are 122 stars for which we were not able to return asteroseismic data or which had bad spectra. We have asteroseismic evolutionary states for 6076 of the remaining objects in \citet{els17}, including 2453 RC stars and 3623 RGB stars. (The calibration set described above is smaller because we required asteroseismic parameter measurements from all pipelines for calibration, but report catalog values if any pipeline returned measurements.) For the 478 stars without asteroseismic evolutionary state assignments from \cite{els17}, we infer DR13 spectroscopic evolutionary states as described in \citet{hol18}. This includes 276 RC stars and 152 RGB stars; only 50 stars had ambiguous evolutionary states given their spectroscopic properties.  Our data for the stars without asteroseismic state data, and for stars with  no seismic parameters, is illustrated in Figure 3. This Kiel diagram is related to the classical HR Diagram, as surface gravity is related to luminosity. The cluster of targets with log g > 3.1 without results are stars where the asteroseismic frequencies are close to, or exceed, the Nyquist sampling frequency from the \textit{Kepler} data.  The remainder are an admixture of stars close to the boundary between the RGB and the RC, where it is most challenging to distinguish RC from RGB stars spectroscopically.  This group of targets also includes a substantial number of higher mass and surface gravity (log g > 2.6) core He-burning stars, and the hotter RGB sample includes a number of very metal-poor targets.  For our remaining analysis we will treat the stars with spectroscopic evolutionary state assignments in a manner similar to the approach taken for targets with asteroseismic states; the sole exception is the group with ambiguous evolutionary states, for which the final mass and radius estimates are more uncertain (see Section 3.2.1).  A more detailed discussion of the evolutionary states of our targets, and a comparison of spectroscopic and asteroseismic methods, can be found in \citet{e18} and \citet{hol18}. 

\begin{figure}


\plotone{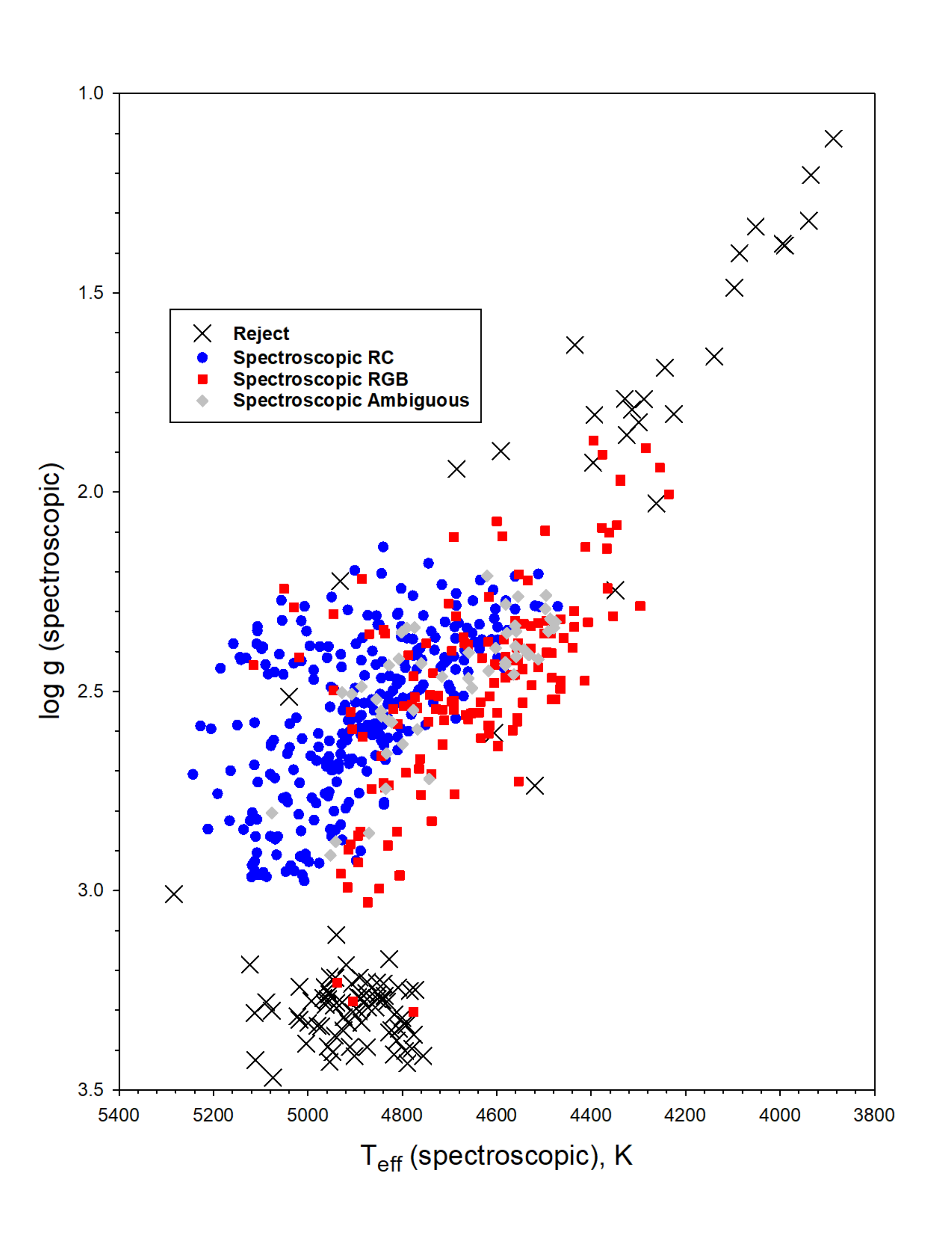}

\caption{Spectroscopic effective temperature versus log g in our sample without seismic evolutionary states.  $\rm T_{\rm eff}$ values (in K) and log g are both the DR14 values. Stars for which we report no asteroseismic data are shown with X symbols and are predominantly RGB stars when classified spectroscopically. The remainder of the sample is primarily composed of RC stars (blue points, 276 total), with some RGB stars (red points, 152 total) and 52 stars that had ambiguous spectroscopic data (gray points).}

\end{figure}

Our main sample is shown in Figure 4, and it illustrates the power of asteroseismic evolutionary state classification.  As one would expect on stellar populations grounds, the RC stars are, on average, hotter than the RGB ones.  Higher mass RC stars had a non-degenerate He flash, however, which produces an interesting population feature. Stars with masses between 2.2 and 3.0 $M_{\sun}$ can have a smaller He core mass at ignition than their low or high mass counterparts, so they show up at lower luminosity in the secondary red clump than the typical old RC star.   The excess of red giants around log g = 2.6 is the RGB bump, where the H-burning shells of ascending RGB stars cross the composition discontinuity produced by the first dredge-up at the maximum depth of the surface convection zone.  More luminous RGB stars are seen to span a wider range of $\rm T_{\rm eff}$ than less luminous ones.  This is because they include a mixture of metal-poor objects (preferentially seen at greater distances) and double shells source, or AGB, stars that are asteroseismically similar to first ascent shell H-burning stars.  There is also an admixture of higher mass RGB stars in the same Kiel diagram position as RC stars; these can only be distinguished asteroseismically and would be missed in traditional survey methods. Scatter plots with many points have a tendency to emphasize outliers.  An alternative visualization (combining red clump and secondary clump) is illustrated in Figure 5, where the data is binned to illustrate the density of points. For a further discussion of spectroscopic evolutionary states we refer to \citet{e18}.

\begin{figure}


\plotone{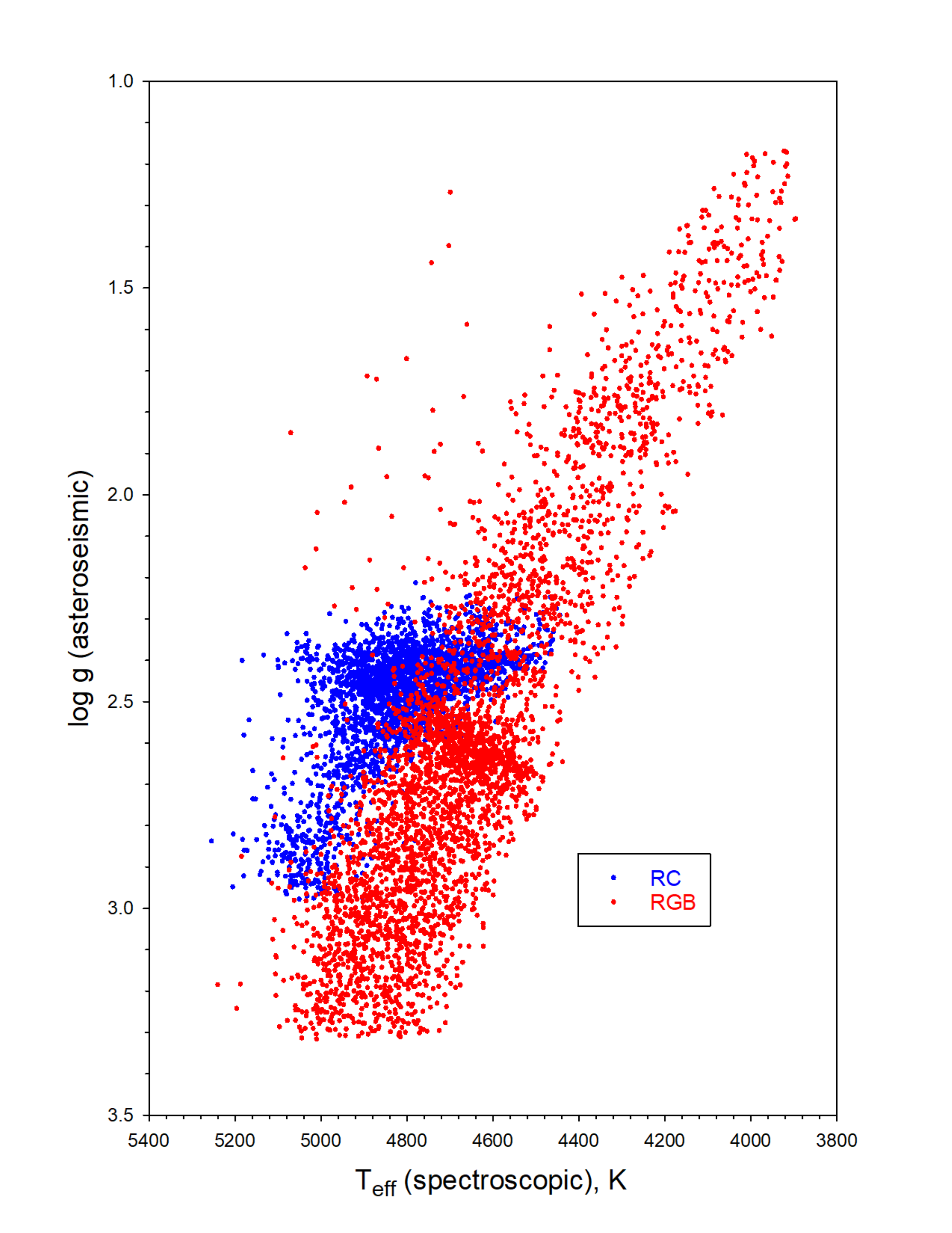}

\caption{Spectroscopic effective temperature versus asteroseismic log g in our sample by asteroseismic evolutionary state.  RC (core He-burning) stars are in blue. RGB (H-shell or double shell burning) stars are in red.  The $\rm T_{\rm eff}$ values (in K) are the DR14 values.  The asteroseismic surface gravities are defined in Sections 3 and 4.}

\end{figure}

\begin{figure}


\plotone{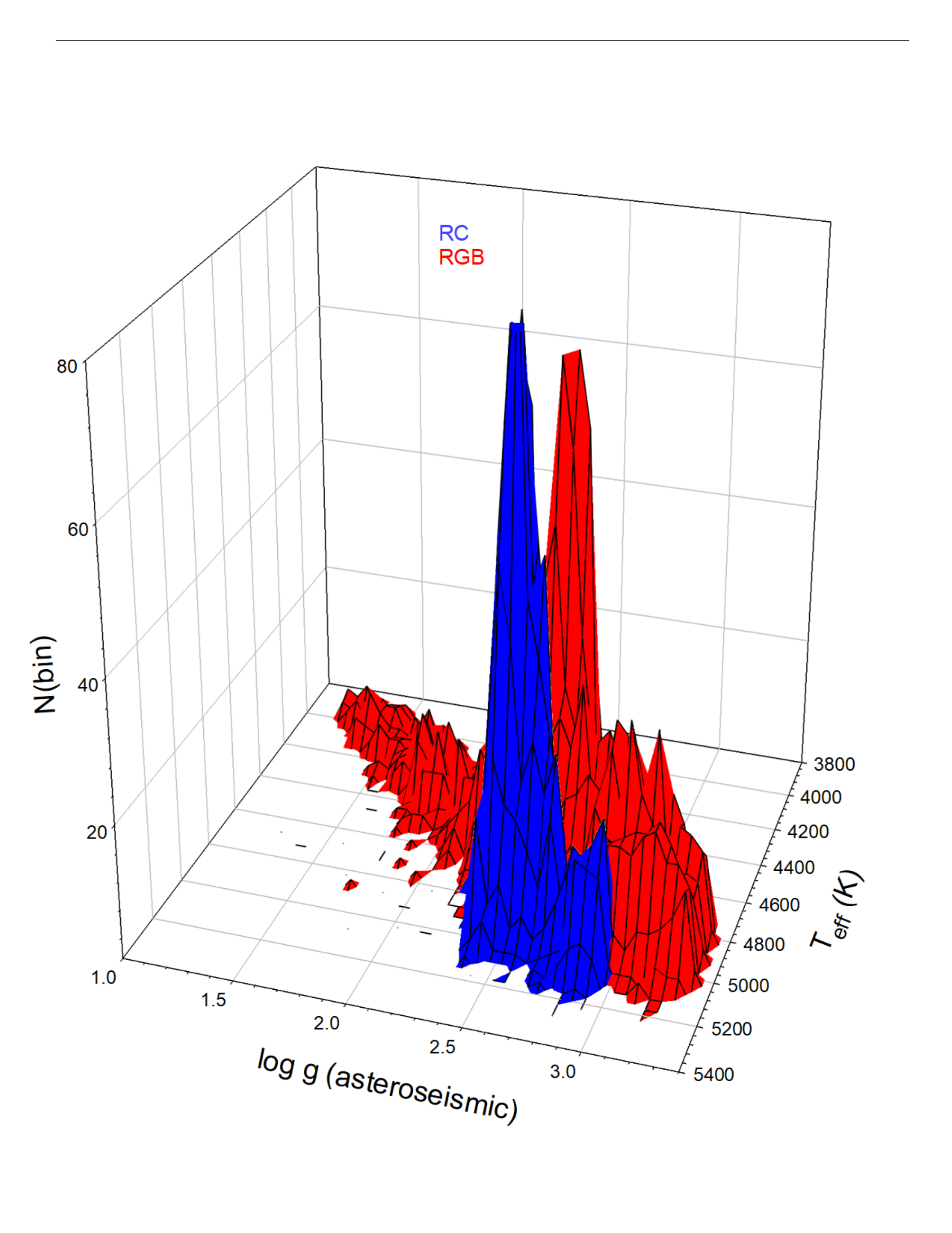}

\caption{As Figure 4, but counting the number of targets as a function of evolutionary state in bins of 50 K in $\rm T_{eff}$ and 0.05 dex in log g.  RGB stars are the red distribution, while RC stars are the blue distribution.}

\end{figure}

\section{Calibrating and Defining The Empirical Asteroseismic Mass and Radius Scales}

Our main goal is to estimate masses and radii of stars using asteroseismic and spectroscopic data. In Section 3.1 we discuss how we combine data from multiple analysis techniques to infer the asteroseismic observables $\Delta \nu$ and $\nu_{\rm max}$, the appropriate ensemble solar reference values, and the associated random and systematic uncertainties. In APOKASC-1 we used equations (3) and (4), which assumed that the scaling relations (1) and (2) were exact ($f_{\nu_{\rm max}}$=$f_{\Delta \nu}=1$). If we relax these assumptions, we can define corrected mass and radius estimates $M_{cor}$ and $R_{cor}$ by

\begin{equation}
M_{cor} = \frac{f_{\nu_{\rm max}}^{3}} {f_{\Delta \nu}^{4}} M_{sc}
\end{equation}

and

\begin{equation}
R_{cor} = \frac{f_{\nu_{\rm max}}} {f_{\Delta \nu}^{2}} R_{sc}
\end{equation}

In Sections 3.2 and 3.3 we present the determination of $f_{\Delta \nu}$ and $f_{\nu_{\rm max}}$ respectively. We adopt a theoretically motivated prescription for $f_{\Delta \nu}$, which is computed star-by-star (and is therefore a function of the stellar parameters and not a scalar).  In the absence of a comparably well-motivated theoretical prescription for changes in the frequency of maximum oscillation power scaling relation, we solve for a scalar factor $f_{\nu_{\rm max}}$ empirically calibrated to be in agreement with fundamental data. Calibrated stellar parameters for the full sample are then given and discussed in the subsequent sections of the paper.

\subsection{Empirical Asteroseismic Parameters}

Inferring masses from asteroseismic scaling relations has rested on two major assumptions: that the measurements from each pipeline can be scaled relative to their solar values and that the uncertainties are the internal values returned by those pipelines.  Both assumptions are, in principle, reasonable; but neither is exact.  Measuring the frequency of maximum oscillation power requires us to first model and account for the background, then to make choices about the smoothing and/or fitting of the power excess.  The asteroseismic frequency spectrum contains a pattern which is close to uniformly spaced, and the large frequency spacing is a theoretically well-posed quantity.  However, there are deviations from uniform spacing, caused for example by the way acoustic glitches perturb the comb structure of the pressure modes \citep{vra15}.  This measurement must also be normalized and defined in a specific method-dependent frequency domain \citep{mos13b}.  Different pipelines can therefore have both random and systematic differences from one another that are non-negligible, and there can also be differential scale factors relative to the individual solar zero-points. In practice, this means that any relative bias in solar measurements from a given pipeline does not necessarily translate into a similar relative bias when it is used for measurements in evolved stars. First principles error estimation is notoriously difficult to perform, and it can be challenging to disentangle systematic and theoretical uncertainties. Fortunately, we can test both assumptions with our data set; we have multiple pipeline results for a large sample of objects.  The \textit{relative} values inferred from different methods for the same targets provide robust constraints on the differential zero points of the various techniques, and the dispersion in values once systematic differences are accounted for yields guidance on errors.  We begin with RGB stars as a calibrating set, and then follow with an analogous study of RC stars. Our asteroseismic values for individual pipelines are given in Appendix A; the values derived from our method are used to derive the stellar observables and are given in the main catalog table (Table 5).

\subsubsection{Relative Pipeline Zero Points}

To motivate the averaging discussion, we will begin with methods used in prior efforts, and then generalize to the current one.  We use $\nu_{\rm max}$ as an example, but the same considerations apply to $\Delta \nu$. In this discussion, $\langle quantity \rangle$ refers to a simple average of multiple measurements of that quantity from different pipelines for a given star. The scaling relations require both a measurement ($\nu_{\rm max}$ in this case) and a solar reference value $\nu_{\rm max}^{\rm ref}$; both are in general different for different methods.

The approach used in \citet{p14} for evolved stars and in \citet{ser17} for dwarfs used the solar-scaled hypothesis. In this case, the quantity to be averaged is the ratio of the measurement to the solar value for each pipeline: $\frac{\nu_{\rm max}^{\rm star}} {\nu_{\rm max}^{\rm ref}} = \langle \frac{\nu_{\rm max}^{\rm pipe}} {\nu_{\rm max,\sun}^{\rm pipe}}\rangle$.  As we will show below, however, there are significant mean offsets between pipelines using this averaging method.
Another logical approach would then be to decouple the averaging of the measurements and the reference values.
In this limit, $\frac{\nu_{\rm max}^{\rm star}} {\nu_{\rm max}^{\rm ref}} = \frac{\langle \nu_{\rm max}^{\rm pipe} \rangle} {\langle \nu_{\rm max,\sun} \rangle}$.  The mean value would then be a simple average of the absolute measurements, and the solar reference could be averaged in the same way. The different pipelines also show significant average differences in the absolute measurements, unfortunately.  We therefore treat the relative normalization of the different pipelines, and the choice of reference values, as quantities to be solved for empirically.

 We define the ensemble solar reference values as the average of the individual pipeline solar values: $\nu_{\rm max,ref} = \langle \nu_{\rm max,\sun} \rangle$ and $\Delta \nu_{\rm ref} = \langle \Delta \nu_{\sun} \rangle$ respectively.  This choice is not fundamental; if another method were included the mean would shift.  These averaged solar reference values, and the individual values on which they are based, are included in Table 1. We note that the COR pipeline \citep{mos13a} has a published correction term for asteroseismic scaling relations, implying a different solar normalization; as we are correcting for this physical effect separately (see Section 3.2), we use the solar values instead. However, in our final results we calibrate the overall zero point of the $\nu_{\rm max}$ scaling relation to reproduce fundamental data, as described in Section 3.2; because the different $\Delta \nu$
 methods have very similar solar reference values, and our empirical data constrains only the ratio of the solar reference values, we did not attempt separate empirical adjustments for both solar reference values.  
 
We are searching for scale factors for each pipeline such that they all, on average, return the same mean values over the full sample.  Once these scale factors are defined, we then scale and average the results to obtain our star-by-star measurements, and we use the dispersion between the scaled values to estimate measurement uncertainties. We proceed as follows.  For each star (index j), we have measurements from five pipelines (index i) and can define mean values   
\begin{equation}
\langle \nu_{\rm max}^{j} \rangle = \frac{1}{\rm N_{pipe}} \sum_{I=1}^{\rm N_{pipe}} \frac {\nu_{\rm max}^{i,j}}{X_{\nu_{\rm max}}^{i}}
\end{equation}
and 
\begin{equation}
\langle \Delta \nu^{j} \rangle = \frac{1}{N_{\rm pipe}} \sum_{I=1}^{N_{\rm pipe}} \frac {\Delta \nu^{i,j}}{X_{\Delta \nu}^{i}}
\end{equation}
where $N_{\rm pipe}$ is the total number of pipelines available for that star.
We have 2833 targets classified as RGB or AGB/RGB with data returned from all five pipelines; we use this sample to compute the X scale factors. The scale factor is defined in a two step process. We first determine the factor by which an individual seismic value for a given pipeline differs from the unweighted average over all the returns for that star by defining the normalization factors $Y_{\nu_{max}}^{i,j}$ and $Y_{\Delta \nu}^{i,j}$ for each pipeline i and star j by
\begin{equation}
Y_{\nu_{\rm max}}^{i,j} = N_{\rm pipe} \frac{\nu_{\rm max}^{i,j}}{\sum_{i=1}^{N_{\rm pipe}}\nu_{\rm max}^{i,j}}
\end{equation}
and
\begin{equation}
Y_{\Delta \nu}^{i,j} = N_{\rm pipe} \frac{\Delta \nu^{i,j}}{\sum_{i=1}^{N_{\rm pipe}}\Delta \nu^{i,j}}
\end{equation}
For each star j we can also compute the absolute measurement dispersions $\sigma_{\nu_{\rm max}}^{j}$ and $\sigma_{\Delta \nu}^{j}$. The second stage in the determination of the scale factors ($X^i$) is to use the $Y^{ij}$ together with the $\sigma^j$ values to form a weighted average for a given pipeline. The overall normalization factors $X_{\nu_{\rm max}}^{i}$ and $X_{\Delta \nu}^{i}$ for each pipeline i are then defined by 
\begin{equation}
X_{\nu_{\rm max}}^{i} = \frac{\sum_{j=1}^{N_{\rm star}} \frac{Y_{\nu_{\rm max}}^{i,j}}{(\sigma_{\nu_{\rm max}}^{j})^2}}{\sum_{j=1}^{N_{\rm star}}\frac{1}{(\sigma_{\nu_{\rm max}}^{j})^2}}
\end{equation}
and
\begin{equation}
X_{\Delta \nu}^{i} = \frac{\sum_{j=1}^{N_{\rm star}} \frac{Y_{\Delta \nu}^{i,j}}{(\sigma_{\Delta \nu}^{j})^2}}{\sum_{j=1}^{N_{\rm star}}\frac{1}{(\sigma_{\Delta \nu}^{j})^2}}
\end{equation}
with uncertainties defined by standard error propagation. The use of the $\sigma^j$ terms ensures that stars for which there is a large spread in the determinations are given a lower weight in the formation of the average.

Our approach, by construction, ensures that all pipelines return the same mean values when averaged over the full sample, but this approach does not account for how these differences change as a function of the mean values themselves. To quantify trends in the pipeline means, we rank-ordered our data in $\nu_{\rm max}$ and then broke it up into non-overlapping bins of 100 targets.  For each bin and pipeline i, we then computed the average $X_{\nu_{\rm max}}^{i}$ that we would have obtained.  To test the impact of adopting a solar normalization, we can define an analog of X, $X^{',i}_{\nu_{\rm max}}$, where the quantity being averaged is not the absolute measurement $\nu_{\rm max}^{i}$ but $\frac {\nu_{\rm max}^{i}} {\nu_{\rm ax,\sun}^{i}}$.  We would expect $X_{\nu_{\rm max}}^{',i}=1$ for all pipelines if the solar-scaled hypothesis were correct.  We then repeated the rank-ordering and binning exercises for these alternate values.  The results are shown in Figure 4, where the top panel shows the solar-normalized ratio $X^{',i}_{\nu_{\rm max}}$ for bins of 100 stars and the bottom panel compares the absolute ratio $X_{\nu_{max}}^{i}$ for the same bins as a function of the mean $\nu_{\rm max}$ of the bins.  To place both panels on the same scale, we multiplied the average $X^{',i}_{\nu_{\rm max}}$ values for the bins by the pipeline mean solar reference value $\nu_{\rm max,\sun}$ = 3103.266 $\mu$Hz. Because we are defining the value relative to the mean for all five pipelines, the important feature here is the spread between the pipelines, not the absolute position of any one pipeline.  However, the usage of solar-normalized values increases the dispersion between them in Figure 6 rather than decreasing it, disfavoring the scaling of each pipeline output relative to the results it would have obtained for the Sun. To quantify this effect we compute the dispersion in the ratio of individual values to the total; we obtain $\sigma$ = 0.0131 for the solar-normalized values and $\sigma$ = 0.0088 for the absolute values.   

\begin{deluxetable}{ccccccc}
\tabletypesize{\scriptsize}
\tablecaption{Solar Reference Values}
\tablewidth{0pt}
\tablehead{
\colhead{Quantity} & \colhead{A2Z} & \colhead{CAN} & \colhead{COR}\tablenotemark{a} & \colhead{OCT} &
\colhead{SYD} & \colhead{Average} 
}
\startdata
$\nu_{max,\sun}$ & 3097.33 & 3140 & 3050 & 3139 & 3090 & 3103.266 \\
$\Delta \nu_{\sun}$ & 135.2 & 134.88 & 135.5 & 135.05 & 135.1 & 135.146 \\
\enddata


\tablecomments{Solar reference values for individual pipelines. All measurements are in $\mu$Hz. Uncertainties are not included when computing the mean, as the zero point is ultimately inferred empirically.}

\tablenotetext{a}{Does not include \citet{mos13a} scaling relation corrections}

\end{deluxetable}

\begin{figure}


\plotone{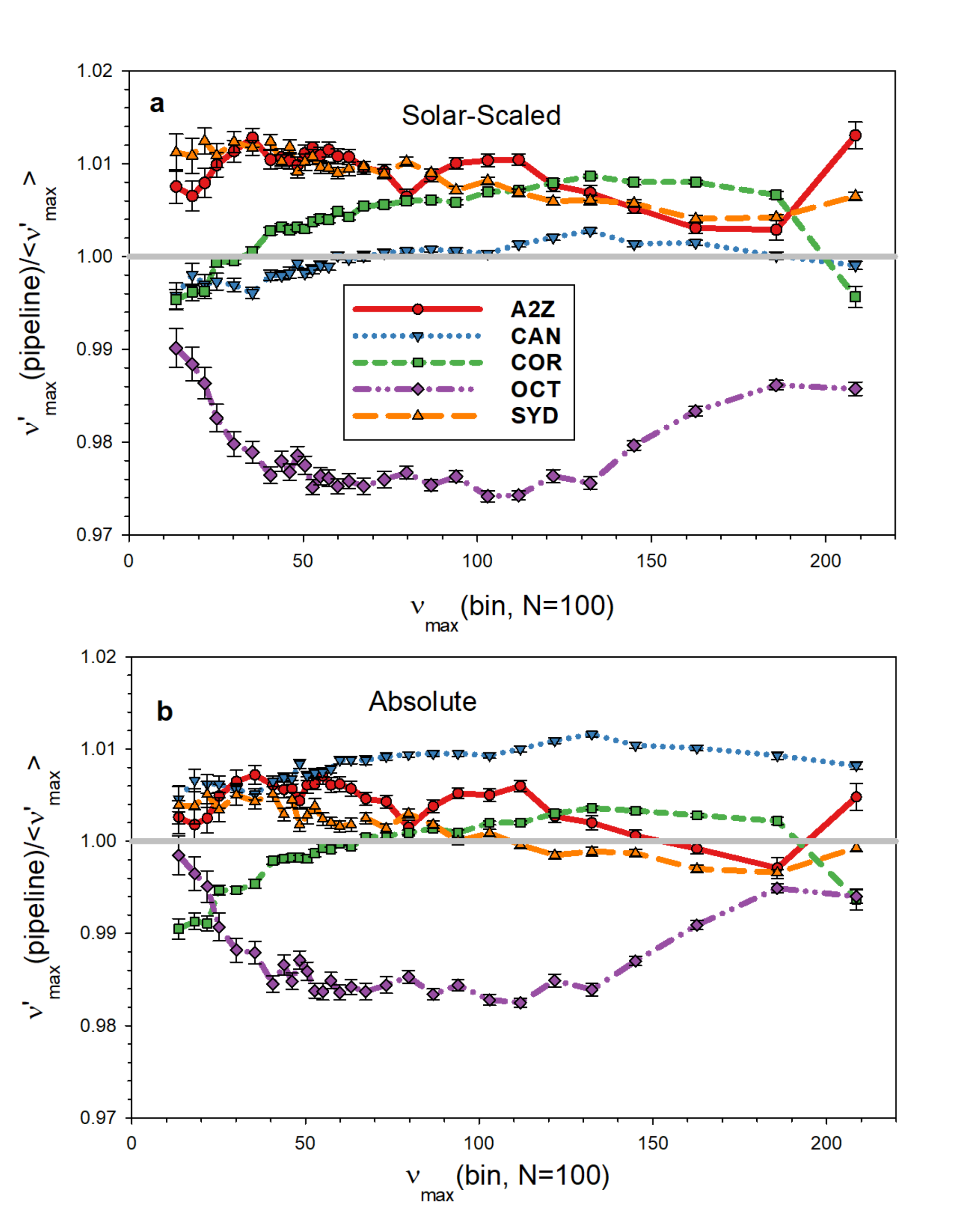}

\caption{The ratio between individual pipeline measurements of $\nu_{\rm max}$ and the average for each star as a function of $\nu_{\rm max}$ for RGB stars.  Data points are mean values of these ratios for rank-ordered bins of 100 targets between in absolute terms (b, bottom) and in solar-normalized terms (a, top), with error bars reflecting the standard error of the mean.}

\end{figure}

We repeat this exercise for $\Delta \nu$, and the results are illustrated in Figure 7. In contrast to the $\nu_{\rm max}$ case, there is a slight improvement in the agreement between pipelines in the solar-normalized case ($\sigma$ = 0.0041 for solar normalized ratios, as opposed to $\sigma$ = 0.0047 for absolute ones); however, there are still significant systematic trends between pipeline values as a function of $\Delta \nu$. 

\begin{figure}


\plotone{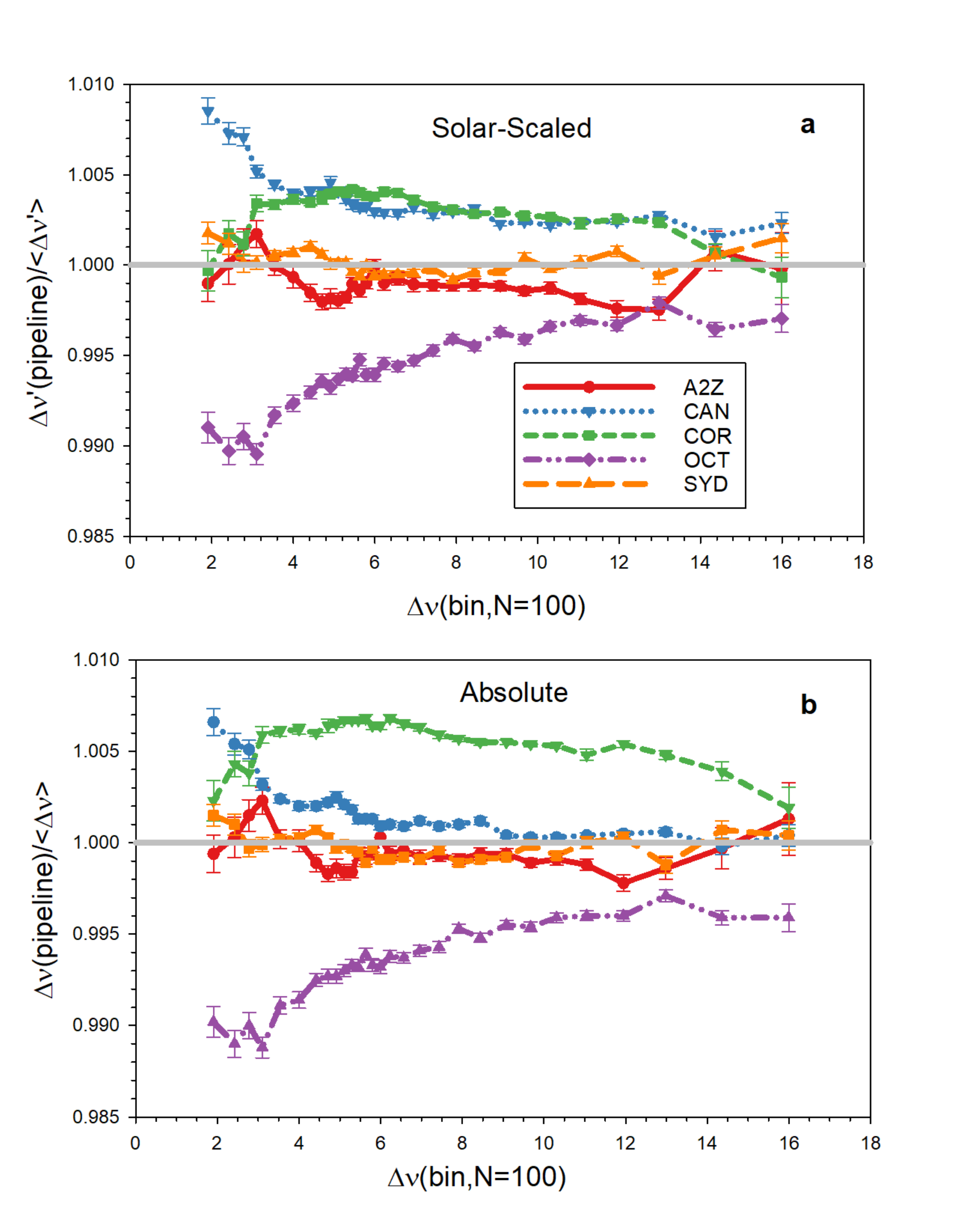}

\caption{As for Figure 6, except using $\Delta \nu$ rather than $\nu_{\rm max}$.}

\end{figure}

Our final mean values are taken using the scale factors presented in Table 2, which also contains a statistical description of our data.  In this normalized framework, trends in the mean values for different pipelines as a function of $\nu_{\rm max}$ and $\Delta \nu$ are shown in Figure 8.  For intermediate values of the asteroseismic parameters, the normalized values of the pipelines agree well; for the most luminous targets (with small frequency spacings and low frequency of maximum power) pipeline to pipeline systematics are apparent at the $\pm 1\%$ percent level for $\nu_{\rm max}$ and the $\pm 0.5\%$ percent level for $\Delta \nu$.  Pipeline to pipeline differences for $\nu_{\rm max}$ are also somewhat larger than the norm in the least luminous targets as well.  These trends are consistent with well-understood data analysis challenges: for high-luminosity giants the limited frequency resolution yields less reliable measurements of asteroseismic parameters, while oscillations near the long-cadence Nyquist frequency cause systematic differences in estimating the background noise level.

These differences illustrated here are a systematic error source. For example, if we had adopted the OCT pipeline as our reference, similar to APOKASC-1, then the purple dashed lines in this figure would be straight lines at 1 on the Y axis, and the entire bundle of lines would have been shifted up or down accordingly as a function of $\Delta \nu$ and $\nu_{\rm max}$.  By construction, the mean value would be unchanged; but, especially for luminous giants, the relative values would be different, and so would the derived masses. This figure also illustrates an important point about the nature of our empirical normalization procedure: it is specific to this sample and this data set, rather than being an absolute and universal calculation.  For example, if the fraction of intrinsically luminous targets in our sample were larger, the systematic differences between methods at low $\nu_{\rm max}$ and $\Delta \nu$ would have had more weight in our solution, and we would have inferred different absolute scale factors.

\begin{deluxetable}{cccccc}
\tabletypesize{\scriptsize}
\tablecaption{Relative Pipeline Zero Points}
\tablewidth{0pt}
\tablehead{
\colhead{Quantity} & \colhead{A2Z} & \colhead{CAN} & \colhead{COR} & \colhead{OCT} &
\colhead{SYD} }
\startdata
$X_{\nu_{\rm max}, \sun}$ & 0.9981 & 1.0118 & 0.9828 & 1.0115 & 0.9957  \\
$X_{\nu_{\rm max}, RGB}$ & 1.0023(2) & 1.0082(2) & 0.9989(2) & 0.9900(2) & 1.0006(2)  \\
$X_{\nu_{\rm max}, RC}$ & 1.0035(3) & 1.0067(2) & 0.9909(2) & 0.9979(4) & 1.0010(3)  \\
$\sigma_{\nu_{\rm max}, RGB}$ & 0.010 & 0.006 & 0.006 & 0.012 & 0.009  \\
$X_{\Delta \nu, \sun}$ & 1.0004 & 0.9980 & 1.0026 & 0.9993 & 0.9997  \\
$X_{\Delta \nu, RGB}$ & 0.9993(1) & 1.0007(1) & 1.0051(1) & 0.9955(1) & 0.9995(1)  \\
$X_{\Delta \nu, RC}$ & 0.9965(3) & 1.0108(2) & 0.9960(1) & 0.9935(2) & 1.0032(2)  \\
$\sigma_{\Delta \nu, RGB}$ & 0.006 & 0.004 & 0.004 & 0.005 & 0.003  \\
\enddata


\tablecomments{Error-weighted mean ratios of values from individual pipelines to the ensemble average.}


\end{deluxetable}

\begin{figure}


\plotone{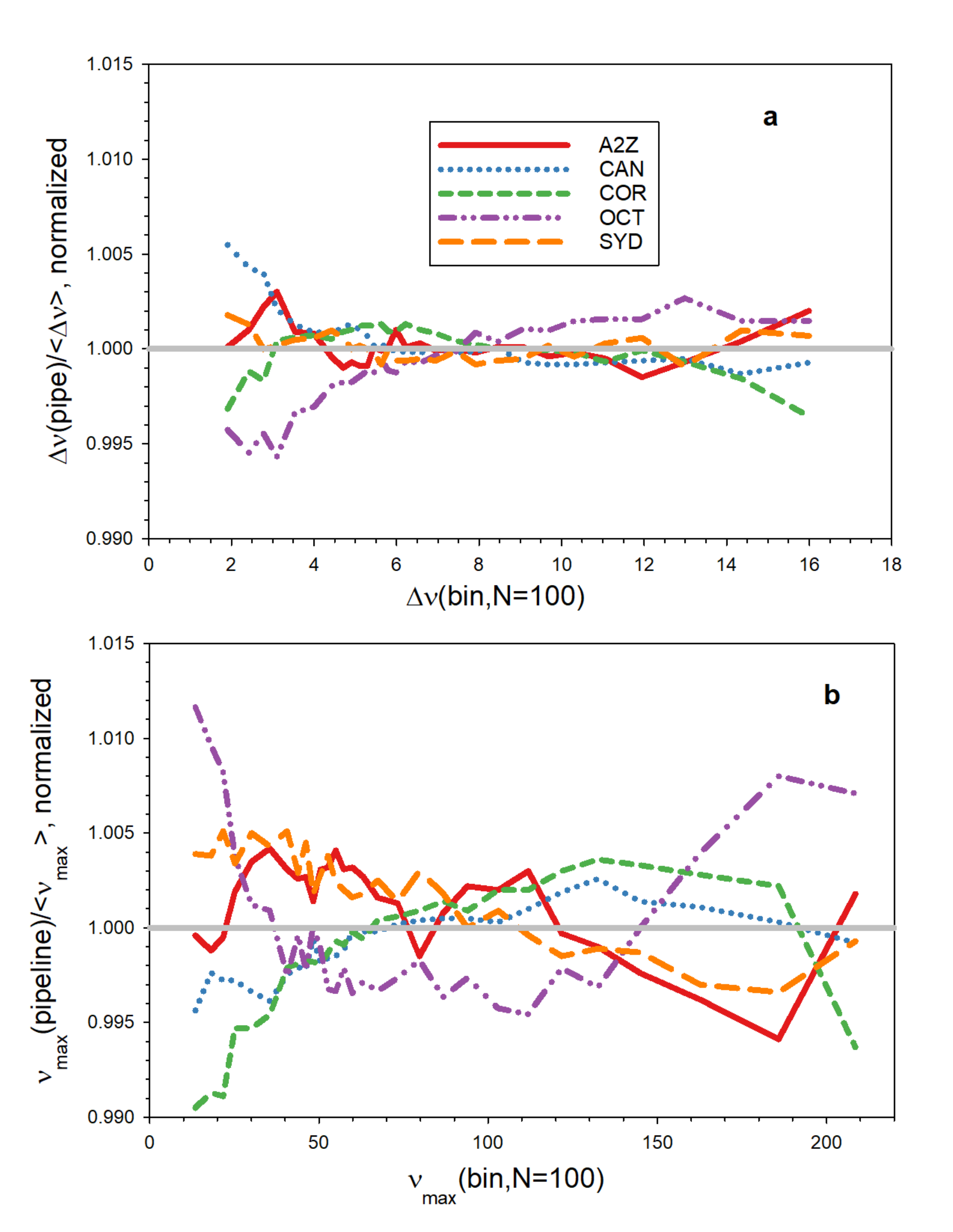}

\caption{Pipeline values for $\Delta \nu$ (top) and $\nu_{\rm max}$ (bottom) compared to the ensemble mean for targets classified as RGB stars after adjusting for the scale factor differences indicated in Table 2.  The lines connect averages of 100 targets in rank-ordered bins of 100, and the fractional dispersion of each pipeline around the mean is also given in Table 2.}

\end{figure}

We can also use our large sample of RC stars to test whether the data for them has the same overall behavior as that of RGB stars.  As the observed frequency spectra are quite different, evolutionary-state dependent differences are certainly possible.
We present the RC results in Figure 9. If we use the same scale factors as those derived for the RGB, it is apparent that there are real pipeline to pipeline differences in the relative RC zero points.  Although the scatter would be reduced if we were to adopt a RC-specific normalization, there are significant trends with seismic parameter ($\nu_{\rm max}$ or $\Delta \nu$) which would still yield significant method-dependent scatter.  As our masses and radii are ultimately calibrated on RGB stars, we choose to adopt the RGB normalization for the relative pipeline values. We will use the systematic differences in Figure 9 as a guide to systematic uncertainties in the relative derived RC masses, which are significantly higher than the corresponding trends on the RGB (as reflected in Figure 8).

\begin{figure}


\plotone{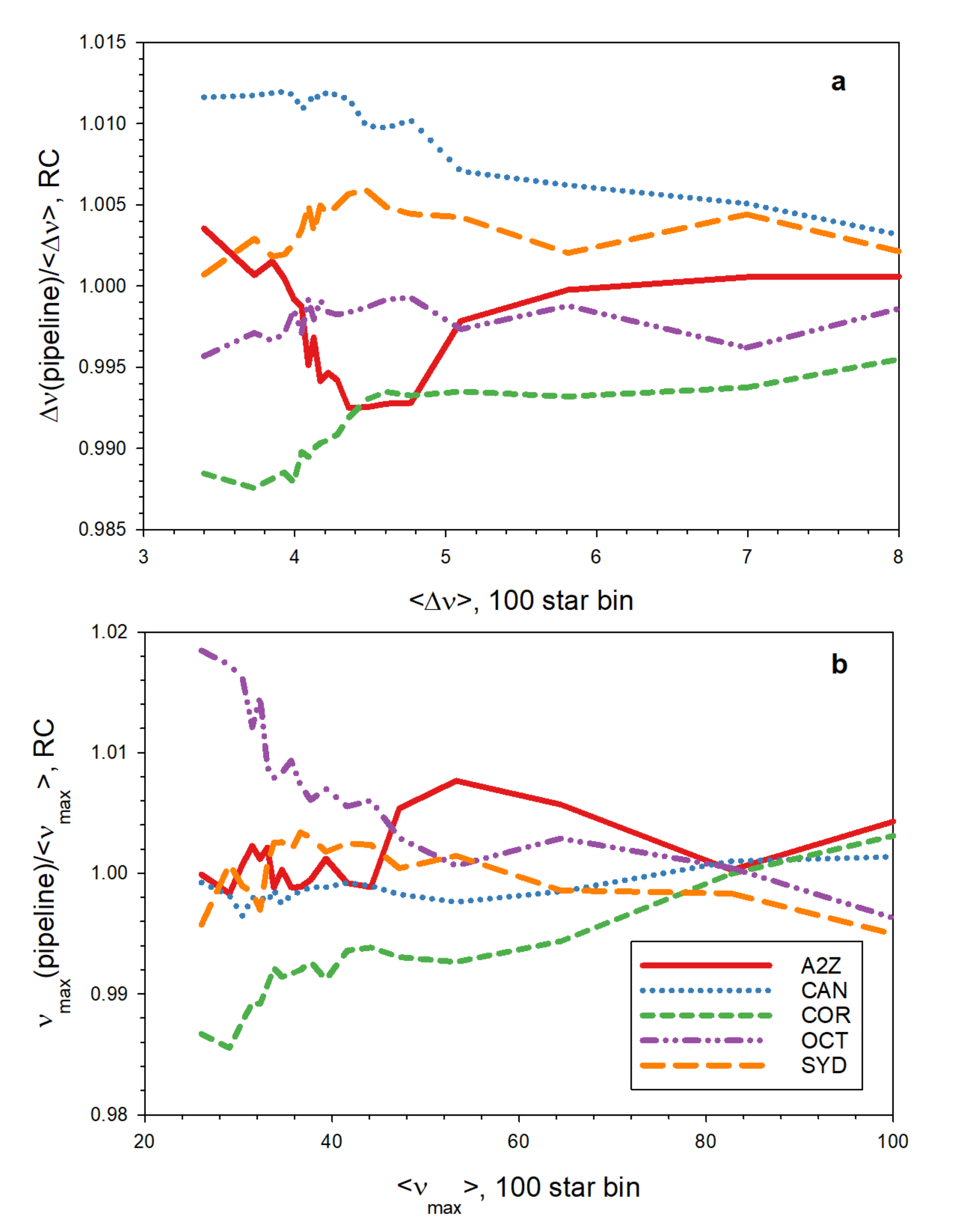}

\caption{Pipeline values for $\Delta \nu$ (a) and $\nu_{\rm max}$ (b) compared to the ensemble mean for targets classified as RC stars after adjusting for the RGB scale factors. The lines connect averages of 100 targets in rank-ordered bins of 100.}

\end{figure}


\subsubsection{Random Uncertainties in $\nu_{\rm max}$ and $\Delta \nu$}

Our treatment of the random uncertainties represents another significant change in the model.  Prior work has used the formal values returned by the pipelines as a measure of random errors and combined these values with pipeline-to-pipeline differences in results to infer total error budgets. To test the pipeline error models, we compare the dispersion of measurements from each pipeline around the normalized mean to the dispersion that would have been predicted by the formal pipeline uncertainties. In effect, this is testing whether the pipeline uncertainties behave as one would expect for random error sources, i.e. that they reflect how well the given method predicts the average measurements returned by all analysis techniques. The results are presented in Figure 10 (for $\nu_{\rm max}$ on the left and for $\Delta \nu$ on the right). Two features become immediately obvious: 1) the pipelines usually (but not always) predict uncertainties much larger than the observed method to method differences; 2) there is no clear mapping between the formal predicted uncertainties for the different pipelines and their true scatter around the ensemble mean.  To take a concrete example, if we were to use the pipeline-predicted uncertainties in an error-weighted mean, we would have assigned very high weight to the CAN measurements of $\Delta \nu$ and lower weight to the COR measurements.  However, the two pipelines are very similar in terms of how well they predict the ensemble mean values; there is no evidence that COR values have larger random measurement scatter.

\begin{figure}


\plotone{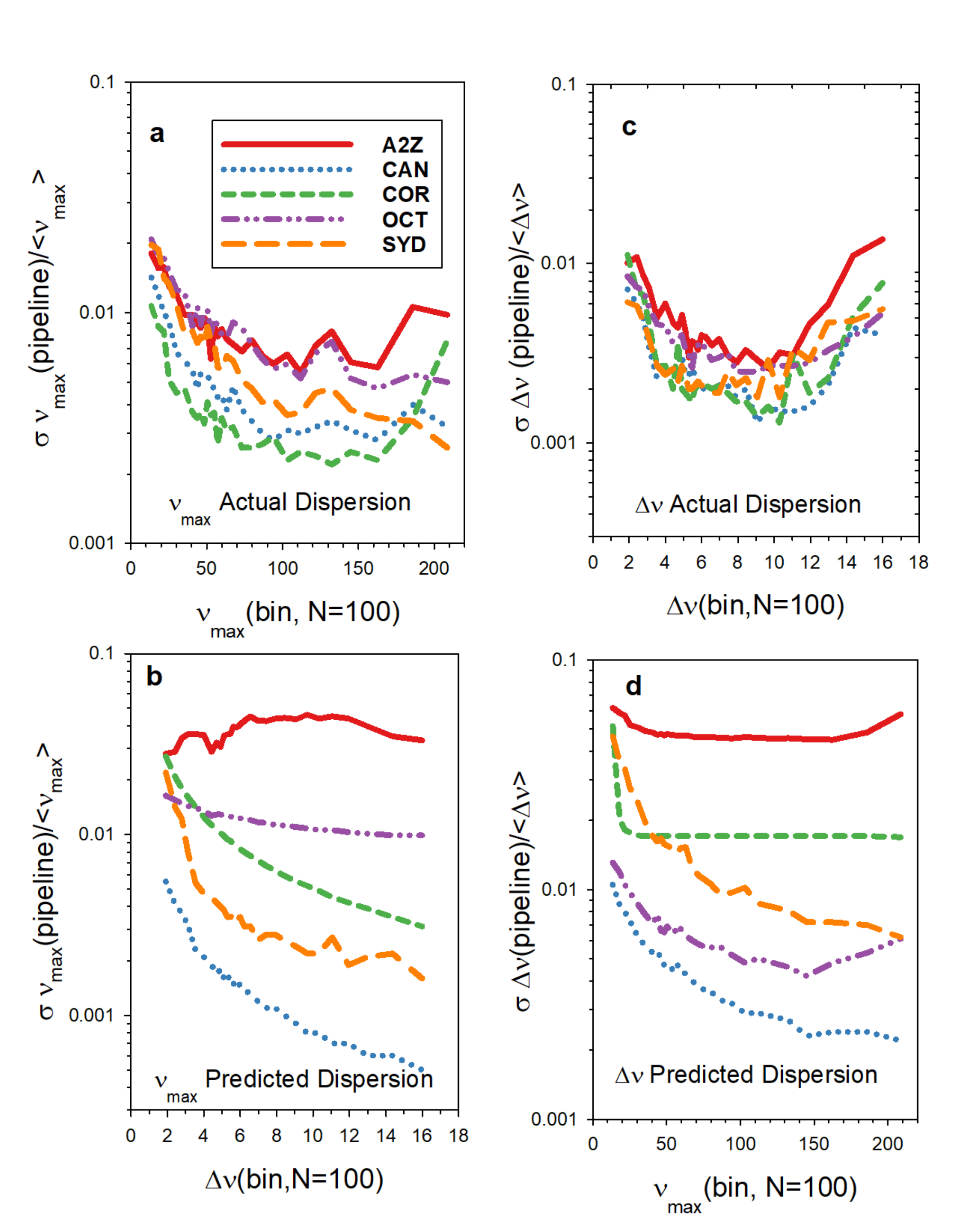}

\caption{Measured fractional dispersion in $\nu_{\rm max}$ (a,b - left) and $\Delta \nu$ (c,d - right) of pipeline values around the ensemble mean (a,c - top) compared with the formal fractional errors (b, d - bottom) for RGB stars.  The pipeline mean values were adjusted for the scale factor differences shown in Figure 6.  The lines connect averages of 100 targets in rank-ordered bins of 100.}

\end{figure}






Some caution is in order, as the dispersion that we are testing here is not necessarily the random error in the underlying data. Rather, we are measuring the dispersion between methods being used to infer the mean asteroseismic parameters.  In the case of the \textit{Kepler} light curves for evolved stars, however, the targets are bright and the time series are extremely long.  As a result, it is at least a plausible hypothesis that our ability to interpret the light curves, rather than signal to noise, is the primary contributor to the error budget.  This would not in general be true for shorter time series or lower signal to noise data, for example in K2 or TESS.  We therefore adopt an empirical random uncertainty based on the concordance between different methods, and test this error model against star cluster data in the next section.

From the results presented here, it is clear that weighting the various pipelines by their formal errors, or comparing method to method based on the formal errors, does not accurately capture how well the different techniques can predict the ensemble mean of any given star.  The data in Figure 10 also collapses the distribution of uncertainties down to a single figure of merit, $\sigma$, which assumes a normal error distribution. We would ideally like to determine whether we can justify a single overall uncertainty estimate or whether we need a star by star measurement. 

We therefore proceed as follows.  For each star we compute an unweighted mean across pipelines of the asteroseismic properties using the relative normalizations illustrated in Figure 8, and we compute the dispersion about the mean.  The results are displayed in Figures 11 and 12 (for $\nu_{\rm max}$ and $\Delta \nu$ respectively).  There are four distinct groups represented on each figure: 1) the calibrating sample (RGB stars with results from all five pipelines, in red); 2) RGB stars with results from fewer than five pipelines, in pink; 3) RC stars with results from all five pipelines, in blue; and 4), RC stars with results from some pipelines, in cyan.  There are some striking trends in the data.  For the RGB sample with results from all pipelines, the distribution of dispersions matches well the expectations from a normally distributed distribution with small uncertainties.  RGB targets where one or more pipelines failed, however, had substantially larger scatter, and the distribution of dispersions is clearly not drawn from a normal distribution.  The dispersions for RC stars are larger than for RGB stars; this is a combination of systematic differences between pipelines and truly larger random differences. The RC stars with partial detections have a larger dispersion than those with detections from all methods, but the differences with RC stars that were measured by all methods are not as stark as they are for the RGB case.  One possible factor is that the most problematic RC stars lacked an evolutionary state classification (see Figure 1), which removed them from our sample.  For $\Delta \nu$, the peak in the dispersion for RC stars around 0.01 in the third panel from the top reflects systematic zero-point differences between pipelines relative to the means for the RGB targets. 

\begin{figure}


\plotone{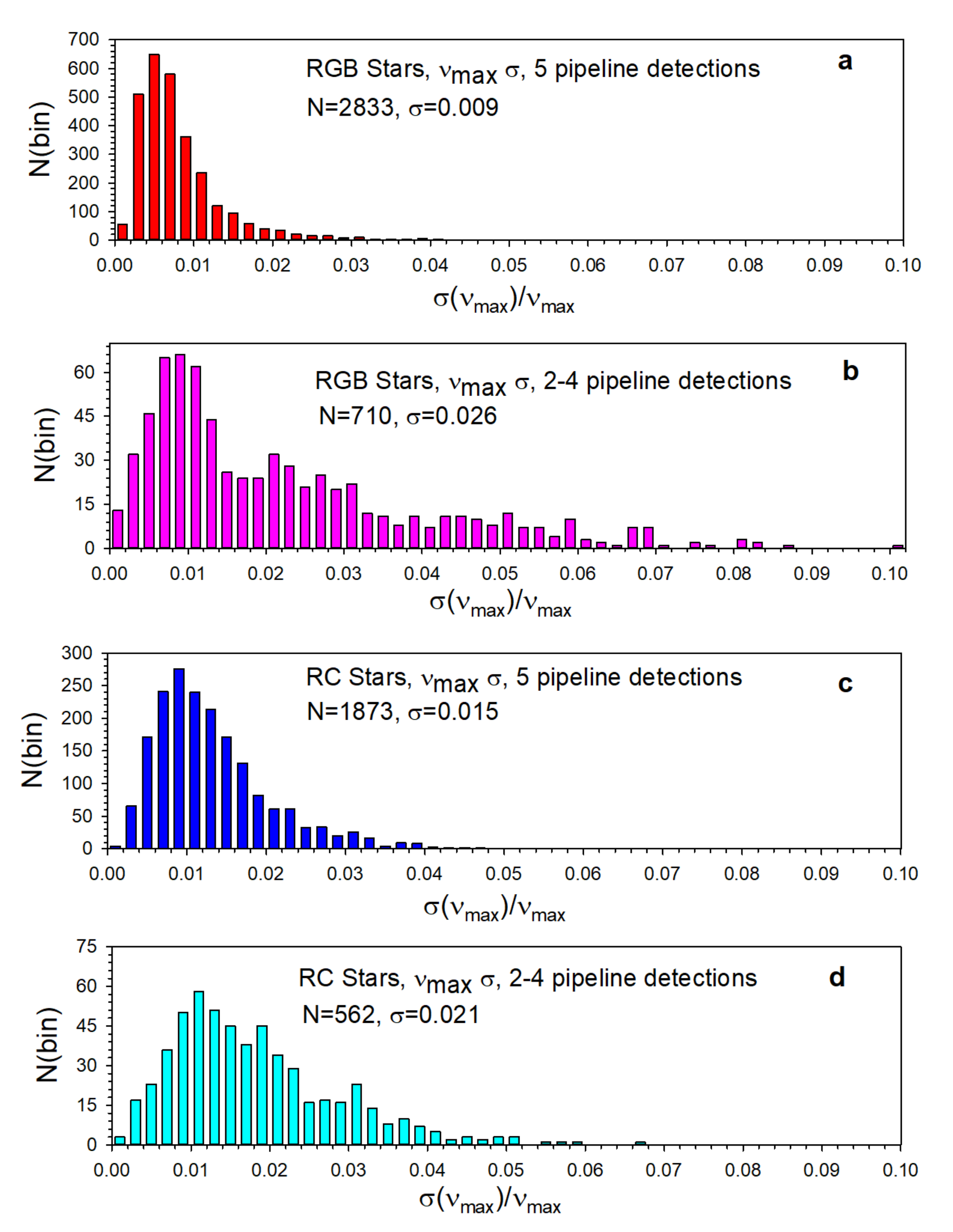}

\caption{Distribution of fractional dispersion in $\nu_{\rm max}$ of pipeline values around the ensemble mean.  The samples are RGB stars with results from all pipelines (red, a); RGB stars with results from 2-4 pipelines (pink, b); RC stars with results from all five pipelines (blue, c), and RC stars with results from 2-4 pipelines (cyan, d.)  The sample sizes and formal dispersions are indicated in the panels.}

\end{figure}

In light of these results, we choose to treat the fractional standard deviation of our sample measurements for the RGB calibrators (0.009 in $\nu_{\rm max}$ and 0.004 in $\Delta \nu$ from the top panels of Figures 11 and 12) as a minimum fractional random uncertainty for the asteroseismic parameters. If the fractional dispersion of the normalized measurements around the mean are larger than these minimum values, we adopt them instead for our error analysis. This conservative approach assigns larger uncertainties to targets where different analysis methods disagree by more than the norm, while avoiding unphysical small formal error estimates for targets with small formal dispersions. It is not uncommon to have multiple measurements agree much better than their formal dispersion would predict if the sample size is small; our approach avoids this pitfall.  

\begin{figure}


\plotone{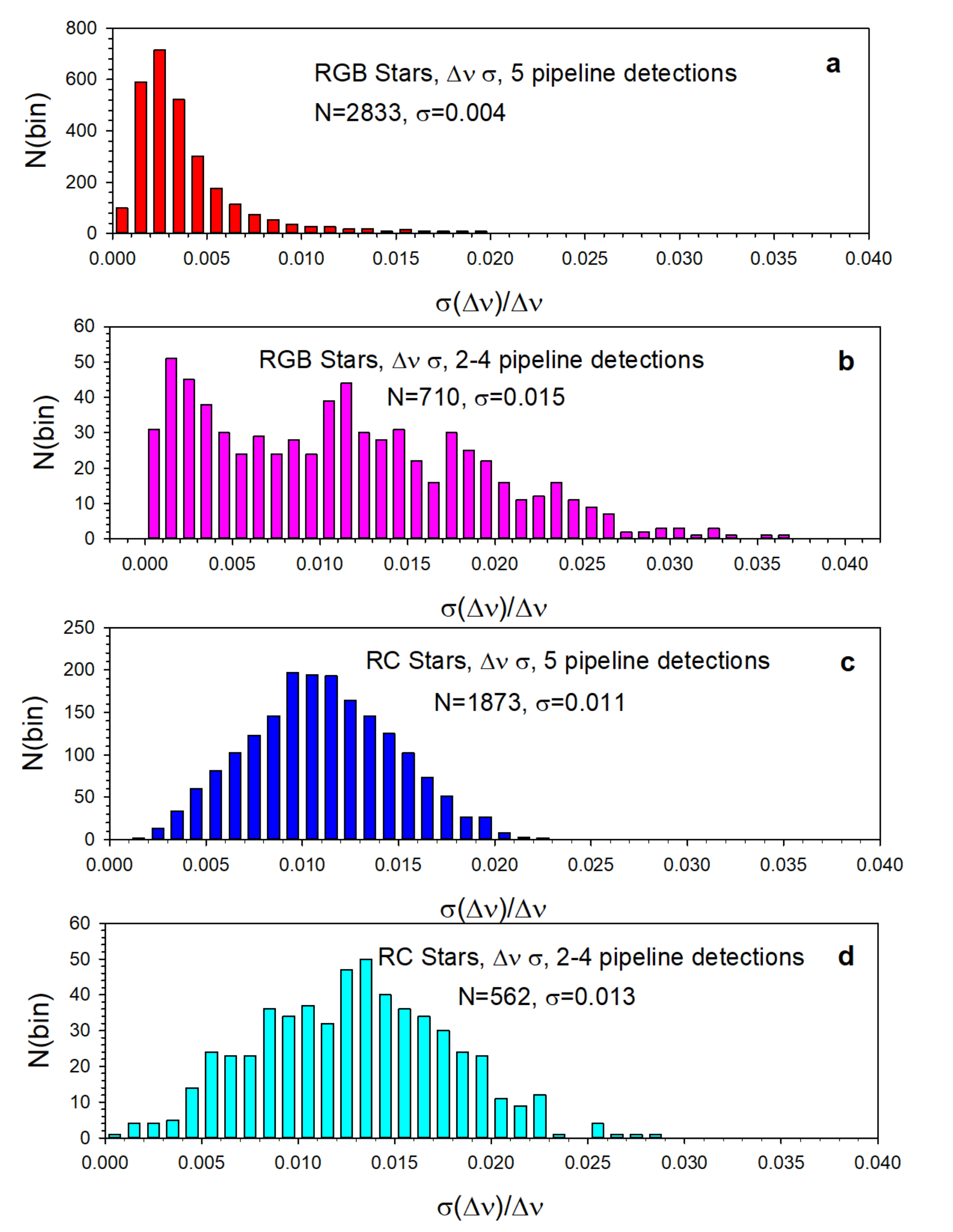}

\caption{Distribution of fractional dispersion in $\Delta \nu$ of pipeline values around the ensemble mean.  The samples are RGB stars with results from all pipelines (red, a); RGB stars with results from 2-4 pipelines (pink, b); RC stars with results from all five pipelines (blue, c), and RC stars with results from 2-4 pipelines (cyan, d.)  The sample sizes and formal dispersions are indicated in the panels.}

\end{figure}

\subsubsection{Systematic Uncertainties in $\nu_{max}$ and $\Delta \nu$}

Our systematic errors include the systematic differences that could arise from the choice of pipeline, the $f_{\nu_{\rm max}}$ scale factor, and the $f_{\Delta \nu}=1$ individual correction factors. Although the $\rm T_{eff}$ scale does have systematic uncertainties, they have a smaller impact on our stars than the other ingredients, so we do not include them as a systematic error source (although they are a random error source.) The effective temperatures that we adopt are tied to the Infrared Flux Method (IRFM) fundamental scale in low extinction fields, which should be reliable across the $\rm T_{eff}$ domain of our data. APOGEE uses the \citet{ghb09} scale. There are modest differences, at the 30 K level, between their system and the more recent \citet{cas10} scale for dwarfs. Unfortunately the latter is not currently available for evolved stars, but the differences there should be comparable to those for main sequence stars.  Our empirical calibration approach makes us less sensitive to absolute zero point shifts, which would simply induce scale shift in the derived $f_{\nu_{\rm max}}$ factor.  


As illustrated in Figures 8 and 9, there are systematic offsets in both asteroseismic observables between our mean scale and the one that would have been obtained for each of our five analysis methods.  These effects dominate our systematic error budget.  In Figure 13 we show the implied shifts in the mass, radius and surface gravity scales that we would have observed if we had chosen each of the pipelines as the preferred scale, rather than an ensemble mean.  Our RC results have substantially more variance than the RGB ones. Masses have larger systematic differences than radius, which in turn has more uncertainty than surface gravities do.

To infer errors, we take the conservative approach of fitting a straight line to the upper envelope of the family of curves in Figure 13 for M and R.  This yields fractional systematic errors in mass of 0.09 - 5x$10^{-4}*\nu_{\rm max}$ for RC stars and 0.04 - 1x$10^{-4}*\nu_{\rm max}$ for RGB stars. The fractional systematic errors in radius of $0.03-2.5x10^{-4}*\nu_{\rm max}$ for RC stars and 0.015 - 5x$10^{-5}*\nu_{\rm max}$ for RGB stars. For log g and $<\rho>$, the systematic errors can be inferred directly from Figures 8 and 9, as they depend solely on $\nu_{\rm max}$ and $\Delta \nu$ respectively, rather than a combination of the two. A corresponding fit to their upper envelopes for log g is 0.01 - 5x$10^{-5}*\nu_{\rm max}$ for RC stars and 0.005 for RGB stars; for $<\rho>, we adopt fractional uncertainties of 0.01 for RC stars and 0.005 - $1.5x$10^{-5}*\nu_{\rm max}$ for RGB stars.

The uncertainty in our mean calibrator mass scale of ~2\% induces a systematic error for our $f_{\nu_{\rm max}}$ of 0.007, which we treat as a systematic error source.  Our derived $f_{\Delta \nu}=1$ factors, depend on the stellar parameters, and are larger for RGB stars than RC ones. As a result, systematic errors in the \textit{differential} values for the two groups exist. We discuss this systematic error source, which is larger for RC stars than RGB ones, at the end of Section 3.2.


\begin{figure}


\plotone{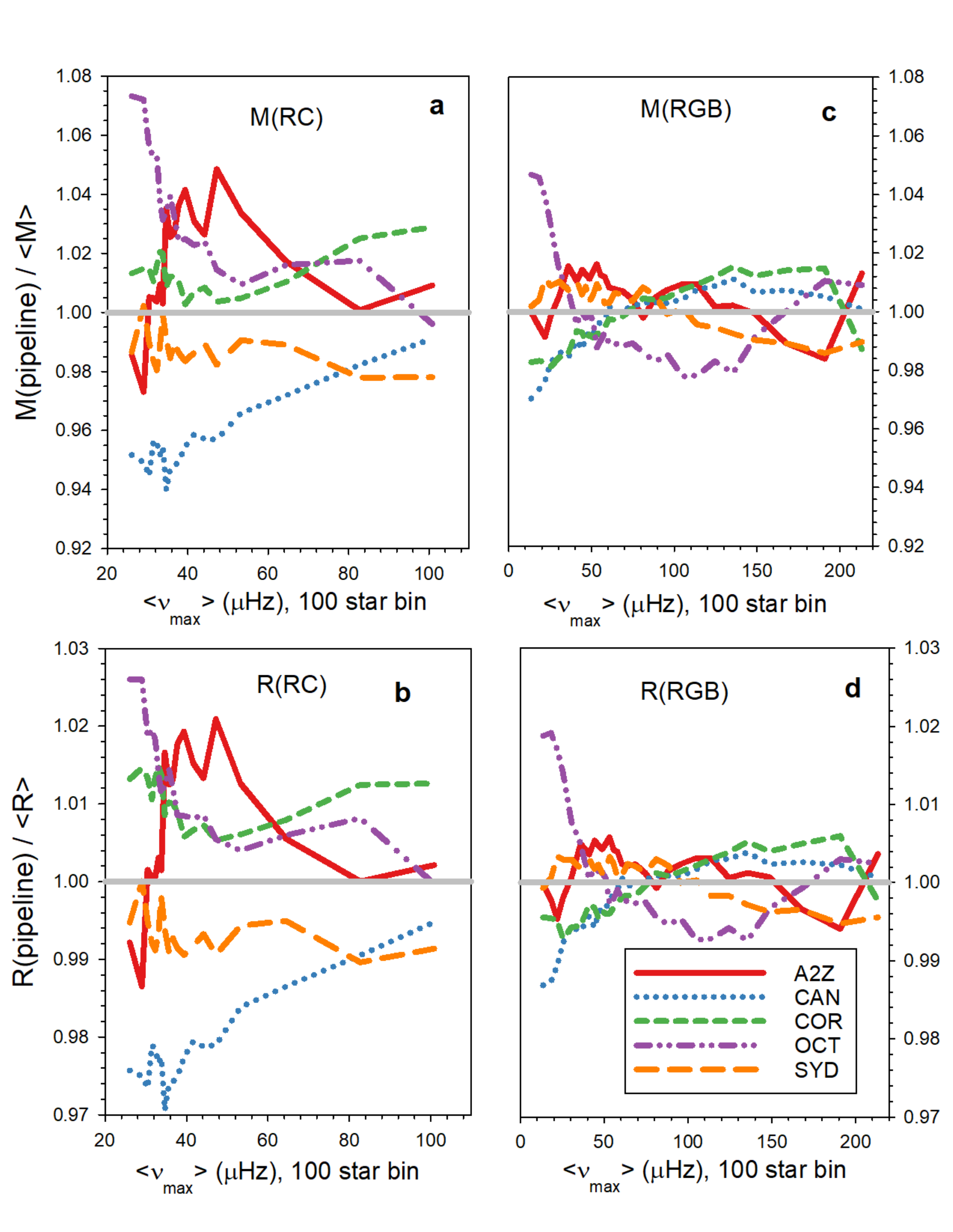}

\caption{Systematic uncertainties in mass (a,c), and radius (b, d) for RC stars (a, b) and RGB stars (c, d).  The different lines represent the changes in the mean value that would have been obtained had we adopted each of the five input pipelines as the reference value rather than adopting the ensemble mean.}

\end{figure}

As an illustration of the potential systematic effects that could be present, we examined all stars with data from all five pipelines. We used the scale factors in Table 2 to place the $\Delta \nu$ and $\nu_{\rm max}$ measurements from all pipelines on the same average scale.  We then constructed mean asteroseismic parameters $<\Delta \nu>$ and $<\nu_{\rm max}>$ by averaging these corrected values.  For each star, we then used equations (1) and (2) to infer the masses that we would have obtained using each pipeline alone, as well as the mass implied by the ensemble average.  The ratio of the masses that we would have obtained from each pipeline to the ensemble mean are shown for RGB and RC stars in Figures 14 and 15 respectively.  Note that the pipeline values for $\Delta \nu$ and $\nu_{\rm max}$ were placed on the same mean system before computing masses, so zero-point shifts have been suppressed. This exercise uses simple scaling relations.  We see well-behaved errors in the RGB case, consistent with method-dependent systematics being well-controlled there, and larger offsets in the RC case.  For a fuller discussion of these systematics, we refer the interested reader to \citet{s18}.

\begin{figure}


\plotone{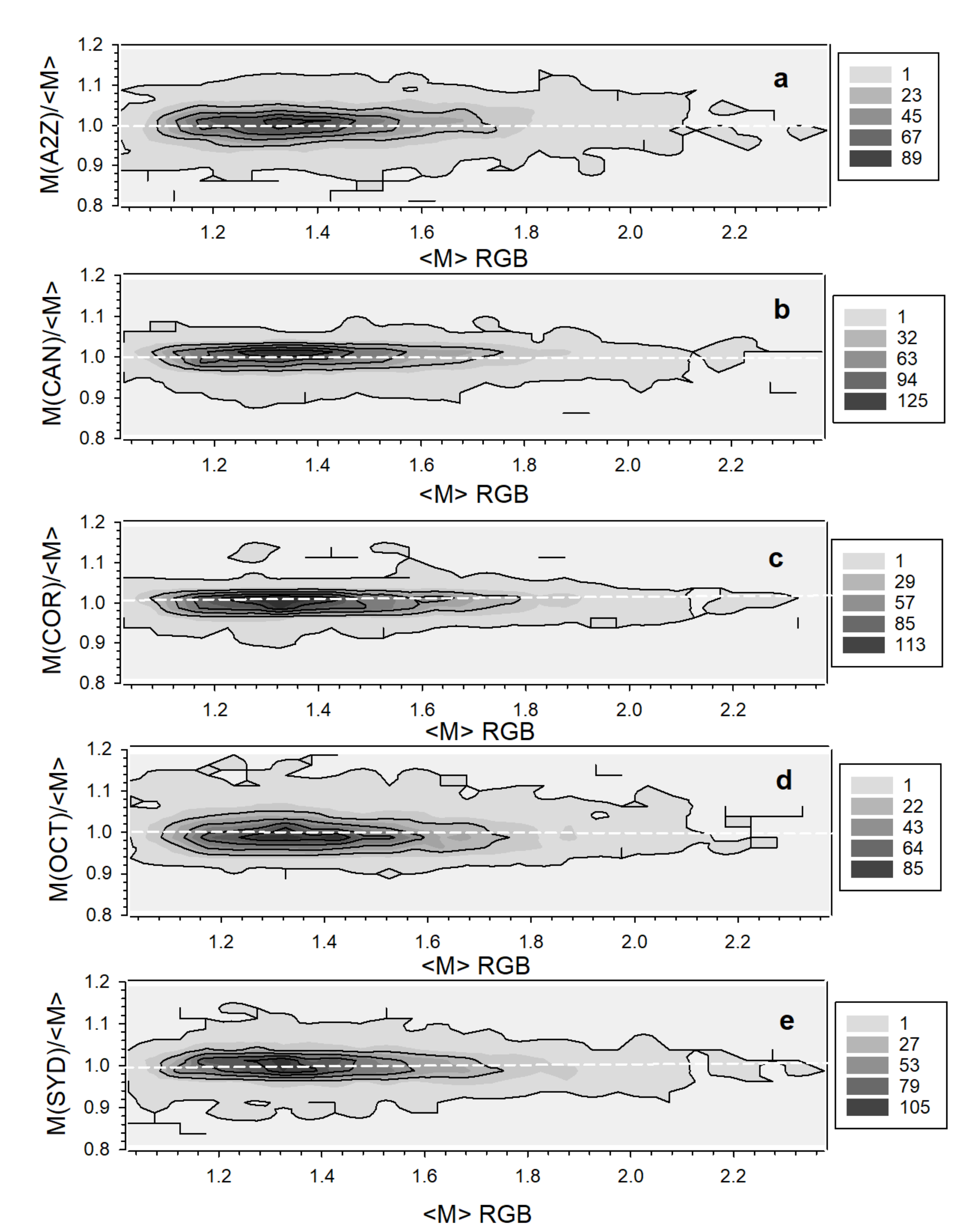}

\caption{The ratio of the masses using data obtained separately from each of our five pipelines (A2Z, a; CAN, b; COR, c; OCT, d; and SYD, e) to the ensemble average mass as a function of mean mass.  Data plotted are for RGB stars with parameters returned from all methods. Masses were computed from equation 3 using simple scaling relations and averaged solar reference values. Color indicates the density of objects in the bin.}

\end{figure}

\begin{figure}


\plotone{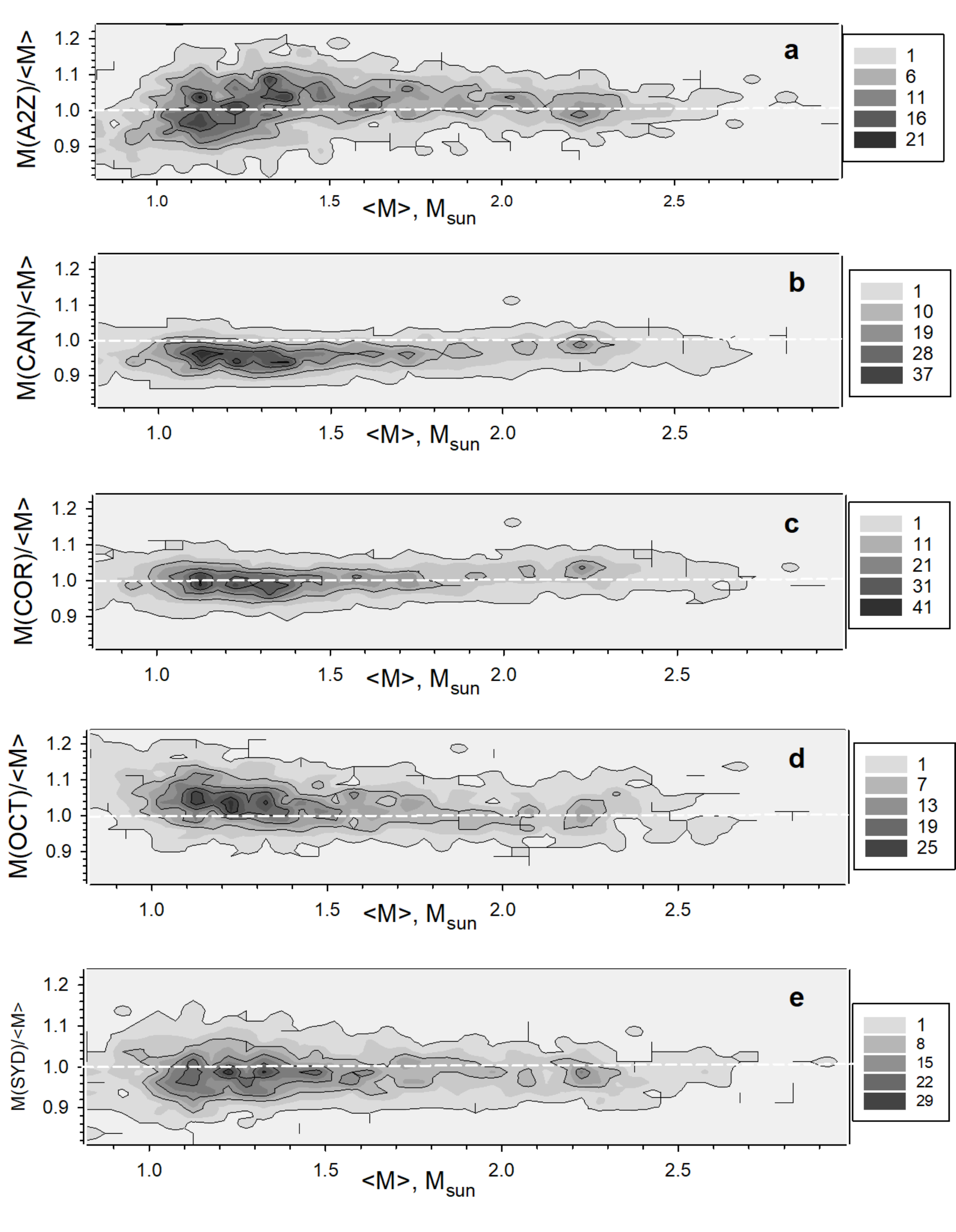}

\caption{As for Figure 14, except for RC stars. Systematic errors are significantly larger in this case than for the RGB one.}

\end{figure}

\subsection{Corrections to the $\Delta \nu$ Scaling Relation}

In a departure from the APOKASC-1 approach, we apply theoretically motivated corrections to the $\Delta \nu$ scaling relation. These corrections can be computed with knowledge of the stellar mass, composition, evolutionary state, surface gravity and effective temperature. We do this by implementing stellar models for which one derives $\Delta\nu$ from adiabatically-derived radial modes on the one hand and from the $\Delta \nu$ scaling relation on the other; see \citet{whi11} for an example. For our open cluster calibrators, we assume a known true mass and do a table lookup to infer the correction using the procedure below.  For the general case, we must iteratively solve for the correction factor and the mass (as our final mass estimate requires knowledge of the correction term itself.) This approach is similar in spirit to that employed by \citet{rod17}.

The determination of the correction to the $\Delta \nu$ scaling relation is done using a modified version of grid-based modeling as implemented in the BeSPP code \citep{s18} in which the input data are stellar mass, surface gravity, [Fe/H], and [$\alpha$/Fe].  The procedure is implemented as follows. We use the average solar reference values defined in Table 1. 
For the asteroseismic measurements, we take the average of the normalized measurements (the raw values divided by the factors given in Table 2).  The uncertainties are the larger of our minimum uncertainty for all stars (0.009 and 0.004 for $\nu_{\rm max}$ and $\Delta \nu$ respectively) and the fractional dispersion in normalized measurements.  

For star cluster members we used the asteroseismic $\nu_{\rm max}$ and the spectroscopic $\rm T_{eff}$ to infer log g.  We then perform a table look-up at that surface gravity using the mean cluster mass, [Fe/H] and [$\alpha$/Fe] in Table 3 to infer $f_{\Delta \nu}$ factors for our open cluster calibration exercise.

For the remainder of the sample we had to adopt an iterative approach. Initial guesses $M_0$ and $g_0$ for the stellar mass and the surface gravity respectively are obtained using the scaling relations in equations 3 and 4 with $\rm T_{eff}$, $\Delta \nu$, and $\nu_{\rm max}$ as inputs. The uncertainties $\sigma_{M_0}$ and $\sigma_{g_0}$ of these quantities are also determined from the scaling relations by propagating the errors in the input quantities.  Then, an iterative procedure is run by feeding BeSPP with these quantities at step i ($M_i, g_{i}, [Fe/H], [\alpha/Fe], \sigma_{M_i}, \sigma_{g_0}, \sigma_{[Fe/H]}, \sigma_{[\alpha/Fe]}$) to compute a new value $f_{\Delta \nu, i}$ define as

\begin{equation}
f_{\Delta \nu} = \frac {\Delta \nu_{sc}}{\Delta \nu_{l=0}}
\end{equation}

Note that, because of the notation used in this paper, this ratio is the inverse of the value frequently used in the literature.  The mean value is determined from the probability distribution of $f_{\Delta \nu}$ given the quoted uncertainties. For subsequent iterations, $g_0$, [Fe/H], [$\alpha$/Fe] and their uncertainties are held fixed, while the mass estimate is updated based upon applying the $f_{\Delta \nu}$ factor. The iteration procedure continues until the mean value of $f_{\Delta \nu}$ converges to one part in $10^5$. The uncertainty of $f_{\Delta \nu}$ is defined as the standard deviation of the final probability distribution, and we include it as a random error source.

As a result of this procedure $\rm T_{eff}$ is only used to determine the initial values $M_0$ and $g_0$. Moreover, $\rm T_{eff}$  and [Fe/H] are never used simultaneously. This has the positive effect that the GBM scheme is not directly sensitive to the $\rm T_{eff}$ scale in the stellar tracks, and it is thus robust with respect to its calibration. If we had done a table look-up in correction as a function of $\rm T_{eff}$ instead of surface gravity, by contrast, systematic errors in the input stellar tracks would have a much larger impact on the derived corrections because of the steep dependence of log g on $\rm T_{eff}$ in stellar evolution tracks.

As a cross-check on our system, we independently computed $f_{\Delta \nu}$ using the method of \citet{sha16} \footnote{The stellar models along with a code to compute the correction factors are available at http://www.physics.usyd.edu.au/k2gap/Asfgrid}.  We then performed the open cluster calibration exercise (described below, in section 3.3) with this alternate approach, and propagated the masses through for the full sample,  The mass differences between our base approach (BeSPP) and this alternate approach (Sharma) are illustrated for the full sample in Figure 16. For RGB stars and RC stars the BeSPP values are higher on average by a scale factor of 1.006 and 1.002 respectively. The dispersion in the correction factor between methods is modest for the majority of cases (0.003 and 0.007 for RGB and RC stars respectively, for objects with $f_{\Delta \nu}$ between 0.99 and 1.04).  By comparison, in our reference BeSPP method, the formal uncertainty for RGB stars is 0.001 and that for RC stars is 0.012. We use the scale shifts to estimate the magnitude of the induced systematic uncertainties, which we include in our error model. In our mass calibration, which scales as $\frac{f_{\nu_{\rm max}}^3}{f_{\Delta \nu}^4}$, a scale shift of 1.006 for RGB stars would have implied a compensating scale shift of 1.008 in $f_{\nu_{\rm max}}$.  The net impact would be systematic shifts in M, R, log g and $<\rho>$ of 0.016, 0.004, 0.004, and 0.002 for RC stars.  RGB stars would have systematic shifts in M, R, log g and $<\rho>$ of 0 (by construction), 0.004, 0.004, and 0.006. 


\begin{figure}


\plotone{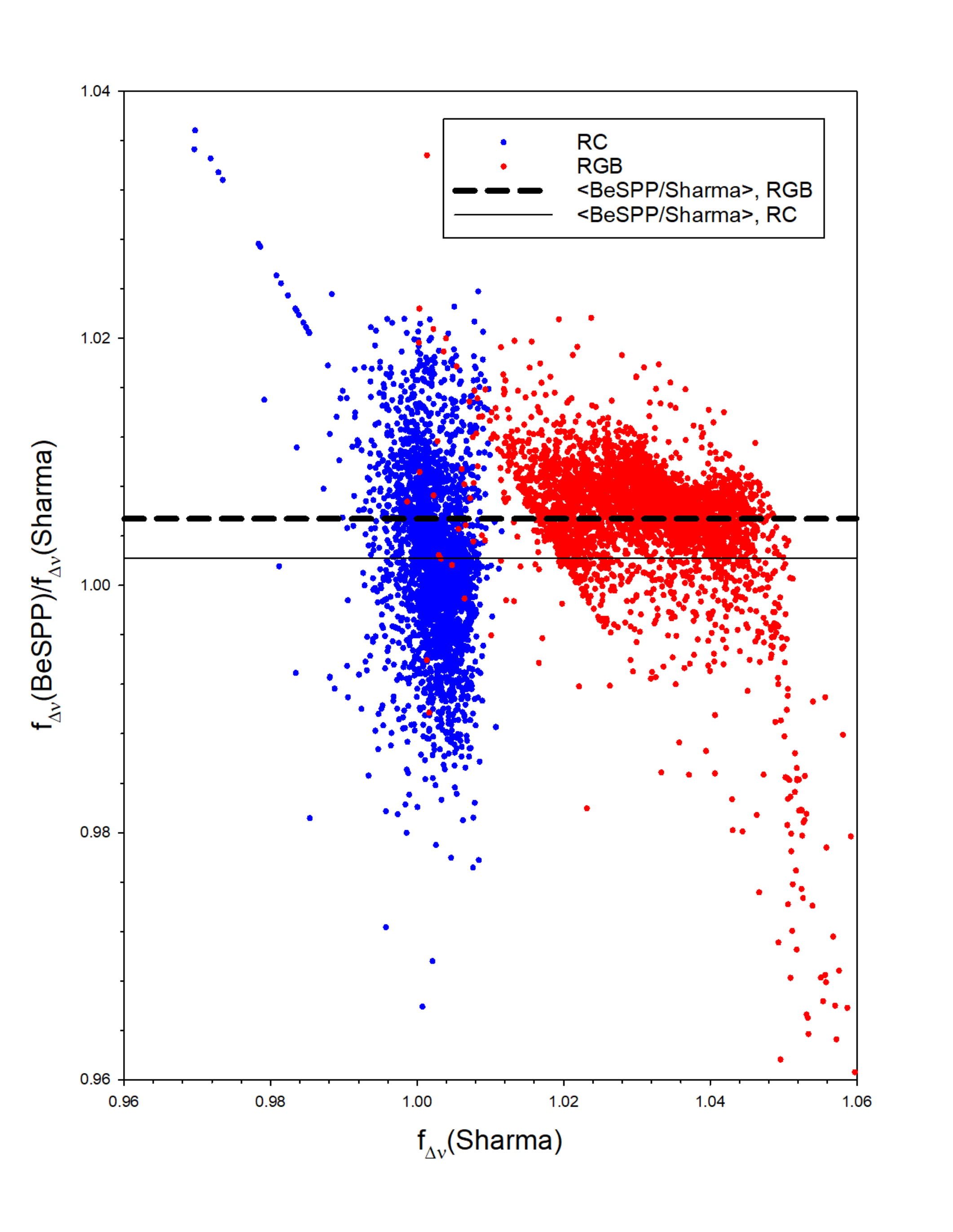}

\caption{The ratio of the $\Delta \nu$ correction factors from two different calculation methods.  Data is shown for RGB stars (red) and RC stars (blue) a, and differences are defined in the sense $f_{\Delta \nu}^{BeSPP}$/$f_{\Delta \nu}^{Sharma}$, where $f_{\Delta \nu}^{BeSPP}$ uses the \citet{s18} correction factors and $f_{\Delta \nu}^{Sharma}$ uses the \citet{sha16} correction factors.  Masses and mass differences are in solar units. Mean values for RC stars (solid line) and RGB stars (dashed line) are shown; there is a clear differential offset between the two.}

\end{figure}

\subsection{Open Cluster Calibration of the $\nu_{\rm max}$ Scaling Relation}

Our calibration for the scale factor $f_{\nu_{\rm max}}$ comes from requiring that the masses derived for RGB members of star clusters be in agreement with fundamental measurements. For this purpose, we treat the zero point of the $\Delta \nu$ scaling relation as being fixed. In a technical sense,  that there is a degeneracy between the ratio of the adopted solar reference values $(\nu_{\rm max,\sun}^3/\Delta \nu_{\sun}^4)$ and the $f_{\nu_{\rm max}}$ factor that we define here, so the latter factor can also be thought of as defining an effective or calibrated solar reference value.

Mass loss complicates the expected initial masses of RC stars, so our mass calibration uses only stars asteroseismically classified as RGB. In both NGC 6791 and NGC 6819, there are eclipsing binary stars near the turnoff with well-measured masses.  The relative masses expected on the red giant branch are only weakly model dependent. For NGC 6791, \citet{bro12} did a comprehensive study of the expected masses of lower RGB stars, and we adopt their value of $1.15\pm0.02 M_{\sun}$.  NGC 6819 has a higher uncertainty, largely because of complications in interpreting the eclipsing binary system closest to the turnoff there; see \citet{bre16} for a discussion.  For NGC 6819, we used the isochrones and ages in \citet{bre16} to infer a mean predicted RGB mass of $1.55\pm0.04 M_{\sun}$.  


For each such star j, we can compute $<\Delta \nu>^j$, $<\nu_{\rm max}>^j$, $M^j$, and their associated errors, using the procedure outlined in Section 3.1. We define a mean cluster mass derived from scaling relations by

\begin{equation}
M_{sc}^{\rm cluster} = \frac{\sum_{j=1}^{N_{\rm rgb}} \frac{M_{sc}^{j}}{(\sigma_{M}^{j})^2}}{\sum_{j=1}^{N_{\rm rgb}}\frac{1}{(\sigma_{M}^{j})^2}}
\end{equation}

We can then define a trial mean cluster mass that we would have obtained after applying the star by star corrections to the $\Delta \nu$ scaling relation, but setting $f_{\nu_{\rm max}}=1$, by 

\begin{equation}
M_{\rm trial}^{\rm cluster} = \frac{\sum_{j=1}^{N_{\rm rgb}} \frac{M_{sc}^{j}}{(f_{\Delta \nu}^j)^4(\sigma_{M}^{j})^2}}{\sum_{j=1}^{N_{\rm rgb}}\frac{1}{(\sigma_{M}^{j})^2}}
\end{equation}

If we require that the corrected scaling relation mass $M_{cor}$ from equation 5 equal the fundamental mass, we can then solve for $f_{\nu_{\rm max}}$ by

\begin{equation}
f_{\nu_{\rm max}}^{3} = \left( \frac{M_{\rm fund}^{\rm cluster}} {M_{\rm trial}^{\rm cluster}} \right)
\end{equation}

We have 17 RGB stars in NCG 6791 and 23 in NGC 6819; of these, 10 stars in each cluster have asteroseismic parameters from all pipelines, while the remainder have measurements from a minimum of two pipelines.  Our cluster data in a color-magnitude diagram is shown in Figure 17, with the RGB and RC stars used in this analysis illustrated as large colored solid and open symbols respectively.  For reference, we compare both data sets to isochrones. 

As we are using ensemble averages to infer mean masses, we have to be alert to biases induced by stars with an unusual evolutionary origin - for example, evolved over-massive blue stragglers. None of our NGC 6791 targets have statistically unusual masses.  However, 5 stars in NGC 6819 were flagged as problematic and not used in computing mean sample properties, and all were flagged as outliers in \citet{han17}.  The RGB star KIC 5113061, visible in the upper left corner of Figure 17, is in an unusual part of the CMD, even though its mass is in a reasonable range. The RGB star KIC 5024272 is a high mass outlier at $2.65\pm0.41$. The RC star KIC 4937011 is highly under-massive (below 0.6 solar masses formally), while KIC 5023953 and KIC 5024476 are statistical outliers on the high mass side (at $2.05\pm0.11$ and $2.47\pm0.16$ solar masses respectively.) Two other overmassive stars in \citet{han17}, KIC 5112880 and KIC 5112361, are statistically consistent with the mean cluster trend and were retained in our analysis. We caution that our uncertainties for some of these stars are large, and their unusual masses may reflect measurement difficulties rather than an anomalous origin; because of large uncertainties, they also would not have significantly impacted our mean properties had we included them.

In Figure 18 we show the scaling relation masses prior to any corrections for NGC 6791 and NGC 6819.  As known from prior work, the classical scaling relation masses are over-estimated, with formal mean RGB masses of $1.294 \pm 0.017 M_{\sun}$ for NGC 6791 and $1.772 \pm 0.034 M_{\sun}$ for NGC 6819, both ruled out at high statistical significance.

\begin{figure}


\plotone{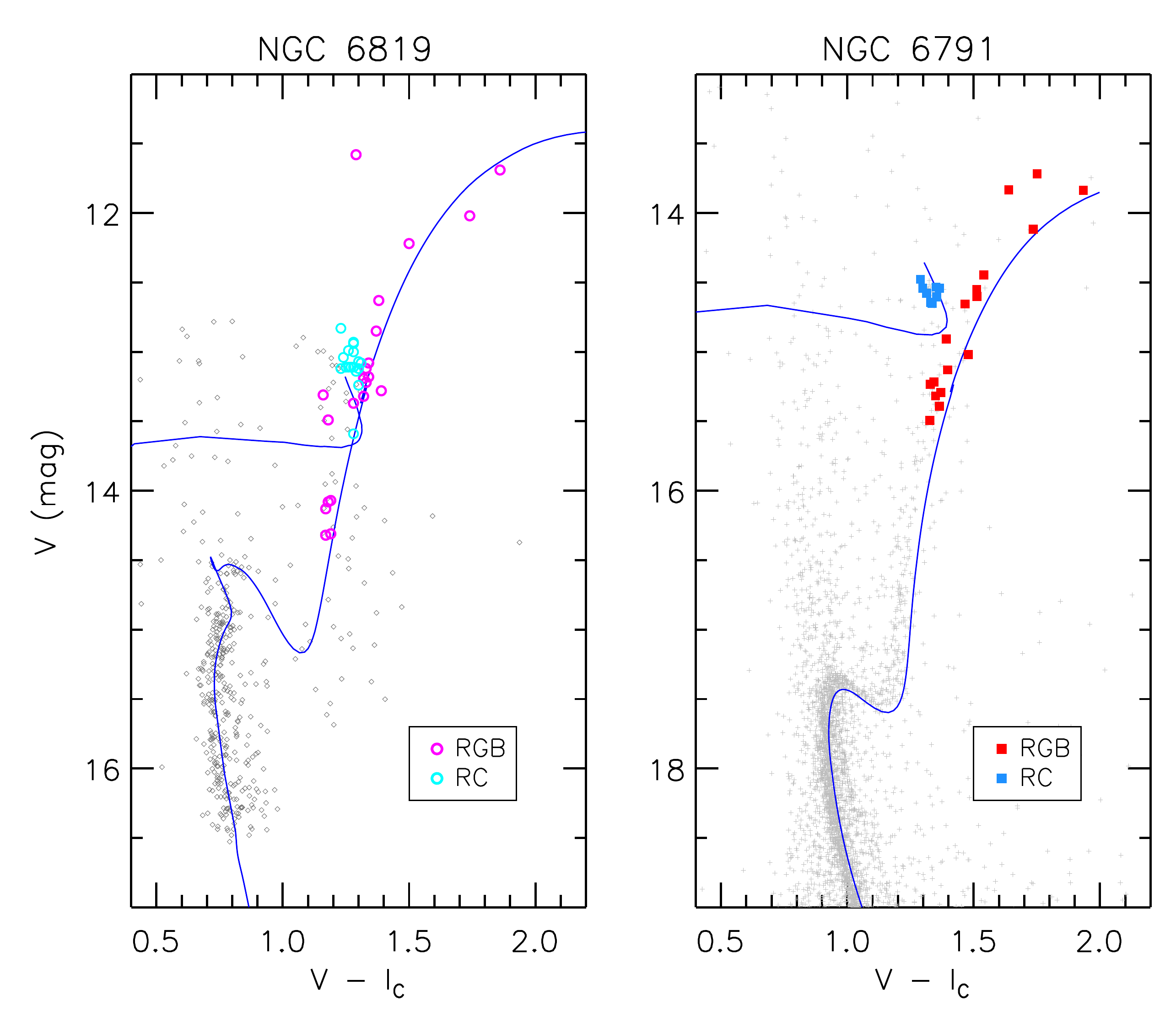}

\caption{CMDs of NGC 6819 (left), with data from \citet{hol09}, and NGC 6791 (right), with data from \citet{bro12}. RGB and RC stars with asteroseismic data are highlighted with red circles and blue triangles (NGC 6791) and with pink circles and cyan triangles (NGC 6819), respectively. Radial velocity members (prob >= 50\%) in Hole et al. are shown in NGC 6819. Isochrones from the Dartmouth Stellar Evolution Database, or DSEP,  \citep{dot08} are overlaid taking the CMD-based cluster parameters adopted in this paper.
%
%
The zero-age horizontal branch models are obtained from evolutionary tracks in the Dartmouth Stellar Evolution Database, with broad-band colors derived using \citet{gir02}. Extinction and reddening coefficients in \citet{an07} are adopted.
}

\end{figure}  

\begin{figure}


\plotone{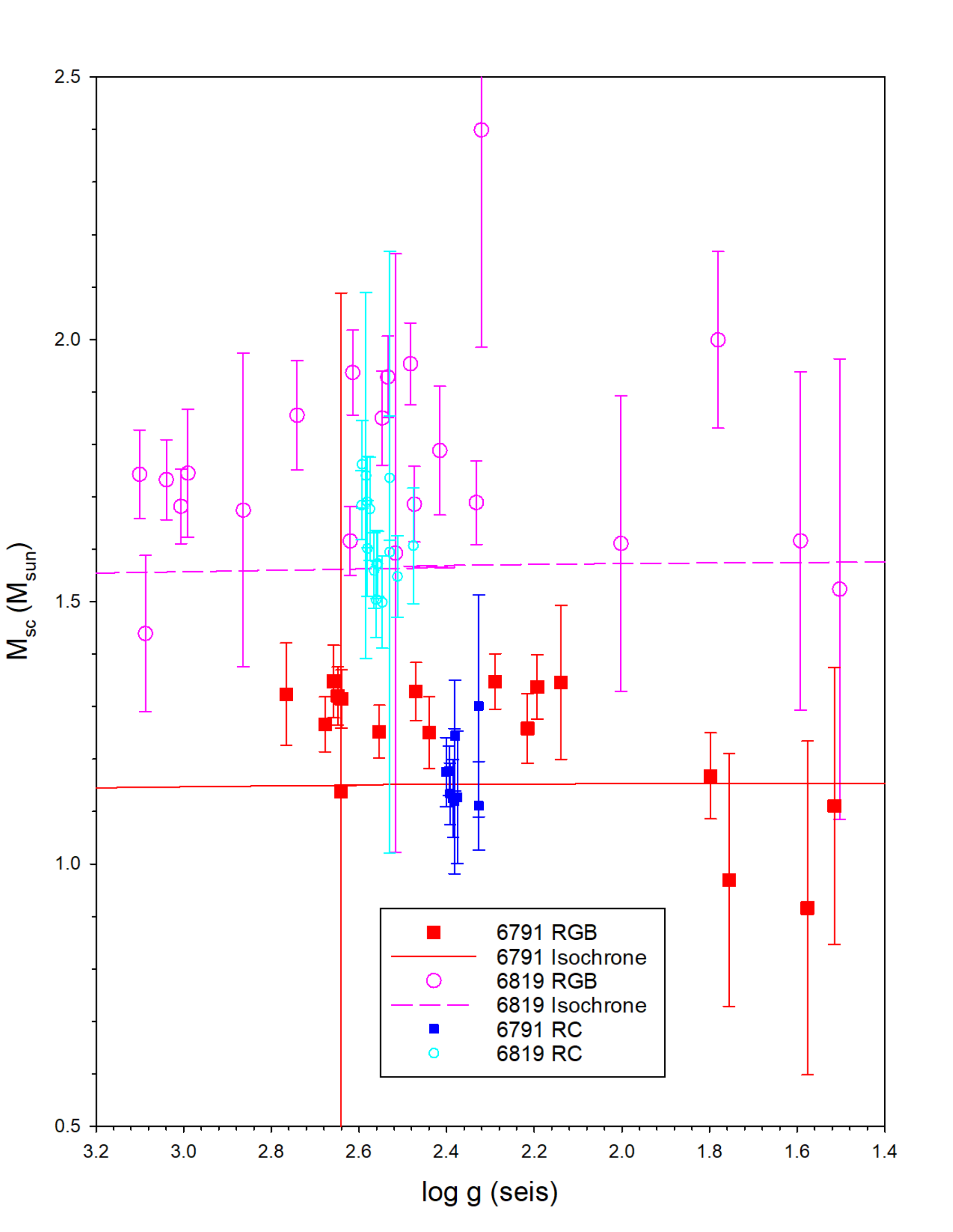}

\caption{Our open cluster samples in the log g - mass plane for NGC 6791 (red) and NGC 6819 (pink).  Masses are inferred from simple scaling relations for stars asteroseismically classified as RGB.  Surface gravities come from a combination of the asteroseismic $\nu_{\rm max}$ and the spectroscopic $\rm T_{eff}$.  The lines reflect the expectations from our adopted cluster parameters.}

\end{figure}

We then apply corrections to the $\Delta \nu$ scaling relation and solve for the best fit $f_{\nu_{\rm max}}$ values for both clusters.  We illustrate our resulting fits for the cluster stars in Figure 19. The mean masses for both systems that we would have derived using equations (14) and (15), and the associated random uncertainties, are given in Table 3, as are the $f_{\nu_{\rm max}}$ factors derived from equation (16). There are several features here worth discussing.  We can achieve concordance between the fundamental and asteroseismic mass scale with small adjustments to the $\nu_{\rm max}$ zero point.  If we combine the two clusters we obtain a final $f_{\nu_{\rm max}}$ factor of $1.009\pm 0.007$, corresponding to a reference "effective solar" $\nu_{\rm max}$ value of $3076\pm21 \mu$Hz.

\begin{figure}


\plotone{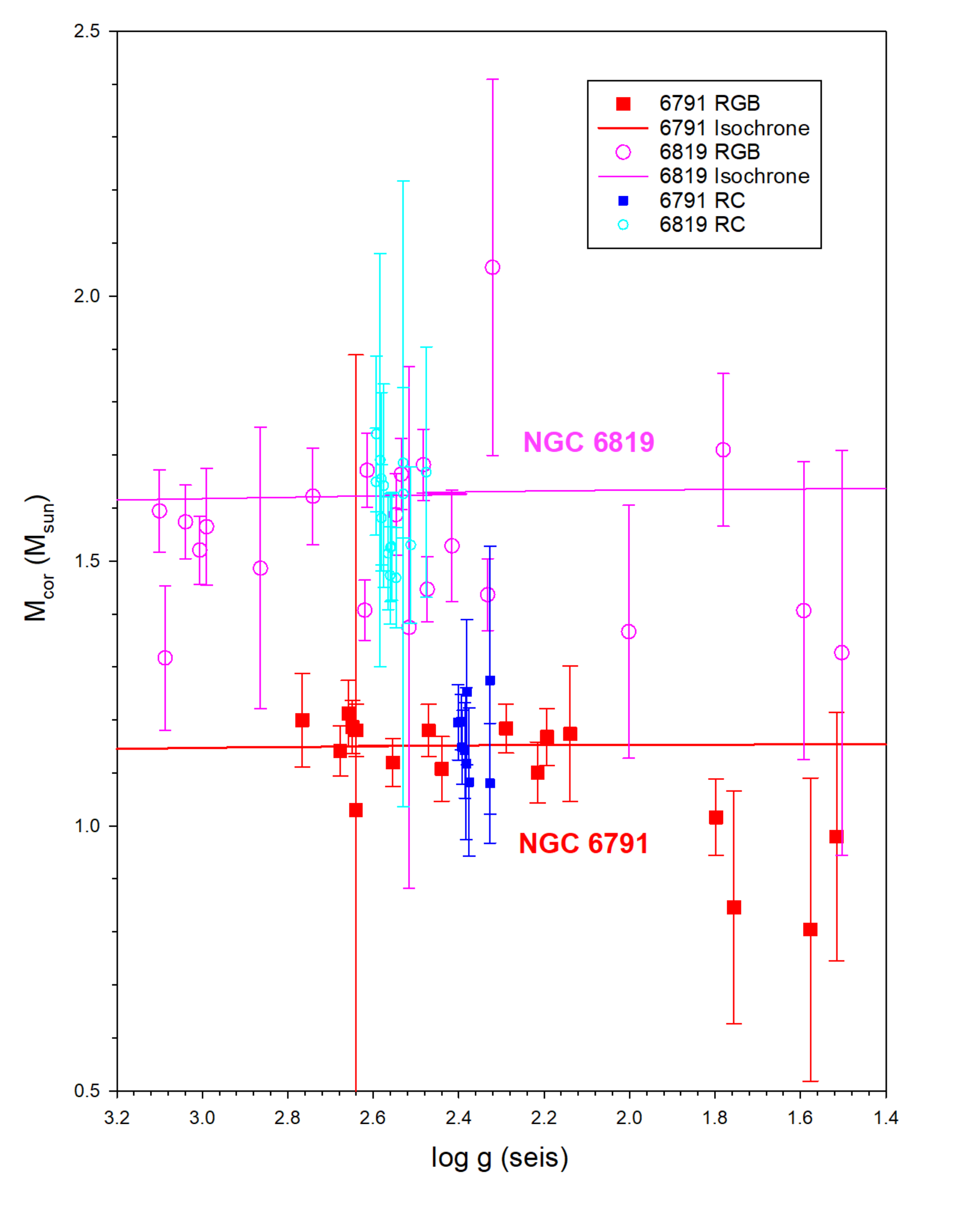}

\caption{Our open cluster samples for RGB stars in the log g - mass plane for NGC 6791 (a, red) and NGC 6819 (b, pink) using our calibrated parameters.  Masses include theoretically motivated corrections to the $\Delta \nu$ scaling relations and the indicated $\nu_{\rm max}$ zero point.  Surface gravities come from a combination of the asteroseismic $\nu_{\rm max}$ and the spectroscopic $\rm T_{eff}$.  The lines reflect the expectations from our adopted cluster parameters.}

\end{figure}

We can also use our data to check on trends with mass, metallicity, and surface gravity over a limited domain. The NGC 6819 and NGC 6791 mass ranges (1.55 and 1.15 solar masses respectively) roughly bracket the mass range for the majority of RGB stars in our field sample; there is no statistically significant evidence for a mass trend in the $\nu_{\rm max}$ relationship once the $f_{\Delta \nu}$ corrections have been applied to the $\Delta \nu$ scaling relationship.  Across most of the surface gravity domain (2.0 < log g < 3.2) we see no strong trends in the asteroseismic masses relative to expectations from isochrones.  This is significant because the predicted corrections are themselves functions of surface gravity, and if there was an error in the adopted functional form we could have seen strong residual trends.  However, as is visually apparent in Figures 18 and 19, the most luminous RGB stars (low surface gravity) have mass estimates systematically lower than the cluster mean, albeit with large random uncertainties.  This represents a possible caution about the scaling relations at low surface gravities, but it is only of marginal statistical significance. A similar result was obtained by \citet{mos13b} in M giants. 

The dispersion of the points around the mean predicted by our error model for both clusters is close to that seen in the real data (with a reduced $\chi^2$ per degree of freedom of 0.79 and 0.60 for NGC 6819 and NGC 6791 respectively). We take this as confirmation that our uncertainty estimates are conservative but reasonable relative to astrophysical expectations. In NGC 6819 in particular the higher mass scatter has been noted before and a potential explanation has been the presence of a large number of stars experiencing binary mass interactions, and we did have to remove several stars from our sample because of either highly discrepant masses or anomalous HR diagram position. However, the excess mass scatter in NGC 6819 is present \textit{for the majority of targets}, which may indicate that some outlier values may simply reflect the difficulty in inferring asteroseismic properties in this mass domain. A more detailed examination of the possibilities will be covered in \citet{zin18}.


\begin{deluxetable}{ccccccc}
\tabletypesize{\scriptsize}
\tablecaption{Open Cluster Global Properties For Mass Calibration}
\tablewidth{0pt}
\tablehead{
\colhead{Cluster} & \colhead{$M_{\rm fund}^{\rm cluster}$} & \colhead{[Fe/H]} & \colhead{$[\alpha/Fe]$} & \colhead{$M_{sc}^{\rm cluster}$} &
\colhead{$M_{\rm trial}^{\rm cluster}$} & \colhead{$f_{\nu_{\rm max}}$}}
\startdata
NGC 6791 & $1.15(2) M_{\sun}$ & 0.42 & 0.04 & 1.294(17) & 1.103(15) & 1.014(8)\\
NGC 6819 & $1.55(4) M_{\sun}$ & 0.11 & 0.00 & 1.771(22) & 1.582(20) & 0.993(14) \\
Mean & --- & --- & --- & --- & --- & 1.009(7) \\
\enddata

\tablecomments{The benchmark masses are derived from the sources described in the text, and the cluster abundances are the means of the DR14 values for the targets with asteroseismic measurements.  $M_{sc}^{\rm cluster}$ is the mean mass for RGB cluster members without any corrections to the scaling relations ($f_{\nu_{\rm max}}=f_{\Delta \nu}=1$). $M_{\rm trial}^{\rm cluster}$ is the mean mass for RGB cluster members including a correction to the $\Delta \nu$ scaling relation only ($f_{\nu_{\rm max}}=1$). The final column is the $f_{\nu_{\rm max}}$ factor needed to reproduce the calibrating mass for RGB members relative to an assumed solar $\nu_{\rm max}$ reference value of 3103.266 $\mu Hz$.  The bottom row gives the adopted mean $f_{\nu_{\rm max}}$ factor, corresponding to an effective solar $\nu_{\rm max}$ reference value of $3076\pm21 \mu$Hz.}


\end{deluxetable}

\subsection{Checks Against Other Fundamental Data}

Eclipsing binary systems with asteroseismic detections and well-characterized masses can also be used to test the mass scale. There are interesting samples of such stars \citep{gau16}. Unfortunately, the overlap between the EB sample and our data set is small, with only KIC 9970396 and KIC 10001167 being stars in common with precise measurements (KIC 4663613, has partial data with large parameter scatter and is not a useful calibration point). The fundamental radii are in good agreement with the radii derived using our method ($R_{cor}/R_{EB} = 1.005\pm 0.038$ and $1.013\pm 0.031$ respectively for KIC 9970396 and KIC 10001167.  The masses are in tension, $M_{cor}/M_{EB} = 1.162\pm 0.106$ and $1.092\pm 0.053$ respectively, but not with high statistical confidence. This finding is consistent with recent comparisons using other analysis methods by \citet{bro18}, and indicates either a caution on the overall mass scale or complications in the analysis of the eclipsing binary stars. 


A more significant concern is the validity of our asteroseismic stellar parameters for RC stars. We cannot assume a calibrating mass for core He-burning stars because of the possibility of significant mass loss on the first ascent red giant branch; in fact, the derived masses of such stars are a science result, not a calibration point \citep{m12}. Our methodology, when applied to RC stars in open clusters, yields masses equal to their RGB precursors (but with systematic errors large enough that we cannot rule out the possibility of either no mass loss or significant mass loss.) 

An alternate approach, using comparisons to theoretical models of the core He-burning phase, can also be employed to infer RC masses.  This approach when applied to NGC 6791 suggests moderate mass loss on the RGB, but there are significant model dependencies in the results \citep{an18}. We can, however, look at radii, which can be tested for both RGB and RC stars independent of the assumed mass scales.  

To test the radius scale, we take advantage of the fact that members of these star clusters have a small intrinsic range in relative distance to us.  As a result, we can infer the total luminosity of stars by a combination of their apparent brightnesses, the cluster extinction and distance modulus, and their bolometric corrections.  When combined with spectroscopic $\rm T_{eff}$, we can then infer a fundamental radius.  The weakness of this approach is that changes in the cluster distance and extinction cause correlated zero-point shifts in the radius scale, making this method uncompetitive for calibrating the scaling relations relative to using the (relatively precise) masses.  However, we can use the cluster radii to test the \textit{relative} corrections for red clump and red giant stars as described below.

Our V-band photometry is from \citet{bro12} for NGC 6791 and from \citet{kal01} for NGC 6819.  All of our targets have 2MASS JHK photometry.  To constrain the extinction we take advantage of the large number of APOGEE $\rm T_{eff}$ measurements that have been calibrated on the \citet{ghb09} IRFM system.  We derive a mean photometric $\rm T_{eff}$ value as the average of the value derived from the V-J, V-H and V-Ks reddening-corrected colors using the \citet{ghb09} color-temperature relationships.  Our extinction model uses equations (5)-(7) from \citet{an07} to derive color-dependent extinction terms, following the model of \citet{bes98}.  We then solve for the E(B-V) value required to bring the average difference between the two temperature scales to zero, and treat the dispersion between the values as a representative temperature error.  As indicated in Table 4, our derived extinctions (0.123 and 0.144 for NGC 6791 and NGC 6819 respectively) are in good agreement with literature values. There is some discussion of differential extinction in the literature, and we did check the impact of the proposed differences to our answer; they reduced the dispersion in temperature differences somewhat but did not impact our mean values. We adopt V-band bolometric corrections from \citet{flo96} with the erratum from \citet{tor10}, using the spectroscopic $\rm T_{eff}$ values from DR13.  In combination with an adopted cluster distance, we can then predict stellar luminosities and, from knowledge of the effective temperatures, radii.

\begin{deluxetable}{ccccccc}
\tabletypesize{\scriptsize}
\tablecaption{Open Cluster Global Properties For Radius Calibration}
\tablewidth{0pt}
\tablehead{
\colhead{Cluster} & \colhead{$(m-M)_{o,cmd}$} & \colhead{$(m-M)_{o,seis}$} & \colhead{$E(B-V)_{cmd}$} & \colhead{$E(B-V)_{sp+ph}$} & \colhead{$\frac{R_{RC}}{R_{RGB},sc}$} & \colhead{$\frac{R_{RC}}{R_{RGB},cor}$}} 
\startdata
6791 & $13.06(8)$ & $13.03(1)$ & $0.122(17)$ & $0.153(5)$ & $0.954(23)$ & $1.011(24)$ \\
6819 & $11.91(4)$ & $11.88(2)$ & $0.160(7)$ & $0.165(5)$ & $0.917(22)$ & $0.9741(23)$ \\
\enddata


\tablecomments{NGC 6791 distance and extinction are the average of \citet{an15} and \citet{bro12}. NGC 6819 distance modulus is the apparent distance modulus of $12.38\pm0.04$from \citet{bre16} modified by the extinction, taken from \citet{ant14} and $Rv=3.26$. The seismic distance is derived from requiring concordance between the asteroseismic radius and that derived from the combination of L, extinction and spectroscopic $\rm T_{eff}$.  The seismic extinction is derived from requiring concordance between the spectroscopic and photometric $\rm T_{eff}$. The final two columns are the ratios of RC to RGB radii with scaling relations alone (fifth column) and after corrections to the $\Delta \nu$ scaling relation (sixth column).}


\end{deluxetable}

Because our method is calibrated to reproduce known masses, concordance between these radii and those predicted by asteroseismology can be achieved with a suitable choice of the cluster distance.  This technique for inferring "asteroseismic distances" has been successfully used in the literature, for example by \citet{ste16}.
For our purposes, we adopt the $\nu_{\rm max}$ zero point derived from the mass constraints, and adjust the cluster distance modulus (m-M) such that the average radii inferred for RGB stars from the calibrated asteroseimic scaling relations were in agreement with the fundamental average radii inferred from our combination of photometric and spectroscopic constraints.  The inferred cluster distance moduli (13.03 and 11.88 for NGC 6791 and NGC 6819 respectively) are given in Table 4, and are well within the observational uncertainties.  We can then examine the concordance between the radii inferred for RC stars and those inferred for RGB ones; in the absence of corrections to the $\Delta \nu$ scaling relation, for example, \citet{m12} found that the relative radii of the two populations were discordant at the 5\% level.  The results are shown in Figure 20.  In panel a, we show the relative radii that we would have obtained without corrections to the $\Delta \nu$ scaling relation.  The RC radii are too small relative to the RGB radii, and the radius system overall is inflated relative to expectations.  Once $\Delta \nu$ corrections are applied, the RGB properties are significantly affected, while the RC properties change only minimally.  As a result, the RC radii are \textit{inflated} relative to the RGB radii.  This is tentative evidence that the $\Delta \nu$ corrections adopted here may be somewhat too large, although we caution that this result is not highly statistically significant.

\begin{figure}


\plotone{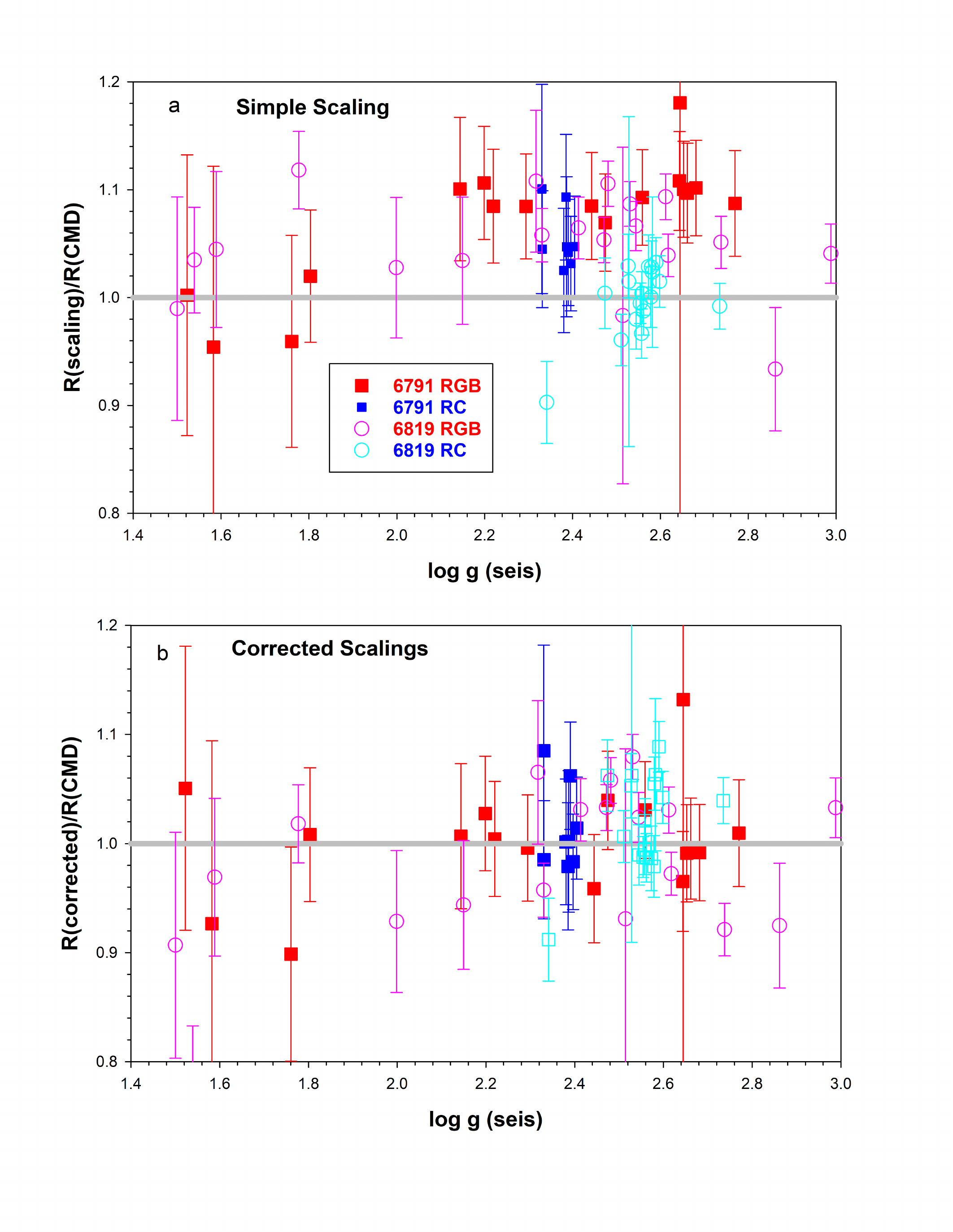}

\caption{The ratio of the asteroseismic to the fundamental radii of open cluster members as a function of the asteroseismic log g values.  In panel a (top), values are shown for uncorrected scaling relations ($f_{\nu_{\rm max}}=f_{\Delta \nu}=1$), while in panel b (bottom), we show the values for our calibrated parameters including theoretically motivated corrections to the $\Delta \nu$ scaling relations and the indicated $\nu_{\rm max}$ zero point.  Surface gravities come from a combination of the asteroseismic $\nu_{\rm max}$ and the spectroscopic $\rm T_{eff}$. RGB star properties are significantly changed with respect to RC properties by the corrections; RC radii are systematically below RGB stars in panel a and above the RGB stars in panel b.}

\end{figure}

An independent check on this result can be obtained using $\rm T_{eff}$ values and extinctions derived from SED fitting \citep{sta17, hub17, sta18}.  The results are consistent with those derived from spectroscopy, but with somewhat larger uncertainties (RC radii relative to RGB of 1.022 and 1.021 in NGC 6791 and 6819 respectively.)  This represents welcome confirmation of the overall results. 


\section{The Second APOKASC Catalog}

\subsection{How the Values Were Generated}
We generate our final stellar parameters as follows.

\begin{enumerate}

\item{We use the DR14 spectroscopic $\rm T_{eff}$, as described in \citet{dr14} and  \citet{hol18}.}

\item{We use measured asteroseismic evolutionary states where available.  If there is no consensus evolutionary state, we used spectroscopic evolutionary states from DR13.  In cases where the spectroscopic states were ambiguous, we derived asteroseismic parameters assuming that the target had both states and treated the range as an additional systematic error source.}

\item{The raw asteroseismic parameters from each pipeline with data are divided by the scale factors in Table 2, and these normalized values are then averaged (see Section 3.1.)}

\item{We adopt solar reference values of $135.146 \mu$Hz and $3076\pm 21 \mu$Hz for the $\Delta \nu$ and $\nu_{\rm max}$ scaling relations respectively, as discussed in Section 3. This choice incorporates the $f_{\nu_{\rm max}}$ factor as a correction to the average solar reference value for $\nu_{\rm max}$ in Table 1.}

\item{Fractional random errors for asteroseismic observables are taken as the larger of the normalized measurement dispersion $\sigma$(parameter)/<parameter> and our 'best case' uncertainties (0.009 and 0.004 for $\nu_{\rm max}$ and $\Delta \nu$ respectively). Random uncertainties in these observables and $f_{\Delta \nu}$ are propagated in quadrature to obtain the tabulated values.}

\item{Systematic fractional uncertainties are taken from Sections 3.1, 3.2, and 3.3. These include systematic shifts in derived quantities that would have resulted from adopting single pipeline values rather than the ensemble mean, discussed in 3.1.3; the measurement technique for $f_{\Delta \nu}$, discussed in Section 3.2; and the zero-point of the open cluster calibrating mass scale, discussed in Section 3.3, which induces a global uncertainty in $f_{\rm \nu_{max}}$.}

\item{We compute the star by star correction factors $f_{\Delta \nu}$ iteratively as discussed in Section 3.2.  In addition to the $\rm T_{eff}$ and asteroseismic parameters discussed above, this procedure also uses evolutionary state and the DR14 [Fe/H] and [$\alpha$/Fe] values.}

\item{We then infer asteroseismic mass, radius, mean density and surface gravity measurements using equations 1-6 combined with standard error propagation.}

\item{Ages are estimated using our derived mass, surface gravity, [Fe/H] and [$\alpha$/Fe] and uncertainties, while extinctions are computed using our data and additional photometry.}

\end{enumerate}

Our results are presented in Table 5, available electronically; a sample is reproduced here.  For the remainder of this section we present some comparisons of the current work to the prior catalog and briefly illustrate some of the main properties of the catalog itself. For some stars we were not able to obtain full results, and the reason is given in the Notes column.  As noted in Section 2, there were a number of stars for which we were unable to get asteroseimic parameters; they are flagged as No Seis. Our mass corrections require interpolation in a model grid, and for 25 targets we could not obtain internally consistent parameters; these are flagged as No Fdnu. We also flag stars with unusually large measurement uncertainties in their asteroseismic parameters (as SeisUnc) and those with formal mean ages greater than 14 Gyr (as AgeOld), although we do provide data for them.

\begin{deluxetable}{ccc}
\tabletypesize{\scriptsize}
\tablecaption{The APOKASC-2 Catalog of Stellar Properties}
\tablewidth{0pt}
\tablehead{
\colhead{Label} & \colhead{Source} & \colhead{Description}
}

\startdata
KIC & \textit{Kepler} Input Catalog \\
2MASS & 2MASS Catalog \\
Teff & DR14 & $\rm T_{eff}$ in K\\
S Teff  & DR14 & $\sigma(\rm T_{eff})$ in K\\
FeH & DR14 & [M/H] \\
S FeH & DR14 & $\sigma(\rm [Fe/H])$ \\
AFe & DR14 & $[\alpha/\rm Fe]$ \\
S AFe & DR14 & $\sigma([\alpha/\rm Fe])$\\
Nmax & This paper & $<\nu_{\rm max}>$, $\mu$Hz\\
S Nmax & This paper & $\sigma \frac{<\nu_{\rm max}>}{<\nu_{\rm max}>}$ \\
Dnu & This paper & $<\Delta \nu>$, $\mu$Hz\\
S Dnu & This paper & $\sigma \frac{<\Delta \nu>}{<\Delta \nu>}$ \\
ES & \citet{els17}, DR13 & Evolutionary State \tablenotemark{a} \\
Fdnu & This paper & $f_{\Delta \nu}$, $\Delta \nu$ correction factor\\
S Fdnu & This paper & $\sigma(f_{\Delta \nu})$ \\
M(cor) & This paper & Corrected Mass, $M_{\sun}$\\
S Mran & This paper & $\frac{\sigma_{\rm ran}(M)}{M}$ \\
S Msys & This paper & $\frac{\sigma_{\rm sys}(M)}{M}$\\
R(cor) & This paper & Corrected Radius, $R_{\sun}$\\
S Rran & This paper & $\frac{\sigma_{\rm ran}(R)}{R}$ \\
S Rsys & This paper & $\frac{\sigma_{\rm sys}(R)}{R}$ \\
logg(seis) & This paper & Log asteroseismic surface gravity \\
S Gran & This paper & $\frac{\sigma_{\rm ran}(log g)}{\rm log g}$ \\
S Gsys & This paper & $\frac{\sigma_{\rm sys}(log g)}{\rm log g}$ \\
<Rho> & This paper & $<\rho>$ in $g cm^{-3}$ \\
S Rhoran & This paper & $\frac{\sigma_{\rm ran}(<\rho>)}{<\rho>}$ \\
S Rhosys & This paper & $\frac{\sigma_{\rm sys}(<\rho>)}{<\rho>}$ \\
Log Age & This paper & Log age in Myr \\
S LogageP & This paper & $\sigma_{\rm ran}(log age)+$ \\
S LogageM & This paper & $\sigma_{\rm ran}(log age)-$ \\
Av & This paper & Extinction $A_V $($R_V = 3.1$) \\
S Av & This paper & $\sigma_{\rm ran} A_V$ \\
Notes & This paper & \tablenotemark{a} \\
\enddata


\tablecomments{Contents of the Main APOKASC-2 Data Table. Stars without parameters requiring asteroseismic inputs were cases where the time series data was analyzed but no asteroseismic parameters were returned.  The asteroseismic observables are derived as described in the text from the raw pipeline values.  The $\Delta \nu$ correction factor $f_{\Delta \nu}$ for each stat was derived iteratively from the combination of the spectroscopic, evolutionary state and asteroseismic data with the models of \citet{s18}. Masses, radii and uncertainties were derived from the combined asterseismic and spectroscopic data as described in the text. Ages were derived from the \citet{s18} models using the catalog mass, radius, [Fe/H] and [$\alpha$/Fe]. The procedure for the extinction map is described in the text.}

\tablenotetext{a}{Derived from asteroseismology if available. Otherwise, uses DR13 spectroscopic evolutionary states, noted by (s) after class. RGB is a shell-burning source only (first ascent red giant or asymptotic giant); RC is a star which has a core He-burning source; Amb(s) is a star where the spectroscopic state was ambiguous. Reject stars are ones with suspect or absent asteroseismic data, while Bad Teff stars did not have reliable DR14 effective temperatures.}

\tablenotetext{b}{The notes in the final column are included for stars whose reported properties are either incomplete or uncertain. No Seis flags all stars without reported reliable asteroseismic values (and, by extension, reported masses, radii, log g or mean density.) No Fdnu flags stars where we could not obtain a $f_{dnu}$ correction factor; the only derived quantity is the asteroseismic surface gravity. SeisUnc flags all stars with fractional uncertainties in $\nu_{\rm max} >0.05$ or $\Delta \nu$ > 0.025. AgeOld flags all stars with formal mean ages greater than 14 Gyr.}


\end{deluxetable}


\subsection{Catalog Properties and Comparisons with Prior Work}

In Figure 21 we compare the current asteroseismic results with values from our 2014 catalog for stars in common.  For the asteroseismic parameters, surface gravities are very well-behaved, and for the large majority of cases the changes in the inferred $\Delta \nu$ and $\nu_{\rm max}$ are small.  We see some excess scatter for RC stars and a handful of cases where the mean parameters changed significantly between the two efforts.

\begin{figure}


\plotone{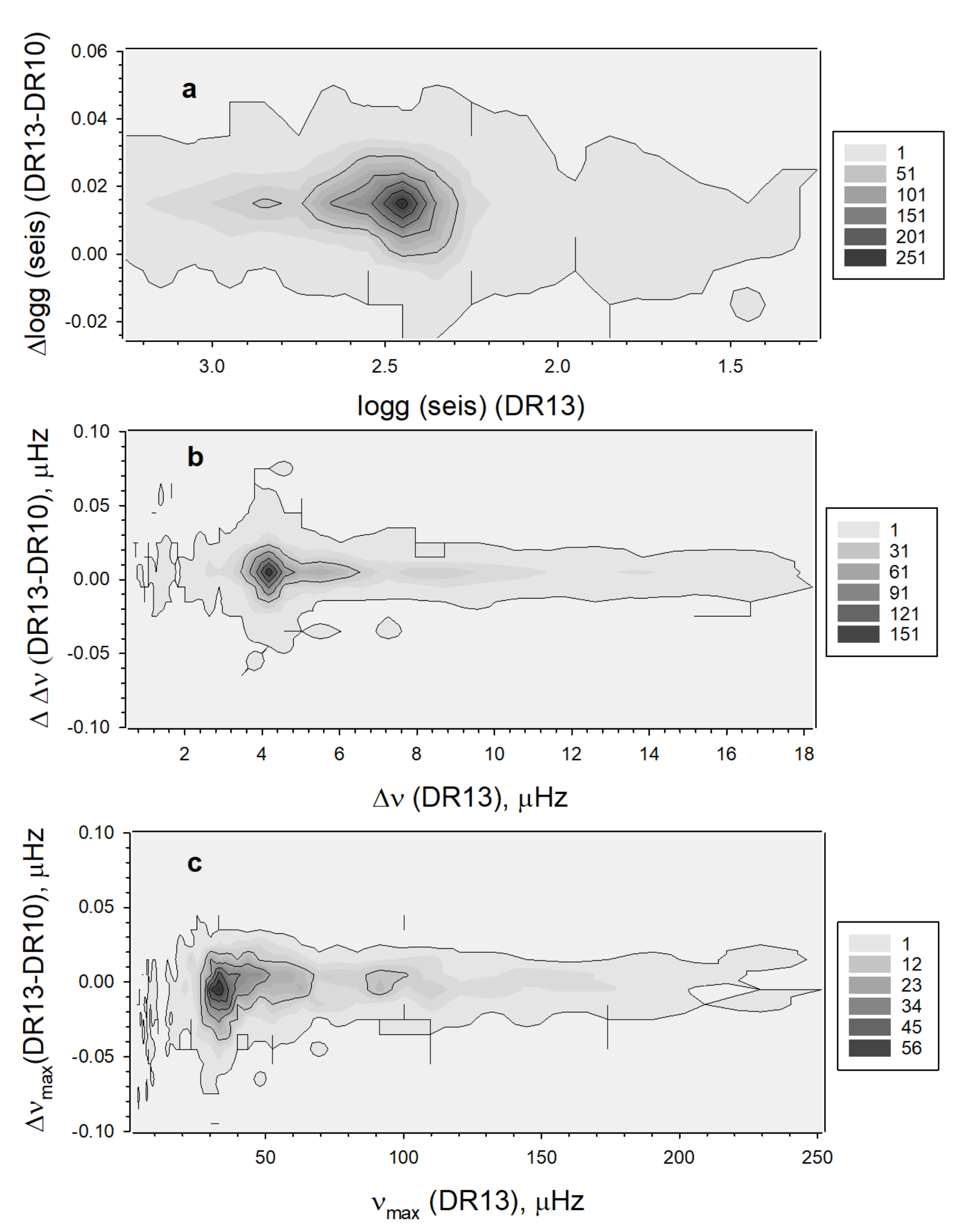}

\caption{Asteroseismic  properties in our 2014 catalog compared with the current values for stars in common between the two data sets. We compare log g in panel a, $\nu_{\rm max}$ in panel b, and $\Delta \nu$ in panel c.}

\end{figure}

Our distributions of uncertainties are illustrated in Figure 22.  The error distribution is skewed, with the high scatter cases corresponding to the high uncertainty tails in the measurements of the asteroseismic parameters.  As a result, median errors are a better guide to the overall performance in the catalog of our data.  The median random and systematic mass uncertainties for RGB stars are at the 4 percent level, while the corresponding median uncertainties for RC stars are higher (at 9 and 8 percent respectively.)

\begin{figure}


\plotone{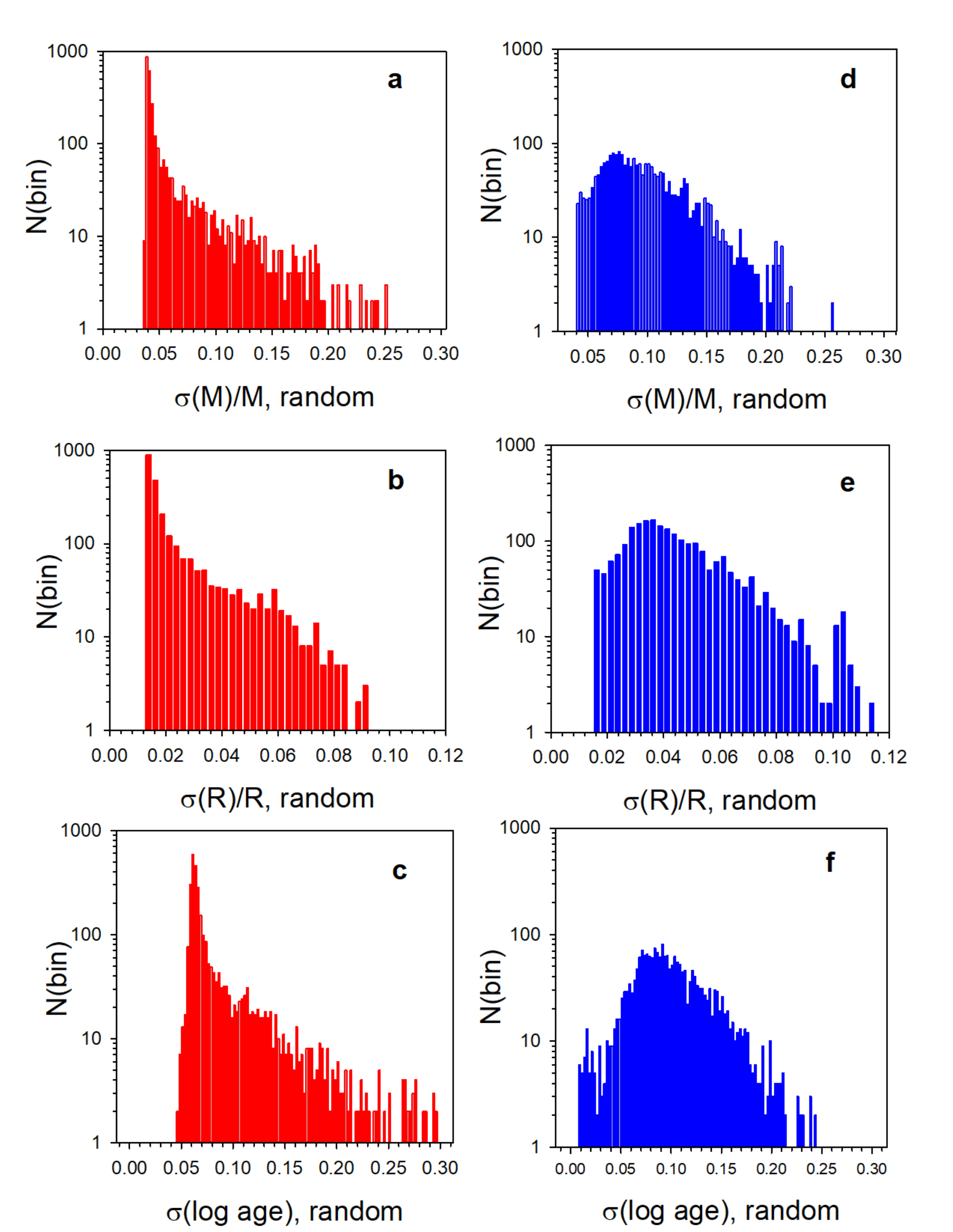}

\caption{Distribution of random uncertainties for RGB stars (panels a to c, left) and RC stars (panels d to f, right. Fractional mass uncertainties (a, d), fractional radius uncertainties (b,e) and logarithmic age uncertainties (c,f) are shown.}

\end{figure}

In Figure 23, we also present two slices of the full data set to illustrate the mass and abundance trends within our sample in a manner analogous to the APOKASC-1 presentation.  We selected all RGB stars with -0.1 < [Fe/H] < +0.1  and color-coded their position in a Kiel diagram as a function of corrected asteroseismic mass on the left panel. In the right panel, we selected all RGB stars with $1.1 M_{\sun} < M < 1.3 M_{\sun}$ and color-coded their position in a Kiel diagram as a function of metallicity.  Relative to APOKASC-1, we can see that these diagrams are now much more richly sampled.

\begin{figure}


\plottwo{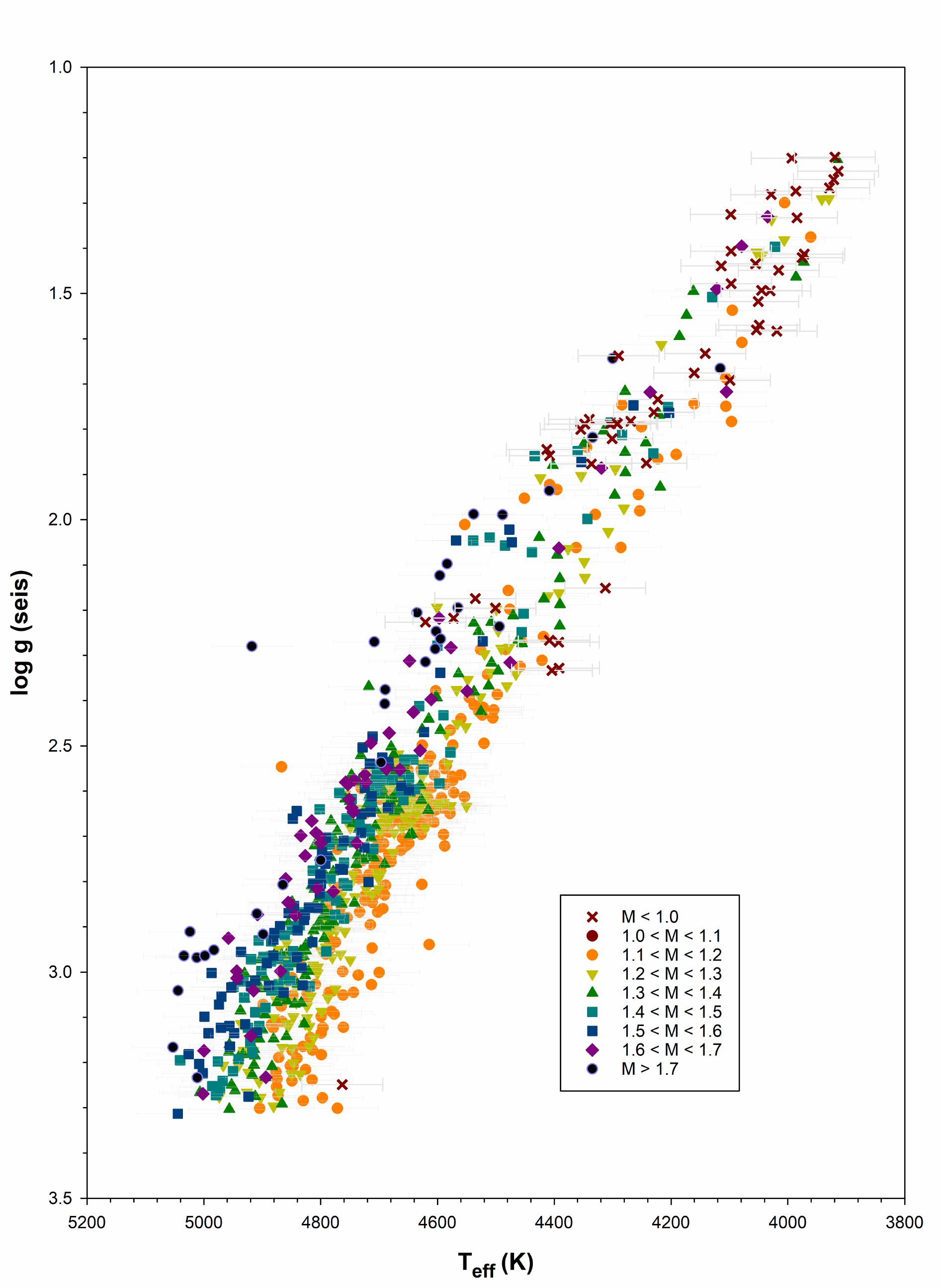}{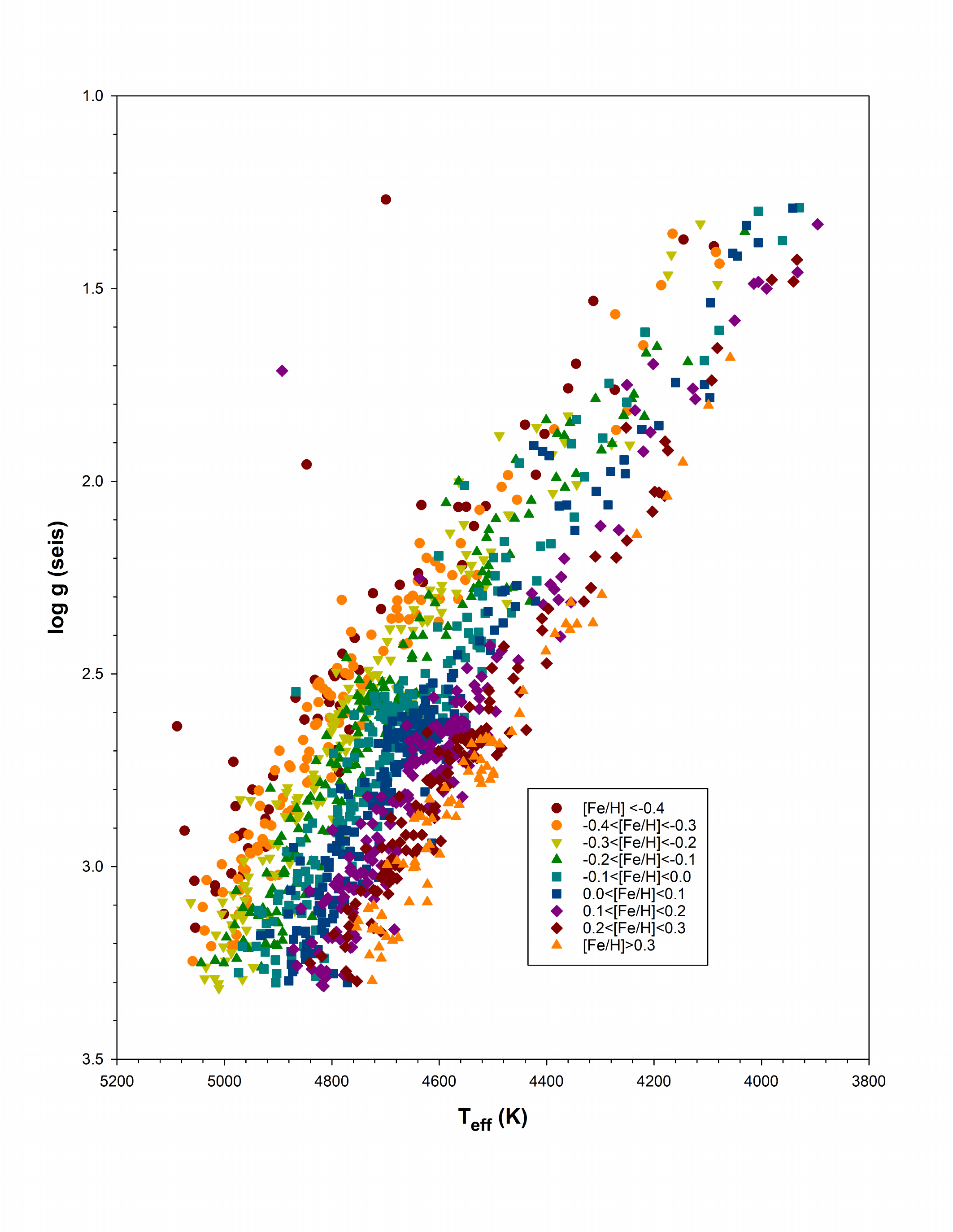}

\caption{Mass trends (left) and metallicity trends (right) in our RGB sample are illustrated here. Mass data is for stars with DR14 [Fe/H] values between -0.1 and +0.1, while metallicity data is for stars with asteroseismic masses between 1.1 and 1.3 $M_{\sun}$.}

\end{figure}

Our data also show distinct mass patterns for different populations, and strong correlations between the surface [C/N] and masses.  To illustrate these points, we have subdivided our sample into two cohorts: stars with [$\alpha$/Fe]>+0.12 and those with [$\alpha$/Fe]<+0.12. We show the density of points in RGB stars for both cohorts as a function of mass and metallicity in Figure 24.  The $\alpha$-rich stars are generally low mass and relatively metal-poor, but there is a clear population of high-mass stars seen in our sample.  The $\alpha$-poor stars, by contrast, show a broad distribution in mass and metallicity, with weak (if any) mean mass-metallicity relationship seen.

\begin{figure}


\plotone{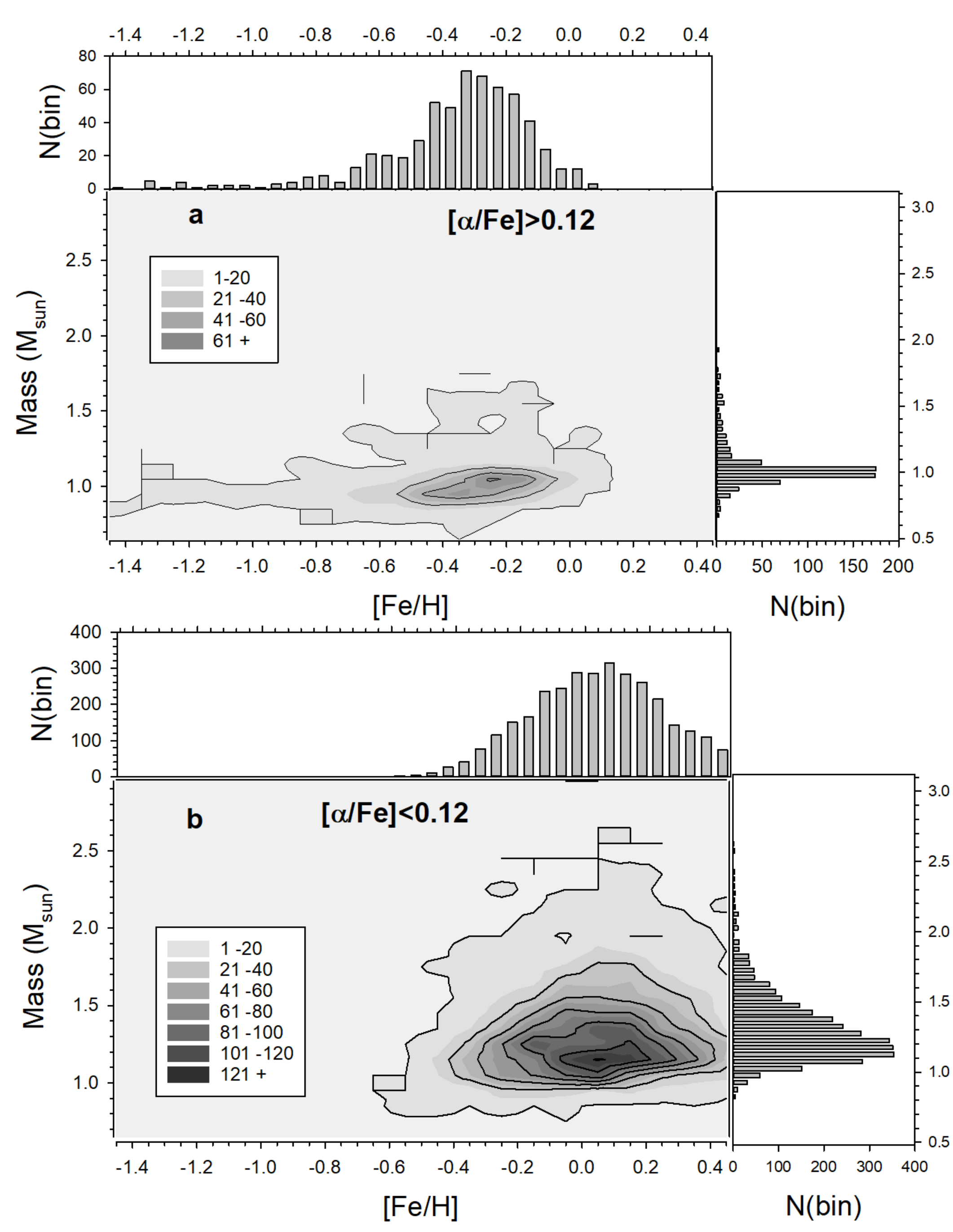}

\caption{Mass-metallicity trends in our RGB sample are illustrated here for $\alpha$-rich stars (panel a) and $\alpha$-poor stars (panel b). The color reflects the density of points within a bin. Distinct populations can clearly be seen in these chemical abundance groups.}

\end{figure}

The surface [C/N] is correlated with mass, and by extension age.  In metal-poor stars there is extra mixing on the RGB which complicates the interpretation of the [C/N] ratio, so we plot the [C/N]-mass relationship for $\alpha$-poor stars in our sample in Figure 25.  Interestingly, there is a strong trend in both the RC and the RGB sample, with a flattening of the slope at high mass (or young age.)

\begin{figure}


\plotone{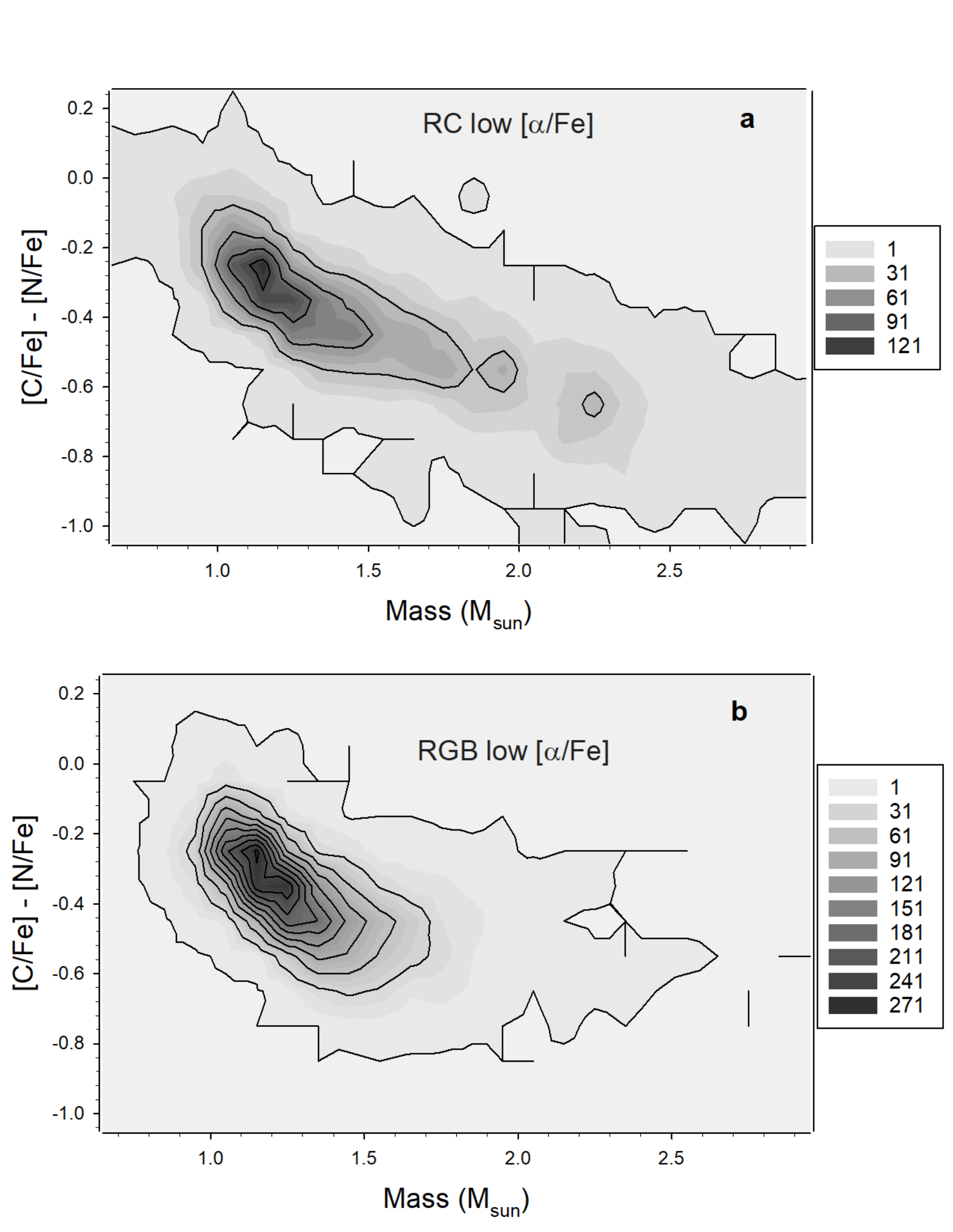}

\caption{Mass-[C/N] trends in our RC sample (panel a) and RGB sample (panel b). Only stars with [$\alpha$/Fe]<+0.12 are shown. The color reflects the density of points within a bin. Even before accounting for correlations with [Fe/H], strong trends with mass are clearly visible.}

\end{figure}

\subsection{Age and Extinction Estimates}

As a byproduct of our catalog we also present ages and extinction values for our targets, as described below.  The central values for ages were inferred by interpolation in the BeSPP grid at the input mass, surface gravity, [Fe/H] and [$\alpha$/Fe] presented in Table 5, and uncertainties around the mean were computed using the tabulated uncertainties.  We stress that the age uncertainties are random only and reflect neither systematic errors in the inputs nor systematic errors in the theoretical age inferences.  For details on the adopted input physics see \citet{s18}. 

Extinctions are derived using the direct method of stellar parameter estimation as implemented in the PARAM code \citep{das06,rod14}.
As a first step, the code estimates the intrinsic luminosity from knowledge of $R_{sc}$ and $\rm T_{eff}$. The luminosity is then transformed into absolute magnitudes in several filters using bolometric corrections inferred from the library of ATLAS 9 synthetic spectra from \citet{cas03} 
at the spectroscopic $\rm T_{eff}$ and metallicity and the asteroseismic surface gravity. Probability density
functions (PDF) of the absolute magnitudes are then generated. 

As a second step, the absolute magnitude PDFs are combined with apparent magnitude PDFs, resulting in a joint PDF of
the apparent distance modulus. We then assume a single interstellar extinction
curve \citep{ccm89}, with $R_V$ = 3.1, 
and infer the PDF for the true distance modulus as as function of $A_V$ for all available passbands. Our best-fitting $A_V$ and its uncertainty are inferred from requiring consistency between values for different filters. For a more detailed description, see \citet{rod14}.

The apparent magnitudes adopted are SDSS griz as measured by the KIC
team \citep{bro11} and corrected by \citet{pin12}; JHKs from 2MASS \citep{skr06}; and WISE photometry from \citet{wri10}.

The median uncertainty in $A_V$ is 0.08 mag. Only 3 percent of the stars
have uncertainties greater than 0.2 mag, typically because they have data from fewer filters. Since this is a statistical method, a small fraction of negative extinction values exist and are expected \citep{rod14}.  In our sample, 
approximately 2 percent of the stars have their
PDFs better matched with slightly negative values of extinctions. For these stars we adopt $A_V$ = 0.0. Figure 26 shows
our extinction map.

\begin{figure}


\plotone{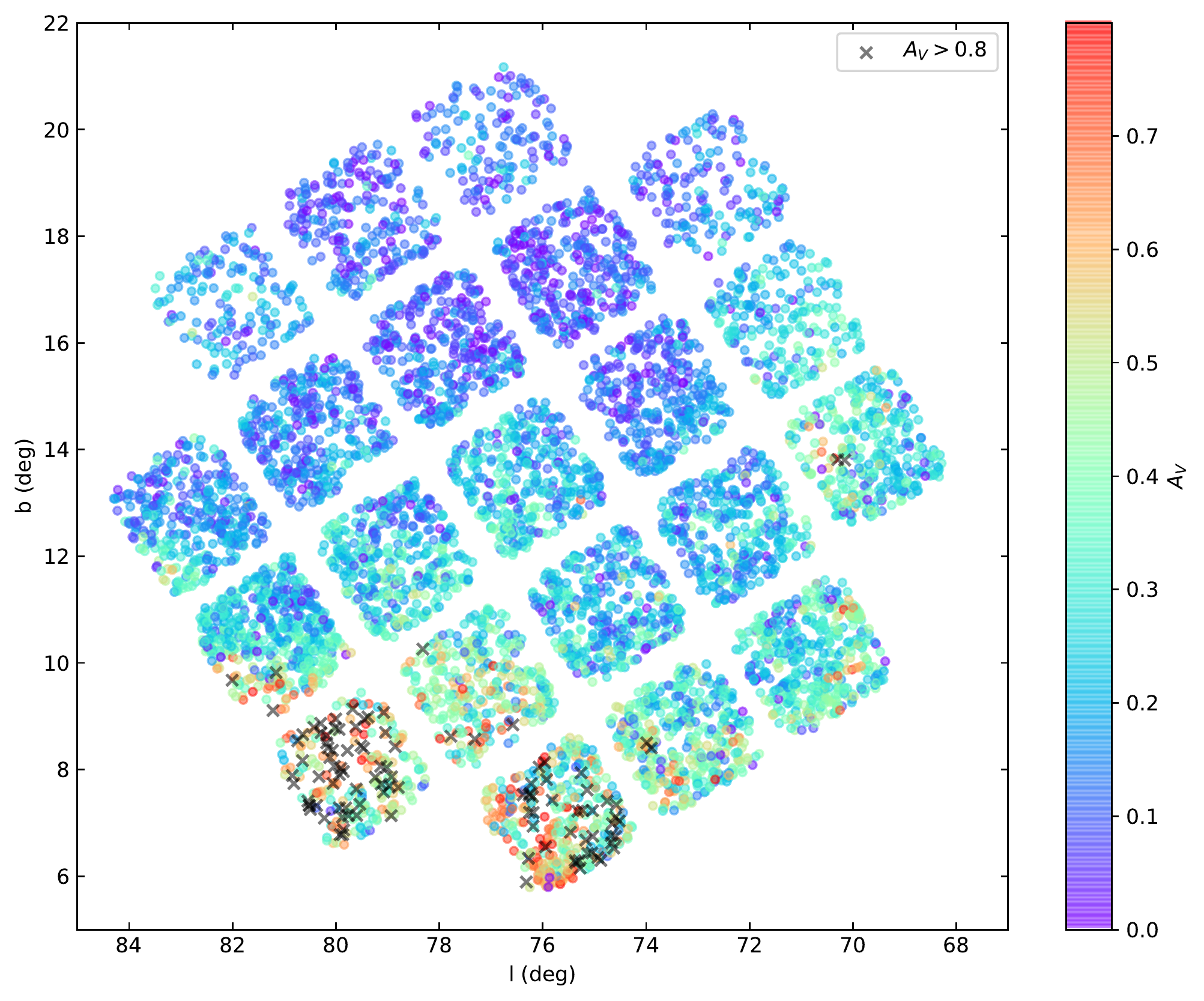}

\caption{Extinction map for our full sample.  Stars with extinctions greater than 0.8 mag are indicated by black crosses.}

\end{figure}

\section{Summary}

The potential for asteroseismology in stellar population studies has been clear.  In this paper we have made substantial progress towards realizing this potential.  At the same time, our method has some limitations and areas where the results need to used cautiously. Here we address both domains.

This catalog is not the first paper reporting asteroseismic estimates of stellar masses, ages and radii. Prior studies, including our own, adopted a forward-modeling approach to the problem: take a set of seismic observables and reference values, add in spectroscopic data, and produce stellar parameters such as mass, radius and age. However, these values were basically delivered "as is", rather than having them be tied to a fundamental scale. A key new step taken in this paper is to require that both the parameters themselves and their uncertainties be calibrated and tested against fundamental data. This is similar in spirit to the methods used for solar models and stellar isochrones.

Unlike \citet{p14}, the asteroseismic parameters described in this paper are calibrated on an absolute scale relative to benchmark stars in open clusters, and we have explicitly adopted theoretically motivated corrections to the $\Delta \nu$ scaling relation.  These two changes go a long way towards addressing the discrepancies between fundamental stellar properties and asteroseismic ones that have been identified in the literature.  In fact, using corrections to $\Delta \nu$ in a large grid, citet{rod17} found good agreement with NGC 6819, and a similar level of agreement with peak-bagging measurements of individual frequencies was found by \citet{han17}. Our zero-point for $\nu_{\rm max}$ is similar to that inferred by \citet{sha16} as well. 

Another advance concerns the interpretation of the asteroseismic measurements themselves. Using multiple analysis techniques allows us to perform outlier rejection, test measurement uncertainties, and explicitly separate out systematic and random effects in a calibrated framework. This is particularly helpful for the error model; with only the individual pipeline results, it can be challenging to quantify how well asteroseismic parameters are actually measured. We also allow the data, rather than theoretical priors, to set the relative zero points for different methods.  In this exercise we have also identified systematic differences that will have to be understood and resolved to achieve higher precision in mass and age. As a concrete example, excess mass scatter in NGC 6819 RGB stars has been noted in the literature \citep{rod17, han17}; with our method, we can see that at least one component is a significant measurement scatter. By contrast, we predict (and observe) a much smaller dispersion in mass in the open cluster NGC 6791. 

However, we still see clear evidence for areas needing improvement.  There are lingering differences between the masses presented here for very metal-poor stars and astrophysical expectations, in the sense that the inferred masses of halo stars are too high.  The magnitude of the offsets is reduced relative to the results reported in \citet{eps14}, however, which is an encouraging sign. Our systematic RC mass uncertainties are large; in fact, large enough that they have a substantive impact on our ability to infer mass loss between the RC and the RGB. We also see tentative evidence that masses for luminous solar-abundance RGB stars are under-estimated.  We therefore urge caution when employing this data for low surface gravity and metallicity, where we lack direct calibrators.  We have only limited constraints on the functional form of corrections to the $\Delta \nu$ scaling relation as well, and different methods have been proposed in the literature; even within a calibrated framework, this could yield systematic changes that are not captured by our current approach.  Similarly, identifying the origin of the method-dependent differences in asteroseismic measurements is an important task and could yield more precise relative and absolute stellar parameters.

The  empirical approach here does not include grid-modeling constraints from stellar tracks, and as a result there are objects included here with implausible combinations of mass, log g, $\rm T_{eff}$, and abundance; the approach in \citet{s18} does a much better job of controlling for such anomalies, at the cost of a dependence on the soundness of the underlying isochrones.


The most significant cautions about the usage of the current data are tied to the derived properties of red clump stars and selection effects in the underlying sample. There are significant method-dependent offsets in the asteroseismic measurements in red clump stars relative to first ascent red giants, and these effects are (if anything) amplified when grid modeling constraints are included.  When combined with a lack of direct calibrators for their masses, we cannot rule out the possibility of substantive zero-point shifts and stretches in the relative properties of such stars.  

As documented in \citet{p14}, the selection function for our sample is quite complex, and as a result great care should be used before treating our mass distribution as characteristic of the underlying population.  An effort in preparation, including a more unbiased sample of spectroscopic and asteroseismic data, will help alleviate the selection effect problem.  The forthcoming availability of precise parallax constraints from Gaia may help provide calibrators, at least for radius, in red clump stars (in conjunction with work on the extinction, bolometric correction, and effective temperature side.)  Finally, improvements in the precision and accurary of the spectroscopic parameters are also a realistic prospect.

We close on a note of optimism: although the emerging picture is more complex than the simple adoption of scaling relations, we believe that it is now clear that asteroseismic masses (and associated ages) are astrophysically well-motivated, and that employing them for stellar population studies has a bright future.

\acknowledgments
MHP, JAJ, JT and JZ acknowledge support from NASA grants NNX17AJ40G and NNX15AF13G. DA acknowledges support provided by the National Research Foundation of Korea (No. 2017R1A5A1070354). EC is funded by the European Union's Horizon 2020 research and innovation program under the Marie Sklodowska-Curie grant agreement No.664931. YE acknowledges the support of the UK Science and  Technology Facilities Council (STFC). K.C and V.S. acknowledge that their work here is supported in part by NASA under grant 16-XRP16 2-0004, issued through the Astrophysics Division of the Science Mission Directorate.
R.A.G acknowledges support from CNES.
A.G.H. acknowledges support from CNESDAGH and OZ acknowledge support provided by the Spanish Ministry of Economy and Competitiveness (MINECO) under grants AYA-2014-58082-P and AYA-2017-88254-P. LG and TR acknowledge support from PRIN INAF 2014 – CRA1.05.01.94.05. D.H. acknowledges support by NASA under Grant NNX14AB92G issued through the Kepler Participating Scientist Program. HJ acknowledges support from the Crafoord Foundation and Stiftelsen Olle Engkvist Byggm\"astare. SK acknowledges support from the European Research Council under the European Community's Seventh Framework Programme (FP7/2007-2013) / ERC grant agreement no 338251 (StellarAges). SM acknowledges support from NASA grant NNX15AF13G and NSF grant AST-1411685 and the Ramon y Cajal fellowship number RYC-2015-17697. SzM has been supported by the Premium Postdoctoral Research Program of the Hungarian Academy of Sciences, and by the Hungarian NKFI Grants K-119517 of the Hungarian National Research, Development and Innovation Office. D.S is the recipient of an Australian Research Council Future Fellowship (project number FT1400147). T.C.B. acknowledges partial support from Grant No. PHY-1430152 (Physics Frontier Center/JINA Center for the Evolution of the Elements), awarded by the U.S. National Science Foundation.

Funding for the Sloan Digital Sky Survey IV has been provided by the Alfred P. Sloan Foundation, the U.S. Department of Energy Office of Science, and the Participating Institutions. SDSS-IV acknowledges
support and resources from the Center for High-Performance Computing at
the University of Utah. The SDSS web site is www.sdss.org.

SDSS-IV is managed by the Astrophysical Research Consortium for the 
Participating Institutions of the SDSS Collaboration including the 
Brazilian Participation Group, the Carnegie Institution for Science, 
Carnegie Mellon University, the Chilean Participation Group, the French Participation Group, Harvard-Smithsonian Center for Astrophysics, 
Instituto de Astrof\'isica de Canarias, The Johns Hopkins University, 
Kavli Institute for the Physics and Mathematics of the Universe (IPMU) / 
University of Tokyo, Lawrence Berkeley National Laboratory, 
Leibniz Institut f\"ur Astrophysik Potsdam (AIP),  
Max-Planck-Institut f\"ur Astronomie (MPIA Heidelberg), 
Max-Planck-Institut f\"ur Astrophysik (MPA Garching), 
Max-Planck-Institut f\"ur Extraterrestrische Physik (MPE), 
National Astronomical Observatories of China, New Mexico State University, 
New York University, University of Notre Dame, 
Observat\'ario Nacional / MCTI, The Ohio State University, 
Pennsylvania State University, Shanghai Astronomical Observatory, 
United Kingdom Participation Group,
Universidad Nacional Aut\'onoma de M\'exico, University of Arizona, 
University of Colorado Boulder, University of Oxford, University of Portsmouth, 
University of Utah, University of Virginia, University of Washington, University of Wisconsin, 
Vanderbilt University, and Yale University.

\facilities{Sloan, Kepler} 

\appendix

Our main catalog uses asteroseismic measurements from five distinct pipelines, and these values were combined as described in Section 3.1; the averaged measurements and their computed uncertainties were given in Table 5.  In Table 6 we present the individual pipeline values used to compute the means and uncertainties.

\begin{deluxetable}{cc}
\tabletypesize{\scriptsize}
\tablecaption{Individual Asteroseismic Pipeline Measurements}
\tablewidth{0pt}
\tablehead{
\colhead{Label} & \colhead{Description}
}
																				
\startdata
KEPLER ID & \textit{Kepler} Input Catalog ID \\
A2Z NUMAX & A2Z Frequency of Maximum Power, $\mu$Hz \\
A2Z NUMAXERR & A2Z Frequency of Maximum Power Uncertainty, $\mu$Hz \\
A2Z DELTANU & A2Z Large Frequency Spacing, $\mu$Hz \\
A2Z DELTANUERR & A2Z Large Frequency Spacing Uncertainty, $\mu$Hz \\
CAN NUMAX & CAN Frequency of Maximum Power, $\mu$Hz \\
CAN NUMAXERR & CAN Frequency of Maximum Power Uncertainty, $\mu$Hz \\
CAN DELTANU & CAN Large Frequency Spacing, $\mu$Hz \\
CAN DELTANUERR & CAN Large Frequency Spacing Uncertainty, $\mu$Hz \\
COR NUMAX & COR Frequency of Maximum Power, $\mu$Hz \\
COR NUMAXERR & COR Frequency of Maximum Power Uncertainty, $\mu$Hz\\
COR DELTANU & COR Large Frequency Spacing, $\mu$Hz \\
COR DELTANUERR & COR Large Frequency Spacing Uncertainty, $\mu$Hz \\
OCT NUMAX & OCT Frequency of Maximum Power, $\mu$Hz \\
OCT NUMAXERR & OCT Frequency of Maximum Power Uncertainty, $\mu$Hz \\
OCT DELTANU & OCT Large Frequency Spacing, $\mu$Hz \\
OCT DELTANUERR & OCT Large Frequency Spacing Uncertainty, $\mu$Hz \\
SYD NUMAX & SYD Frequency of Maximum Power, $\mu$Hz \\
SYD NUMAXERR & SYD Frequency of Maximum Power Uncertainty, $\mu$Hz \\
SYD DELTANU & SYD Large Frequency Spacing, $\mu$Hz \\
SYD DELTANUERR & SYD Large Frequency Spacing Uncertainty, $\mu$Hz \\
\enddata


\tablecomments{individual pipeline values used in our analysis. The formal uncertainties returned by each analysis method are also given.}



\end{deluxetable}



\end{document}